\documentclass[varenna]{cimento}
\usepackage{epsfig}
\usepackage{graphicx}
\usepackage{amsmath}
\usepackage{amsbsy}
\usepackage{amssymb}
\usepackage{amsfonts}
\usepackage{dsfont}

\def\rank{\mathop\mathrm{rank}}
\newcommand\tr{\mathop\mathrm{tr}\nolimits}
\newcommand{\bm}{\mathbf}
\newcommand{\tm}{\mathrm}
\newcommand{\hs}{\hspace}

\newcommand{\bmg}[1]{\makebox{\boldmath $#1$}}

\title{Quantum Computing with Electron Spins in Quantum Dots}

\author{Robert~Andrzej \.Zak, Beat R\"othlisberger, 
        Stefano Chesi, \atque Daniel Loss}

\institute{Department of Physics, University of Basel\\
Klingelbergstrasse 82, CH-4056 Basel, Switzerland}

\begin{document}

\maketitle

\begin{abstract}
Several topics on the implementation of spin qubits in quantum dots are reviewed. We first provide an introduction to the standard model of quantum computing and the basic criteria for its realization. Other alternative formulations such as measurement-based and adiabatic quantum computing are briefly discussed. We then focus on spin qubits in single and double GaAs electron quantum dots and review recent experimental achievements with respect to initialization, coherent manipulation and readout of the spin states. We extensively discuss the problem of decoherence in this system, with particular emphasis on its theoretical treatment and possible ways to overcome it. 
\end{abstract}

\section{Introduction}\label{sec:introduction}

It was in the 1980s, when the idea of exploiting quantum degrees of
freedom for information processing was envisioned. The central
question at the time was whether and how it was possible to simulate
(efficiently) any finite physical system with a man-made machine.
Deutsch \cite{Deutsch1985} argued that such a simulation is not
possible perfectly within the classical computational framework that
had been developed for decades. He suggested, together with
other researchers such as Feynman \cite{Feynman1982, Feynman1986},
that the universal computing machine should be of quantum nature,
\emph{i.e.}, a quantum computer.

Around the same time, developments in two different areas of
research and industry took a tremendous influence on the advent of
quantum computing. On the one hand, it was experimentally confirmed
\cite{Aspect1982a} that Nature indeed does possess some peculiar
non-local aspects which were heavily debated since the early days of
quantum mechanics \cite{Einstein1935a}. Schr\"odinger
\cite{Schrodinger1935} coined the term `entanglement', comprising
the apparent possibility for faraway parties to observe highly
correlated measurement results as a consequence of the global and
instantaneous collapse of the wave function according to the
Copenhagen interpretation of quantum mechanics. The existence of
entanglement is crucial for many quantum computations. On the other
hand, the booming computer industry led to major progress in
semiconductor and laser technology, a prerequisite for the
possibility to fabricate, address and manipulate single quantum
systems, as needed in a quantum computer.

As the emerging fields of quantum information and nanotechnology
inspired and motivated each other in various ways, and are still
doing so today more than ever, many interesting results have been
obtained so far, some of them we are about to review in this work.
While the theories of quantum complexity and entanglement are being
established (a process which is far from being complete) and fast
quantum algorithms for classically difficult problems have been
discovered, the control and manipulation of single quantum systems
is now experimental reality. There are various systems that may be
employed as qubits in a quantum computer, \emph{i.e.}, the basic unit of
quantum information. Here, we will focus on the idea of using the
spins of electrons confined in quantum dots, a proposal made in 1997 \cite{Loss1998}. 
Most knowledge for realizing qubits in the fashion of the spin-qubit proposal of ref.~\cite{Loss1998} 
has so far been obtained for quantum dots formed in a two-dimensional electron gas
at the interface of a GaAs/AlGaAs heterostructure. Therefore, this
is the main type of quantum dot we will turn our attention to. We would
nevertheless like to mention that many other systems are
investigated intensively at present, such as carbon nanotubes
\cite{Bulaev2008}, nanowires \cite{Trif2008}, molecular magnets
\cite{Leuenberger2001, Meier2003, Lehmann2007, Trif2008a}, quantum
dots in graphene \cite{Trauzettel2007}, and nitrogen-vacancy centers in diamond
\cite{Childress2006, Awschalom2007, Dutt2007, Neumann2008}.

The idea behind this work is to review quantum computing starting at
its very roots, and ending at the current state of knowledge of one
of its many branches, here being the spin-qubit proposal of ref.~\cite{Loss1998} for
electron spins in GaAs quantum dots. We approach the subject in the
second section by beginning with classical computing and complexity
theory, mainly to see the concepts that inspired the so-called
standard model of quantum computing discussed later, and to
encounter examples of problems that are assumed to be too involved
to ever be solved in reasonable time on a classical computer. We
then review a set of requirements that every quantum computing
proposal should fulfill in order to be usable on a large scale
\cite{DiVincenzo2000}. After discussing the actual spin-qubit
proposal of ref.~\cite{Loss1998}, we end the chapter with some recent alternative models of
quantum computing and a section about entanglement measures and
their numerical evaluation.

While the second section is quite general, we focus in the third
section on spin manipulation in GaAs quantum dots. We discuss the
latest experimental achievements and we will see that it recently became possible
to define, initialize, manipulate, and read out single electron spin 
qubits with an already quite remarkable rate of success.

Although anticipated to some extent in the third section, the last
section intensively discusses the various mechanisms of decoherence and
possible ways to reduce its effects, which is a necessary prerequisite
for large-scale quantum computing. In greater
detail, we investigate the role of the spin-orbit and the hyperfine
interaction. We show how these mechanisms cause relaxation of the
electron spin, but also how they can be exploited to perform such
beneficial tasks as all-electrical single spin manipulation or full
polarization of the quantum dot's nuclear bath.

\section{Quantum computing in a nutshell}\label{sec:qc_intro}
\subsection{Classical computers and complexity theory}\label{sec:classical
computers}

Computers are devices that solve problems in an algorithmic fashion.
The Church-Turing hypothesis \cite{Church1936, Turing1936} claims
that every function which we would naturally regard as being
computable by an algorithm (\emph{i.e.}, a procedure that solves a given
problem in finite time) can be computed by the universal Turing
machine (\emph{i.e.}, a hypothetical, mathematically formalized,
programmable discrete machine). Fortunately, the rather cumbersome
notion of a Turing machine turns out to be computationally
equivalent to a more human-friendly description of algorithms,
namely the circuit representation. The basic unit of information,
the bit, is a physical system that can be in exactly one out of two
states, usually denoted by $0$ and $1$. Information (or data) is
stored in binary form using many bits which are represented as lines
in the circuit. The data is processed by consecutively applying
logical gates to the bits, depicted symbolically as elements acting
between the lines in the circuit, such that the desired algorithm is
\begin{figure}
\begin{center}
\includegraphics[width=0.8\textwidth]{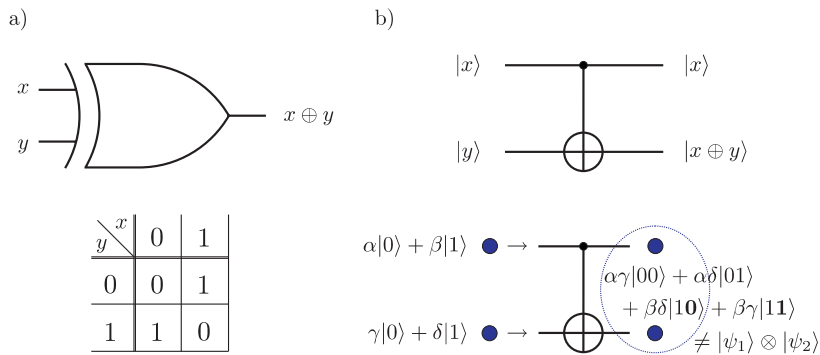}
\vspace{0.5cm} \caption{\label{fig:Gates} a) Classical \textsc{xor}
gate. b) Its quantum analog, \emph{i.e.}, the \textsc{cnot} gate. The truth
table of the \textsc{xor} gate is shown in the lower part of a) and
simply comes from adding the two input bits modulo 2 (this is the
'$\oplus$' operation). By replacing $0$ with $|0\rangle$ and $1$ with
$|1\rangle$ one obtains the truth table for the target qubit (lower wire) 
in \textsc{cnot} if the computational basis states are used as input.
However, the \textsc{cnot} gate is also the source of entanglement
(see also subsect.~\ref{sec:entanglement measures}) in the standard
model of quantum computing, as indicated in the lower part of b) by
the fact that, generally, two states cannot be written as a product
state $|\psi_1\rangle\otimes |\psi_2\rangle$ anymore after
\textsc{cnot} has been applied.}
\end{center}
\end{figure}
performed and the solution to the problem is stored in the bits.
Examples of such gates include the 1-bit \textsc{not} gate which
flips the state of a bit from $0$ to $1$ and vice versa, and the
2-bit \textsc{xor} gate that outputs $1$ if and only if exactly one
of the input bits is $1$ (see fig.~\ref{fig:Gates}). An important
result here is the fact that there exists a set of gates with which
one can implement any algorithm in a circuit, provided one can
freely distribute and copy information. For classical information,
the latter constraint is trivial and the so-called universal set of
gates consists in fact of only the \textsc{nand} gate (yielding $0$
if and only if both input bits are 1).

Having set the playground to implement algorithms tackling
computational problems, one of the most profound questions one can
ask is the following: What is the \textit{most efficient} algorithm
to solve a particular task? It is the field of computational
complexity theory \cite{Papadimitriou1994} that deals with such kind
of issues. We will briefly discuss the most important complexity
classes, focussing on time rather than space complexity. Imagining
that every gate in a circuit requires a finite execution time, one
can study the total time required to run an algorithm as a function
of the input or problem size, \emph{e.g.}, the number of input bits $n$. A
procedure is called efficient (or tractable) if its running time is
upper-bounded by some polynomial in $n$. Roughly speaking, the
complexity class \textsf{P} consists of all problems known to have
efficient algorithms\footnote{We omit here the strict definitions
of complexity classes in terms of decision problems and formal
languages. The interested reader is referred to, \emph{e.g.},
ref.~\cite{Papadimitriou1994}.}. Sorting or searching lists are
examples thereof.

However, there is also a vast amount of problems solved by
algorithms that are not efficient (typically exponential in $n$),
but once a possible solution has been proposed, it can be
efficiently checked for its validity. The class of these types of
problems is called \textsf{NP}. An important example thereof is
integer factorization: Given an $n$-bit integer, the best known
classical algorithm to find its prime factors is exponential in $n$,
but given a proposed factorization one can quickly check whether it
is correct by doing the required multiplications. The observation
that there are problems that can be solved efficiently, and others,
for which the best known solution is still worse than polynomial,
cumulates in the famous \textsf{P = NP?} question: It is clear that
\textsf{P} is a subset of \textsf{NP}, but is this inclusion strict?
In other words, have we just not yet discovered efficient algorithms
for supposedly `hard' problems, or is there something fundamental
within these problems that prevents us from finding such? This
puzzling question has not been answered yet with a formal proof, but
it is widely assumed that \textsf{P} and \textsf{NP} are not equal.

Another important complexity class is \textsf{NP}-complete, a subset
of \textsf{NP} consisting of all problems that are at least as hard
as all other problems in \textsf{NP}. This means that every problem
in \textsf{NP} can be cast into an instance of a problem in
\textsf{NP}-complete in polynomial time. With this transformation
being efficient, a polynomial-time solution to any one of the
problems in \textsf{NP}-complete would render the whole class
\textsf{NP} tractable and it would follow that \textsf{P = NP}.
Integer factorization is widely suspected, but not proven, to be
both outside of \textsf{P} \textit{and} \textsf{NP}-complete.

Finally, we remark that there is a complexity class of high
practical interest, which we include here merely because it has a
quantum analog that we will encounter in subsect.~\ref{sec: quant algo
and quant complexity}. Decision problems in the class \textsf{BPP}
(for bounded-error probabilistic time) have efficient algorithms
that are allowed to make random choices (`coin flipping') during the
computation and yield the correct answer with probability $p > 1/2$,
and a wrong solution with probability $1 - p$. The choice of $p$ is
essentially arbitrary, since the Chernoff bound \cite {Chernoff1952}
guarantees that the error probability in a majority vote drops
exponentially with the number of repetitive executions of the
algorithm. Typically, however, one finds $p = 2/3$ or $p = 3/4$ in
the literature.


\subsection{The standard model of quantum computing}\label{sec:standard model of
qc}

In a seminal work by Deutsch \cite{Deutsch1985}, he proposes to
strengthen the Church-Turing hypothesis into a ``manifestly physical
and unambiguous'' form. His Church-Turing principle reads: ``Every
finitely realizable physical system can be perfectly simulated by a
universal model computing machine operating by finite means'',
arguing that ``it would surely be hard to regard a function
`naturally' as computable if it could not be computed in Nature, and
conversely''. It is further shown that the universal Turing machine
does not fulfill this principle, while the `universal quantum
computer', proposed in the same work, is compatible with the
principle. A maybe less theoretic reasoning for studying computing
machines operating in the quantum regime is the mere fact that
classical computers are governed by Newtonian mechanics, being valid
only in a limiting case of the underlying quantum theory
\cite{DiVincenzo2000}. Quantum computers must therefore have at
least the same, if not a greater, computational power than classical
computers.

Historically understandable, the now so-called standard model of
quantum computing \cite{DiVincenzo2000, Nielsen2000a} follows
closely the circuit model discussed earlier (for alternative
proposals of quantum computing, see subsect.~\ref{sec: alternative QC}).
The basic unit of information is the \textit{qubit}, being a
two-level quantum system with basis states usually denoted by
$|0\rangle$ and $|1\rangle$ according to its classical counterpart.
Qubits are displayed as lines in the circuit to which quantum gates
are applied successively, thereby performing the computation. The
final result is obtained as a readout (a measurement) of the qubits
in the computer's final state.

The following features distinguish quantum from classical computing.
Firstly, quantum states cannot be copied perfectly (this is the
no-cloning theorem \cite{Wootters1982}). Secondly, the nature of a
quantum computer is ultimately an analog one. Quantum gates operate
on amplitudes such as $\alpha$ and $\beta$ in the state
$|\psi\rangle = \alpha |\phi_1\rangle + \beta|\phi_2\rangle$. Apart
from being properly normalized, amplitudes are arbitrary complex
numbers and as such analog. Thirdly, qubits can be in superpositions
and may form intricate entangled states. This fact is heavily
exploited in existing quantum algorithms and is essentially the key
ingredient to the majority of them. These first three points all contribute
to the drawback that error correction, a necessity in the presence
of imperfect gates and decoherence, is a non-trivial thing to do.
Quantum error correction, however, turns out to be possible if the
error probability per gate is smaller than some finite value for all
gates (see criterion 3 in subsect.~\ref{sec:DiVcriteria}). And lastly,
quantum gates must be time-reversal, \emph{i.e.}, unitary operators, in
accordance with basic principles of quantum mechanics. Apart from
the latter, there are no other constraints imposed on quantum gates.

Analogous to the classical case, and quite remarkably, there exist
finite sets of gates which can be used to approximate any unitary
evolution (\emph{i.e.}, the computation) of the quantum machine to
arbitrary precision \cite{Barenco1995}. An example thereof is the
universal set \cite{Nielsen2000a} consisting of the two single-qubit
gates $H = \frac{1}{\sqrt{2}}\left(
\begin{array}{cc}
             1 & 1 \\
             1 & -1 \\
           \end{array}
         \right)$ and
$T = \left(
           \begin{array}{cc}
             1 & 0 \\
             0 & e^{i\pi/4} \\
           \end{array}
         \right)$,
and the two-qubit \textsc{cnot} gate which performs the operation
$|x\rangle|y\rangle \rightarrow |x\rangle|(x + y)\,
\mathop\mathrm{mod} 2\rangle$, $x,y\in \{0, 1\}$. The gates are
represented as usual in the standard computational basis, \emph{i.e.},
$|0\rangle = (1, 0)^T$ and $|1\rangle = (0, 1)^T$. The $H$- and $T$-
gates are used to approximate arbitrary single-qubit rotations
$e^{i\alpha}e^{-i\theta\mathbf{n}\cdot\boldsymbol{\sigma}/2}$, where
$\mathbf{n}$ is a real unit vector and $\boldsymbol{\sigma}$ is the
vector of Pauli matrices. If arbitrary rotations are available on
their own, one can together with the \textsc{cnot} gate implement
any unitary evolution \textit{exactly}.

\subsection{Quantum algorithms and quantum complexity}\label{sec: quant algo and quant
complexity}

Having the new quantum machine at hand, what is its computational
power, and how does it perform compared with classical computers?
Such questions have given birth to the new field of quantum
complexity theory \cite{Bernstein1997, Cleve1999}, resulting in a
plethora of new quantum complexity classes along with the goal of
understanding their relations both between each other, and to
classical complexity classes. This is an active field of research
and the goal mentioned before is far from being reached. Instead of
going into the details of the theory, which would be out of the
scope of this work, we sketch an overview of the subject using the
most prominent examples.

One of the first algorithms demonstrating that quantum computers may
be able to drastically outperform classical computers is the
Deutsch-Jozsa algorithm \cite{Deutsch1992, Cleve1998}. Let $f$ be a
function from $n$ bits to one bit that can either be constant or
balanced, the latter meaning that there are exactly $2^{n-1}$
unknown input strings yielding the output $0$. The function is
supposed to be implemented in an oracle, \emph{i.e.}, a black box without
any further internal specification. The Deutsch-Jozsa algorithm can
determine the function's type (\emph{i.e.}, constant or balanced) with
probability $1$ by querying the oracle just once. A deterministic
classical algorithm requires $2^{n-1} + 1$ queries in the worst
case, as one may coincidentally pick all the $2^{n-1}$ input strings
that yield the same output\footnote{In practice, however, one would
use a randomized algorithm requiring a constant number of $k$
queries and returning the wrong result with arbitrary low
probability (dropping exponentially with $k$).}. The Deutsch-Jozsa
algorithm obtains its power from bringing a register of $n$ qubits
into an equally weighted superposition of all possible bit strings
using a Hadamard transform which is a special case of a so-called
quantum Fourier transform \cite{Nielsen2000a}. Using this state as
the input for the oracle, the function can be evaluated
simultaneously on all strings in the superposition with just one
query. This functionality in quantum computing is referred to as
quantum parallelism. Without going into further details, the
Deutsch-Jozsa algorithm manages to output the correct result with
certainty using these techniques. However, one has to be cautious
not to get the impression that one can calculate and obtain all
function values of an arbitrary function with just one query of an
oracle. In general, one is left with a superposition of results
collapsing upon measurement and yielding just one function value.
The Deutsch-Jozsa algorithm is cleverly designed to work with
certainty for a particular type of problem. These ideas are not
straightforwardly adapted to problems involving other kinds of
functions. The constant/balanced problem is an example from the
class \textsf{EQP} (exact quantum polynomial-time)
\cite{Bernstein1997}, denoting all problems solved in polynomial
time by a quantum algorithm with success probability equal to $1$.
\textsf{EQP} is the analog to the classical complexity class
\textsf{P}.

On the other hand, the set of problems having tractable algorithms
on a quantum computer with error probability smaller than $p = 1/3$
is denoted by \textsf{BQP} (bounded-error quantum polynomial-time)
\cite{Bernstein1997}, its classical counterpart being \textsf{BPP}.
It contains more interesting problems whose solutions are of greater
relevance than the Deutsch-Jozsa algorithm. Famous examples thereof
are Grover's \cite{Grover1996} and Shor's \cite{Shor1997}
algorithms. Grover's algorithm searches an unsorted database in time
upper-bound by a function proportional to $\sqrt{N}$ and has a
probability of failure scaling as $1/N$, where $N$ is the number of
database entries. Although the speed-up with respect to the
classical database search which runs in time proportional to $N$ is
not exponential, it is still of great benefit especially for large
$N$. Grover's algorithm could be used to speed up brute-force search
attempts for finding solutions to computationally hard problems.
However,  probably the most prominent problem in \textsf{BQP} is
integer factorization. Shor \cite{Shor1997} has shown that a quantum
computer can factor an integer in polynomial time, a task which is
supposed to be exponentially hard on a classical computer. These
kind of discoveries have resulted in a tremendous increase of
interest in quantum computing. For example, Shor's algorithm can be
used to break present-day public-key cryptosystems who rely on the
hardness of factorizing integers being a product of two large prime
numbers (such as RSA \cite{Rivest1978}).

A quantum complexity class that has recently called attention is
\textsf{QMA} \cite{Kitaev2002, Aharonov2002} (quantum
Merlin-Arthur), which can be seen as a quantum analog of
\textsf{NP}. Equivalently to the definition in
subsect.~\ref{sec:classical computers}, \textsf{NP} can be characterized
in terms of decision problems (having either `yes' or `no' as
answer): \textsf{NP} contains all problems for which `yes'-instances
are supplied with a proof that can be checked in polynomial time by
a deterministic verifier\footnote{For example, a decision version
of integer factorization would be stated as: Given integers $N$ and
a (user-specified) $x < N$, is there an integer $1 < d < x$ such
that $d$ divides $N$? The answer `yes' provided with a suitable $d$
can be verified in polynomial time, simply by checking whether $1 <
d < x$ and $d$ divides $N$. One then says that $x$ provides a
`yes'-instance to the problem with proof $d$.}. For quantum
computers, deterministic verifiers are however not meaningful.
\textsf{QMA} is thus defined probabilistically: A decision problem
belongs to \textsf{QMA}, if for every instance $x$ there exists an
efficient (\emph{i.e.}, polynomial in $x$) description of a quantum circuit
$Q_x$ (the verifier), such that for every `yes'-instance $\tilde x$
there exists a proof $|\psi\rangle$ with $p(Q_{\tilde
x}\;\mathrm{accepts}\;|\psi\rangle)
> 2/3$, and for every `no'-instance $\bar x$ $p(Q_{\bar
x}\;\mathrm{accepts}\;|\chi\rangle) < 1/3$ holds for all input
states $|\chi\rangle$. Here,
$p(Q_x\;\mathrm{accepts}\;|\psi\rangle)$ denotes the probability to
measure, \emph{e.g.}, $|0\rangle$ (if this is defined to mean `accept') as
the output of a quantum computation described by the circuit $Q_x$
and started with the initial state $|\psi\rangle$.

An important problem known to be in \textsf{QMA} is
$k$-\textsc{local Hamiltonian} which is specified by the following
decision problem: Given a Hamiltonian $H$ acting on $n$ qubits with
interactions that do not involve more than $k$ particles ($k$-body
interactions, where $k$ is a constant independent of $n$) and two
real numbers $a$ and $b$, such that $b - a > 1/\mathrm{poly}(n)$. Is
the ground state energy of $H$ smaller than $a$ (`yes'), or are all
energies larger than $b$ (`no')\footnote{Note that this is a
promised problem: It is guaranteed that either of the two cases will
occur, and we are not interested in Hamiltonians that have energies
between $a$ and $b$.}? A `yes' instance can be verified by providing
an eigenstate with energy smaller than $a$. Furthermore, one can
show that polynomial verifiers can be constructed that accept
(reject) `yes' instances (`no' instances) with sufficient
probability. It is now known that $k$-\textsc{local Hamiltonian} is
\textsf{QMA}-complete for $k\geq 2$ \cite{Kempe2006}, meaning the
following: Given an instance $x$ of any problem $Q$ in \textsf{QMA},
one can find (in time poly($n$)) an instance of $k$-\textsc{local
Hamiltonian} (by constructing a $k$-local $H$ and specifying
properly chosen parameters $a$ and $b$), such that, if $x$ is a
`yes' instance of $Q$, the ground state energy of $H$ is smaller
than $a$, and if $x$ is a `no' instance, the smallest eigenvalue of
$H$ is larger than $b$. $k$-\textsc{local Hamiltonian} (for $k\geq
2$) is as hard as any other problem in \textsf{QMA}. This `hardness'
suggests that calculating the ground state energy, and possibly
other ground state properties, is intractable even on a quantum
computer. The study of $k$-\textsc{local Hamiltonian} is also
important in the context of adiabatic quantum computing, see
subsubsect.~\ref{sec: adiabatic QC}.

\subsection{General criteria for scalable quantum computing}\label{sec:DiVcriteria}

In this section, we review the five DiVincenzo criteria
\cite{DiVincenzo2000} for the physical implementation of quantum
computing. These are the most fundamental requirements any proposal
for a quantum computer must fulfill in order to work with an
arbitrary number of qubits. Starting from sect.~\ref{sec:spin
manipulation in GaAs}, we will examine the experimental and
theoretical progress toward realizing these criteria for the
spin-qubit proposal of ref.~\cite{Loss1998} (see the next section).

\emph{1. A scalable physical system with well characterized qubits.}

\noindent We have already defined the notion of a qubit as simply
being a two-level quantum system. In this review, we will focus
solely on the electron spin in quantum dots. The word `scalable'
plays an important role: Even if current fundamental experiments are
performed with only few qubits, they must at least in principle be
preparable or manufacturable in large numbers, since only in this
case interesting and useful quantum computations can be performed.

A qubit must also be `well characterized' in the sense that one has
a good theoretical description not only of the qubit itself (in
terms of an internal Hamiltonian, accurate knowledge of all physical
parameters, etc.), but also of all relevant mechanisms that couple
qubits among each other and to the environment. On the one hand,
this is necessary to explore the possibilities of manipulating
qubits and letting them interact, but also, on the other hand, in
order to understand and fight the various forms of decoherence a
qubit may suffer from (see sect.~\ref{sec:relax}).

\emph{2. The ability to initialize the state of the qubits to a
simple fiducial state.}

\noindent It is clear that every computation needs to be started in
an initially known state such as $|000\ldots\rangle$. But this is
not the end of the story. Having a fast initialization mechanism at
hand is crucial for quantum error correction (see next criterion),
typically requiring large amounts of ancillary qubits in known
initial states in order to perform its job properly. If a fast
zeroing of qubits is not possible, \emph{i.e.}, if the initialization time
is long compared to gate operation times, then
ref.~\cite{DiVincenzo2000} proposes to equip the quantum computer
with ``some kind of `qubit conveyor belt', on which qubits in need
of initialization are carried away from the region in which active
computation is taking place, initialized while on the `belt', then
brought back to the active place after the initialization is
finished.''

In spin qubits, initialization could be achieved by either forcing
the spins to align with a strong externally applied magnetic field,
or by performing a measurement on the dot followed by a subsequent
rotation of the state depending on the measurement outcome. The
first approach is somewhat problematic, since natural thermalization
times are always longer than the decoherence time which itself needs
to be much longer than gate operation times (see the next
criterion). In this case, a `conveyor belt' scheme would be
required. The second possibility of measurement and rotation depends
on the specific setup examined, but initialization times might in
principle be much shorter than natural relaxation times. See
subsects.~\ref{sec:Loss DiVi Proposal} and~\ref{sec:initialization} for
more information and recent experimental achievements.

\emph{3. Long relevant decoherence times, much longer than the gate
operation time.}

\noindent Due to the coupling of qubits to their environment in a
thermodynamically irreversible way, quantum coherence is lost. In
other words, quantum states in contact with the outside world
ultimately evolve into fully mixed states. Decoherence is the answer
to why the macroscopic world looks classical. In GaAs quantum dots
where electron spins are used as qubits, the most important
mechanisms of decoherence are the spin-orbit and the hyperfine
interaction, see subsect.~\ref{sec:relax_short} and~ sect.~\ref{sec:relax}.

If it were not for quantum error correction, the duration of a
quantum computation would eventually be determined by the shortest
decoherence time in the setup. This would render longer and more
complex computations impossible. By encoding information not
directly into single qubits, but rather into `logical qubits'
consisting of several single qubits, a certain amount of errors due
to decoherence and imperfect gates may be corrected, depending on
what kind of code is used. There is however still a limit on how
faulty elementary gates are allowed to be: The accuracy threshold
theorem \cite{Preskill1998} states that error correction is possible
if the error probability per gate is smaller than a certain
threshold. This threshold comes about the fact that encoding,
verification, and correction steps require an additional overhead of
quantum gates, which introduces new possible sources of errors.
Error correction is thus only meaningful if encoding reduces the
error probability of an encoded operation compared with its original
(`unencoded') counterpart, \emph{despite} the fact that more
elementary gates are required. By employing the technique of code
concatenation \cite{Preskill1998}, \emph{i.e.}, encoding logical qubits
recursively, where using different codes per concatenation level is
allowed, the effective error on the top level of concatenation can
be made arbitrarily small. The threshold value depends on the error
models studied and on the details of the codes considered. Typical
values are in the range of $10^{-5}$ to $10^{-3}$
\cite{Preskill1998, Aliferis2007}, implying that decoherence times
must be a thousand to a hundred thousand times longer than gate
operation times.

\emph{4. A ``universal'' set of quantum gates.}

\noindent We have mentioned in subsect.~\ref{sec:standard model of qc}
that generic quantum computing is possible in the standard model if
certain one- and two-qubit gates are available. The single qubit
gates may be either implemented directly, or can be approximated to
arbitrary precision using a finite set of gates. The only necessary
two-qubit gate is the controlled-\textsc{not} gate
\cite{Barenco1995}. If these gates are for some reason not
implementable directly (\emph{i.e.}, in the sense that there are no
Hamiltonians that can be switched on which perform exactly the
desired gate operations), then a set capable of synthesizing them
needs to be present. This is, \emph{e.g.}, the case in the spin-qubit
proposal of ref.~\cite{Loss1998}, where the only controllable two-qubit interaction is the
exchange coupling between neighboring spins. The \textsc{cnot} gate
can however be implemented using a series of one-qubit operations
and the exchange interaction alone. The details and requirements for
this to work are discussed in the next section.

It is worth pointing out that one also needs to be able to execute
quantum gates in parallel in order for error correction to work.
This does however not pose a major drawback for solid-state systems
\cite{Gottesman2000}, where usually only two-body nearest-neighbor interactions are realizable.
It is also important to note the fact that faulty gates may
introduce systematic or random errors in a calculation. This can be
viewed as a source of decoherence and can therefore be overcome by means of
quantum error correction if the error rate is sufficiently
small. The same threshold values as discussed in the previous
criterion hold in this case.

\emph{5. A qubit-specific measurement capability.}

\noindent Measuring qubits without disturbing the rest of the
quantum computer is required in the verification steps of quantum
error correction and, not remarkably, in order to reveal the outcome
of a computation. If the measurement procedure does not discard
qubits (which could be the case, \emph{e.g.}, for spin-dependent tunneling
of electrons out of a quantum dot) it may be used in the
initialization step (see 2nd criterion). If it is, additionally,
fast enough, it may also be useful for quantum error correction. A
measurement is said to have $100\%$ quantum efficiency if it yields,
performed on a state $\rho = p|0\rangle\langle 0| + (1 -
p)|1\rangle\langle 1| + \alpha |0\rangle\langle 1| + \alpha^\ast
|1\rangle\langle 0|$, the outcome ``0'' with probability $p$ and
``1'' with probability $(1-p)$ independent of $\alpha$, the states
of neighboring qubits, or any other parameters of the system.
Real measurements cannot have perfect quantum efficiency. But this
is also not required since one can, \emph{e.g.}, rerun the computation several
times. 

\subsection{The Loss-DiVincenzo proposal}\label{sec:Loss DiVi
Proposal}

In this section, we review the spin-qubit proposal of ref.~\cite{Loss1998} for
universal scalable quantum computing. Here, the
physical system representing a qubit is given by the localized spin
state of one electron, and the computational basis states
$|0\rangle$ and $|1\rangle$ are identified with the two spin states
$|{\uparrow\rangle}$ and $|{\downarrow\rangle}$, respectively. In
general, the considerations discussed in ref.~\cite{Loss1998} are
applicable to electrons confined to any structure, such as, \emph{e.g.},
atoms, defects, or molecules. However, the original proposal focuses
on electrons localized in electrically gated semiconductor quantum
dots. The relevance of such systems has become clearer in recent
years, where remarkable progress in the fabrication and control of
single and double GaAs quantum dots has been made (see, \emph{e.g.},
ref.~\cite{Hanson2007} for a recent experimental review). We
postpone the discussion of experimental achievements with respect to
satisfying the DiVincenzo criteria to sect.~\ref{sec:spin
manipulation in GaAs}.

Scalability in the proposal of ref.~\cite{Loss1998} 
is due to the availability of local gating. Gating operations are realized through
the exchange coupling (see below), which can be tuned locally with
exponential precision. Since neighboring qubits can be coupled and
decoupled individually, it is sufficient to study and understand the
physics of single and double quantum dots together with the coupling
mechanisms to the environment present in particular systems
\cite{Coish2007}. Undesired interactions between three, four, and
more qubits should then not pose any great concern. This is in
contrast with proposals that make use of long-ranged interactions
(such as dipolar coupling), where scalability might not be easily
achieved.

\begin{figure}
\begin{center}
\includegraphics[width=0.8\textwidth]{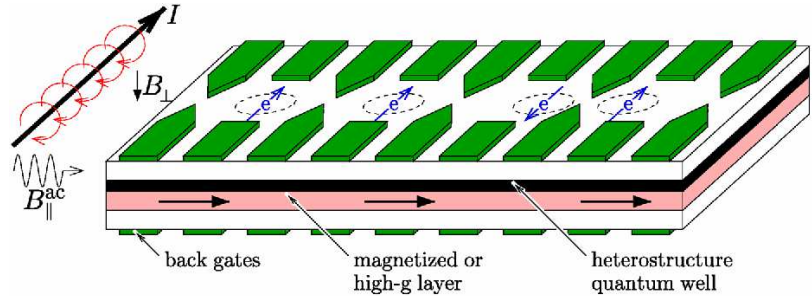}
\caption{\label{fig:Qdots} An array of quantum dot qubits realized
by laterally confining electrons in a two dimensional electron gas
formed at the interface of a heterostructure. The confinement is
achieved electrostatically by applying voltages to the metallic top
gates. Interaction is generally suppressed (as for the two qubits on
the left) but may be turned on to realize two-qubit operations by
lowering inter-dot gates (as for the two qubits on the right).
Single spin rotations may be achieved by dragging electrons down (by
changing back gate voltages) to a region where the Zeeman splitting
in the presence of the external static magnetic field $B_\bot$
changes due to magnetization or an inhomogeneous $g$-factor present
in that layer. A resonant magnetic ac pulse $B_{||}^{\tm ac}$ can
then be used to rotate the spin under consideration, while leaving
all other qubits unaffected due to the off-resonant Zeeman splitting
(ESR). All-electrical single spin manipulation may be realized in
the presence of spin-orbit interaction by applying ac electric
pulses directly via the gates (EDSR). See subsect. \ref{sec:gates} for
more details.}
\end{center}
\end{figure}

Figure~\ref{fig:Qdots} displays part of a possible implementation of a
quantum computer. Displayed are four qubits represented by the four single electron
spins confined vertically in the heterostructure quantum well and
laterally by voltages applied to the top gates. Initialization of
the quantum computer could be realized at low temperature $T$ by
applying an external magnetic field $B$ satisfying $|g\mu_B B|\gg
k_B T$, where $g$ is the $g$-factor, $\mu_B$ is Bohr's magneton, and
$k_B$ is the Boltzmann constant. After a sufficiently long time,
virtually all spins will have equilibrated to their thermodynamic
ground state $|0\rangle = |{\uparrow}\rangle$. As discussed in the
2nd criterion of the last section, this method might be too slow for
zeroing qubits in a running computation. Other proposed techniques
include initialization through spin-injection from a ferromagnet, as
has been performed in bulk semiconductors \cite{Fiederling1999,
Ohno1999}, with a spin-polarized current from a spin-filter device
\cite{Prinz1995, Prinz1998, Loss1998, DiVincenzo1999, Recher2000},
or by optical pumping \cite{Cortez2002, Shabaev2003, Gywat2004,
Bracker2005}. The latter method has allowed the preparation of spin
states with very high fidelity, in one case as high as $99.8\%$
\cite{Atature2006}.

The proposal of ref.~\cite{Loss1998} requires single qubit rotations
around a fixed axis in order to implement the \textsc{cnot} gate
(see below). In the original work \cite{Loss1998} this is suggested
to be accomplished by varying the Zeeman splitting on each dot
individually, which was proposed to be done via a site-selective
magnetic field (generated by, \emph{e.g.}, a scanning-probe tip) or by
controlled hopping of the electron to a nearby auxiliary
ferromagnetic dot. Local control over the Zeeman energy may also be
achieved through $g$-factor modulation \cite{Salis2001}, the
inclusion of magnetic layers \cite{Myers2005} (see also
fig.~\ref{fig:Qdots}) or by modification of the local Overhauser
field due to hyperfine couplings \cite{Burkard1999}. Arbitrary
rotations may be performed via ESR induced by an externally applied
oscillating magnetic field (see subsect.~\ref{sec:gates}). In this case,
however, site-selective tuning of the Zeeman energy is still
required in order to bring a specific electron in resonance with the
external field, while leaving the other electrons untouched (see
also fig.~\ref{fig:Qdots}). Alternative all-electrical proposals
(\emph{i.e.}, without the need for local control over magnetic fields) in
the presence of spin-orbit interaction or a static magnetic field
gradient have been discussed recently. See subsubsect.~\ref{sec:EDSR} for
greater details.

Two-qubit nearest-neighbor interaction is controlled in the
proposal of ref.~\cite{Loss1998} by electrical pulsing of a center gate
between the two electrons. If the gate voltage is high, the
interaction is `off' since tunneling is suppressed exponentially
with the voltage. On the other hand, the coupling can be switched
`on' by lowering the central barrier for a certain switching time
$\tau_s$. In this configuration, the interaction of the two spins
may be described in terms of the isotropic Heisenberg Hamiltonian
\begin{equation}\label{eq:loss divi exchange coupling}
H_s(t) = J(t)\mathbf{S}_L\cdot \mathbf{S}_R,
\end{equation}
where $J(t)\propto t_0^2(t)/U$ is the time-dependent exchange
coupling that is produced by turning on and off the tunneling matrix
element $t_0(t)$ via the center gate voltage. $U$ denotes the
charging energy of a single dot, and $\mathbf{S}_L$ and
$\mathbf{S}_R$ are the spin-$\frac{1}{2}$ operators for the left and
right dot, respectively. Equation~\eqref{eq:loss divi exchange coupling}
is a good description of the double-dot system if the following
criteria are satisfied: (i) $\Delta E \gg k_B T$, where $T$ is the
temperature and $\Delta E$ the level spacing. This means that the
temperature cannot provide sufficient energy for transitions to
higher-lying orbital states, which can therefore be ignored. (ii)
$\tau_s \gg \Delta E / \hbar$, requiring the switching time $\tau_s$
to be such that the action of the Hamiltonian is `adiabatic enough'
to prevent transitions to higher orbital levels. (iii) $U > t_0(t)$
for all $t$ in order for the Heisenberg approximation to be
accurate. (iv) $\Gamma^{-1} \gg \tau_s$, where $\Gamma^{-1}$ is the
decoherence time. This is basically a restatement of the 3rd
DiVincenzo criterion. For recent experimental results on the
decoherence times in lateral GaAs quantum dots, see
subsubsect.~\ref{sec:time_scales}.

The pulsed Hamiltonian eq.~\eqref{eq:loss divi exchange coupling}
applies a unitary time evolution $U_s(t)$ to the state of the double
dot given by $U_s(t) = \mathcal{T}\exp[-i\int_0^t H_s(t')dt'/\hbar]
= \exp[-(i/\hbar)\int_0^t J(t')dt'\mathbf{S}_L\cdot\mathbf{S}_R]$.
If the constant interaction $J(t) = J_0$ is switched on for a time
$\tau_s$ such that $\int_0^{\tau_s} J(t)dt/\hbar = J_0\tau_s/\hbar =
\pi \mod 2\pi$, then $U_s(\tau_s)$ exchanges the states of the
qubits: $U_s(\tau_s)|\mathbf{n}, \mathbf{n'}\rangle = |\mathbf{n'},
\mathbf{n}\rangle$. Here, $\mathbf{n}$ and $\mathbf{n'}$ denote real
unit vectors and $|\mathbf{n}, \mathbf{n'}\rangle$ is a simultaneous
eigenstate of the two operators $\mathbf{S}_L\cdot \mathbf{n}$ and
$\mathbf{S}_R \cdot \mathbf{n'}$. This gate is called \textsc{swap}.
If the interaction is switched on for the shorter time $\tau_s/2$,
then $U_s(\tau_s/2) = U_s(\tau_s)^{1/2}$ performs the so-called
`square-root of swap' denoted by $\sqrt{\textsc{swap}}$. This gate
together with single-qubit rotations about a fixed (say, the $z$-)
axis can be used to synthesize the \textsc{cnot} operation
\cite{Loss1998}
\begin{equation}
U_\textsc{cnot} = e^{i(\pi/2)S_L^z}e^{-i(\pi/2)S_R^z}
U_s(\tau_s)^{1/2}e^{i\pi S_L^z}U_s(\tau_s)^{1/2},
\end{equation}
or, alternatively, as
\begin{equation}
U_\textsc{cnot} = e^{i\pi S_L^z}U_s(\tau_s)^{-1/2}
e^{-i(\pi/2)S_L^z} U_s(\tau_s) e^{i(\pi/2) S_L^z}U_s(\tau_s)^{1/2}.
\end{equation}
The latter representation has the potential advantage that single
qubit rotations involve only one spin, in this case the one in the
left dot. Writing the \textsc{cnot} gate as above, it is seen 
that arbitrary single qubit rotations
together with the $\sqrt{\textsc{swap}}$ gate are sufficient for
universal quantum computing. See subsect.~\ref{sec:gates} for a recent
experimental implementation of the $\sqrt{\textsc{swap}}$ operation.
Errors during the execution of a $\sqrt{\textsc{swap}}$ gate due to
non-adiabatic transitions to higher orbital states
\cite{Schliemann2001, Requist2005}, spin-orbit interaction
\cite{Bonesteel2001, Burkard2002, Stepanenko2003}, and hyperfine
coupling to surrounding nuclear spins \cite{Petta2005, Coish2005,
Klauser2006, Taylor2007} have been studied. Furthermore, realistic
systems will include some anisotropic spin terms in the exchange
interaction which may cause additional errors. Conversely, this fact
might be used to perform universal quantum computing with two-spin
encoded qubits, in the absence of single-spin rotations
\cite{Bonesteel2001, Lidar2001, Stepanenko2004, Chutia2006}.

\subsection{Alternative approaches to quantum computing}\label{sec: alternative
QC}

Although the remainder of the review will mostly be concerned with
the realization of the spin-qubit proposal of ref.~\cite{Loss1998} (and related
decoherence effects), we would nevertheless like to discuss some of
the alternative proposals for quantum computing that have emerged in
recent years. Note that by this we are not referring to the many
alternative physical implementations of qubits which are also
studied extensively in present-day research (see, \emph{e.g.},
ref.~\cite{Cerletti2005} for a review focussing mainly on solid
state qubits). Rather, we would like to review proposals for quantum
computers which fundamentally differ from the standard circuit
model. The schemes we will turn our attention to are
measurement-based and adiabatic quantum computing. We will not
discuss topological quantum computing in greater detail, which
performs computation by braiding non-Abelian anyons. These are
particular quasi-particle excitations predicted to exist in certain
two-dimensional strongly correlated many-body systems such as a
two-dimensional electron gas in the fractional quantum Hall regime.
Topological quantum computing is supposed to be much less
susceptible to gate errors since small deformations of braids do not
change their topology. The interested reader is referred to the
recent reviews refs.~\cite{Nayak2008, Stern2008}. See
ref.~\cite{Zilberberg2008} for a measurement based implementation of
\textsc{cnot} on $\nu = 5/2$ Ising-type anyon qubits.

\subsubsection{Measurement-based quantum computing}\label{sec:measurement based
qc}

Implementing quantum gates, particularly two-qubit gates, with a
precision as required by fault-tolerant error correction is
difficult. Instead of performing gate operations on qubits, there
are proposals that allow for universal quantum computing by
replacing part or all of these gates by measurement. We will mainly
focus on a measurement-based implementation of \textsc{cnot} for
qubits represented by single or multiple electron spins. Afterwards,
we will briefly outline the ideas behind the so-called `one-way
quantum computer'.

\paragraph{Measurement-based implementation of \textsc{cnot}}

When using the polarization state of a photon as a qubit, it is
known that universal quantum computing can be achieved using only
linear optics and single photon measurements \cite{Knill2001}. This
holds similarly for all bosons. For electrons (and, similarly, for
all fermions), there exists a strong no-go theorem
\cite{Terhal2002a, Knill2001a} stating that quantum computing with
single-electron Hamiltonians and single-spin measurements can
efficiently be simulated on a classical computer, thus not
exhibiting the observed exponential speed-up of some algorithms over
their classical analogs. However, the no-go theorem can be
circumvented by exploiting the electron's charge degree of freedom:
It has been shown recently how to build a \textsc{cnot} gate for
single- \cite{Beenakker2004} and multi-electron
\cite{Zilberberg2008} qubits by the ability to perform, apart from
the availability of single-electron operations and single-spin
readouts, charge measurements. Universal quantum computing is
thereby restored. Note that the qubits are still encoded in spin
states of electrons. Since spin and charge are commuting
observables, charge measurements do not alter the information
represented in the spins.

The main idea is to provide parity measurement of two electron spins
via charge detection. The \textsc{cnot} gate is then constructed
from these parity gates. Imagine we had such a device at hand, \emph{i.e.},
for a state in a space either spanned by
$\{|{\uparrow\uparrow}\rangle, |{\downarrow\downarrow}\rangle\}$
(even parity), or by $\{|{\uparrow\downarrow}\rangle,
|{\downarrow\uparrow}\rangle\}$ (odd parity), it could determine
nondestructively which space the state belongs to by detecting the
presence or absence of charge upon a measurement thereof. A rather
abstract notion of a parity gate was described in
ref.~\cite{Beenakker2004}. Further below, we will describe a much
more concrete theoretical proposal in the reach of present-day
experiments. In the following, however, we will first review the
realization of a \textsc{cnot} gate for single-electron qubits as
presented in ref.~\cite{Beenakker2004}. A generalization of this
proposal to multi-electron qubits can be found in
ref.~\cite{Zilberberg2008}. Particularly, a detailed construction of
the \textsc{cnot} gate for two-electron qubits encoded in the
singlet-triplet basis (see subsect.~\ref{sec:initialization}) is
discussed there.

Let a parity gate work as follows. Two electrons can enter the gate
simultaneously, and after the parity was measured via charge
detection, the electrons leave the gate with unmodified spin state
if the latter was in one of the even or odd parity spaces described
above.
\begin{figure}
\begin{center}
\includegraphics[width=0.9\textwidth]{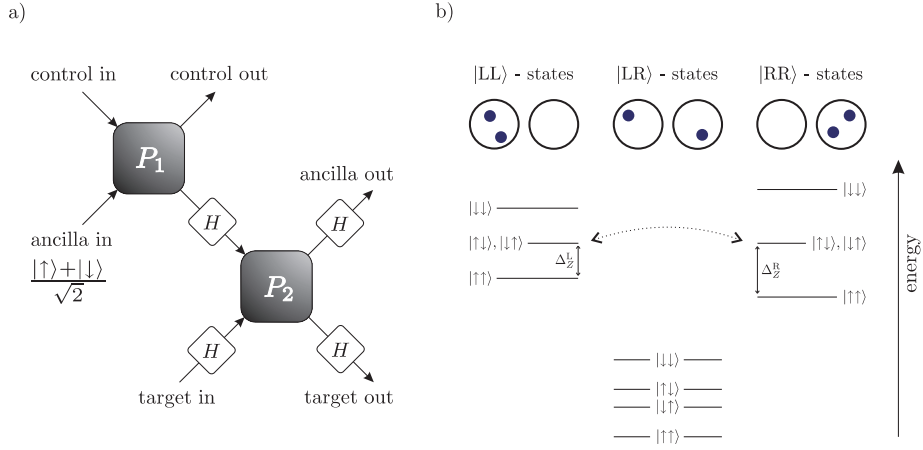}
\vspace{0.5cm} \caption{\label{fig:Parity gates} a) Construction of
a deterministic \textsc{cnot} gate from two parity gates $P_1$ and
$P_2$. Each parity gate has two input and two output arms and can
discriminate whether two incident electrons are in a parallel or
antiparallel spin configuration via charge sensing. The gates $H$
symbolize Hadamard transformations. The \textsc{cnot} operation is
completed by applying Pauli rotations to the control and target
electron depending on the outcomes of the two parity and the ancilla
measurements. See main text for a detailed discussion. (Figure
adapted from ref. \cite{Beenakker2004}.) b) Energy diagram for the
double-dot parity gate proposed in ref. \cite{Engel2005}. The charge
states $|\mathrm{LL}\rangle$, $|\mathrm{LR}\rangle$, and
$|\mathrm{RR}\rangle$ are indicated by the electron configuration in
the double dot. The energy levels of all possible spin states are
shown for each charge configuration. Gate voltages and a magnetic
field are applied such that the $|\mathrm{LR}\rangle$ states are
lowest in energy, and the antiparallel $|\mathrm{LL}\rangle$ and
$|\mathrm{RR}\rangle$ states are on the same energy. This allows for
resonant tunneling between the latter two states (indicated by the
dotted arrow), while tunneling between parallel states is suppressed
due to the Zeeman mismatch $\Delta_Z^\mathrm{R} >
\Delta_Z^\mathrm{L}$. (Figure adapted from ref. \cite{Engel2005}.)}
\end{center}
\end{figure}
Furthermore, let the gate record `no charge' ($p = 0$) for two
antiparallel spins, and `charge' ($p = 1$) for parallel incident
spins. Figure~\ref{fig:Parity gates} a displays the construction of
the deterministic \textsc{cnot} gate using two connected parity
gates $P_1$ and $P_2$. Before the input and after the output arms of
$P_2$, a Hadamard transformation $H$ is applied to each spin,
defined as $|{\uparrow}\rangle \rightarrow (|{\uparrow}\rangle +
|{\downarrow}\rangle)/\sqrt{2}$, and $|{\downarrow}\rangle
\rightarrow (|{\uparrow}\rangle - |{\downarrow}\rangle)/\sqrt{2}$.
The control qubit enters the first gate $P_1$. Its state decides
whether the target qubit, entering the second gate $P_2$, is to be
flipped according to the definition of the \textsc{cnot} operation.
$P_1$ is also provided with an ancilla qubit prepared in the state
$(|{\uparrow}\rangle + |{\downarrow}\rangle)/\sqrt{2}$ which is then
fed back into $P_2$. Upon leaving the second gate, the ancilla is
measured. Conditioned on this result and the outcomes $p_1$ and
$p_2$ of the two parity measurements in $P_1$ and $P_2$,
respectively, a Pauli matrix has to be applied to control and target
qubit in order to complete the \textsc{cnot} operation (see below).
We will now see (following the supplementary appendix of
ref.~\cite{Beenakker2004}) that this setup indeed implements
\textsc{cnot}.

We depart from our usual notation for the spin basis and identify
$|0\rangle \equiv |{\uparrow}\rangle$ and $|1\rangle \equiv
|{\downarrow}\rangle$. Furthermore, all variables represent a number
in $\{0, 1\}$ and addition is performed modulo $2$. We first
consider the action of the second gate $P_2$. After applying the
Hadamard gates on the input arms of $P_2$, but before the parity
measurement, an input state $|a\rangle|y\rangle$ has been
transformed to $(|0\rangle + (-1)^a |1\rangle)(|0\rangle + (-1)^y
|1\rangle)$ (normalization constants will be neglected for the rest
of this section). Here, the first (second) state represents the
qubit entering the upper (lower) arm of the parity gate. After the
parity measurement the state has become
\begin{equation}
|a\rangle|y\rangle \rightarrow \begin{cases} |0\rangle|0\rangle +
(-1)^{(a+y)}|1\rangle|1\rangle &\mathrm{if}\quad p_2 = 1 \\
(-1)^y|0\rangle|1\rangle + (-1)^a|1\rangle|0\rangle &\mathrm{if}
\quad p_2 = 0.
\end{cases}
\end{equation}
In the end, the Hadamard gates on the output arms are performed and
the state of the ancilla is measured, \emph{i.e.},
\begin{align}
|a\rangle|y\rangle &\rightarrow \begin{cases} |0\rangle|a+y\rangle +
|1\rangle |a + y +
1\rangle &\mathrm{if}\quad p_2 = 1 \\
(-1)^a |0\rangle|a+y\rangle - (-1)^a|1\rangle |a + y + 1\rangle
&\mathrm{if} \quad p_2 = 0
\end{cases}\\
&\rightarrow (-1)^{(p_2 + 1)(a + z)}|a + y + z\rangle,
\end{align}
where $z$ is the outcome of the ancilla measurement. The action of
the first parity gate $P_1$ on the control and the ancilla qubit is
given by $|x\rangle(|0\rangle + |1\rangle) \rightarrow |x\rangle|x +
p_1 + 1\rangle$. The second state is transmitted to the upper arm of
$P_2$, yielding the total action of the setup on a control-target
pair $|x\rangle|y\rangle$:
\begin{equation}
|x\rangle|y\rangle \rightarrow (-1)^{(p_2 + 1)(x + z + p_1 +
1)}|x\rangle|x + y + z + p_1 + 1\rangle.
\end{equation}
Post-correction depending on $p_1$, $p_2$ and $z$ has to be
performed now in order to obtain the correct \textsc{cnot} operation
defined by $|x\rangle|y\rangle \rightarrow |x\rangle|x+y\rangle$.
The phase factor $(-1)^{(p_2 + 1)(z + p_1 + 1)}$ is irrelevant since
it does not depend on $x$ and $y$. If $p_2 = 0$, a $\sigma_z$ gate
has to be applied to the control qubit. This eliminates the
remaining phase (since $\sigma_z|x\rangle = (-1)^x|x\rangle$). In
order to obtain the correct target $|x + y\rangle$, a $\sigma_x$
gate needs to be applied if $z + p_1 = 0$ (since $\sigma_x|y\rangle
= |y + 1\rangle$). This completes the description of the
\textsc{cnot} gate in terms of parity measurements.

We now qualitatively describe a concrete proposal due to
ref.~\cite{Engel2005} of a parity gate exploiting charge
measurement. The device consists of two coupled quantum dots
containing the two electrons whose parity is to be determined. The
dots are assumed to have different Zeeman splittings
$\Delta_Z^\mathrm{L}$ and $\Delta_Z^\mathrm{R}$. This could be
realized, \emph{e.g.}, by locally different magnetic fields or with an
inhomogeneous $g$-factor. By applying suitable gate voltages and a
perpendicular magnetic field, one can achieve the energy
configuration shown in fig.~\ref{fig:Parity gates}b. The gate
voltages are set such that all states $|\mathrm{LL}\rangle$ (two
electrons in the left dot) and $|\mathrm{RR}\rangle$ (two electrons
in the right dot) are higher in energy than $|\mathrm{LR}\rangle$
(one electron in each dot), independent of the spin configuration.
The strength of the external magnetic field is chosen such that the
zero-field singlet-triplet splitting in each dot is removed
(see subsubsect.~\ref{sec:singledot_states}). This leads to the degeneracy of all
spin states in the odd-parity space $\{|{\uparrow\downarrow}\rangle,
|{\downarrow\uparrow}\rangle\}$ with charge configuration
$|\mathrm{RR}\rangle$ or $|\mathrm{LL}\rangle$. The gate voltages
can further be properly tuned to align these degenerate levels of
the two dots. However, due to the different Zeeman splittings in the
left and the right dot, parallel spin configurations with charge
state $|\mathrm{RR}\rangle$ are detuned by $\epsilon = \pm
2(\Delta_Z^\mathrm{R} - \Delta_Z^\mathrm{L})$ from the corresponding
states in $|\mathrm{LL}\rangle$. The energy spectrum hereby achieved
allows for elastic tunneling between the states
$|\mathrm{LL}\rangle$ and $|\mathrm{RR}\rangle$ (through the
intermediate state $|\mathrm{LR}\rangle$) with antiparallel spins,
whereas resonant tunneling for parallel spins is suppressed due to
the Zeeman mismatch. The transition to the ground state occurs only
inelastically. A quantum point contact near the neighboring dots is
used as an electrometer to detect the presence (indicating
antiparallel spins, as opposed to the abstract gate described above)
or absence (indicating parallel spins) of tunneling events (see also
subsect.~\ref{sec:readout}).
A microscopic model of the double dot system is
further studied in ref.~\cite{Engel2005}, where it is shown that
elastic tunneling (if present) strongly dominates over inelastic
tunneling, and that the device still works with high fidelity even
if the measurement parameters cannot be controlled perfectly.

\paragraph{The one-way quantum computer}

A proposal for quantum computing that requires nothing but single
qubit measurements during computation is briefly outlined in the
following. The one-way quantum computer \cite{Raussendorf2001}
requires a so-called cluster state \cite{Briegel2001} to begin with.
This is a certain highly entangled state that can be realized on a
two- or three-dimensional array of qubits interacting through
externally controllable nearest-neighbor Ising- \cite{Briegel2001}
or Heisenberg-type \cite{Borhani2005} interactions. After this state
is initialized, a network, \emph{i.e.}, an entangled state among qubits
forming a grid-like structure, is realized by discarding undesired
qubits through measurements of $\sigma_z$. The computation is then
performed by measurements in the $x$-$y$ plane of qubits in the
network. The choice of future measurement bases may depend on past
measurement outcomes. It can be shown that universal quantum
computing is possible, whereby the computation proceeds spatially
from left to right with quantum information flowing on horizontal
branches on the network and two-bit interactions implemented by
measurements on vertical branches. The interested reader is referred
to refs. \cite{Raussendorf2001, Briegel2001, Raussendorf2003} for
greater detail.

Small-scale instances of Grover's algorithm \cite{Walther2005,
Prevedel2007} and the Deutsch-Jozsa algorithm \cite{Tame2007} have
recently been demonstrated within quantum optics.

\subsubsection{Adiabatic quantum computing}\label{sec: adiabatic QC}

Adiabatic quantum computing appeared first in the context of a novel
approach to solve classical optimization problems \cite{Farhi2000}.
It has then evolved into a general approach to quantum computation
now known to be polynomially equivalent to the standard model
\cite{Aharonov2007}, implying that standard and adiabatic quantum
computers have the same computational power. While it was found
rather quickly that a standard quantum computer can efficiently
simulate arbitrary adiabatic computations \cite{Dam2001, Farhi2001},
hence proving one direction of the equivalence, it took several
years to show the opposite, \emph{i.e.}, that an adiabatic quantum computer
can simulate any standard computation with only a polynomial
overhead \cite{Aharonov2007}. We will review the basic ideas behind
adiabatic quantum computing and the original approach towards it
\cite{Aharonov2007}. Afterwards, we will point out some quite recent
developments in the field.

The basis of adiabatic quantum computing is the adiabatic theorem
\cite{Kato1951, Messiah1958}: Given a system initially prepared in
an energy eigenstate and undergoing an externally induced time
evolution, the theorem says that the system's state will remain
arbitrarily close to the corresponding instantaneous eigenstate, if
there is a nonzero energy gap all along the evolution and if the
latter is carried out `slow enough'. Hereby, the time scale for
`slow enough' depends on the desired closeness accuracy of the
system's state to the respective instantaneous eigenstate and on the
size of the minimal gap along the evolution. The smaller the minimal
gap, the slower the process has to be performed in order to suppress
transitions to states higher in energy. In adiabatic quantum
computing, one starts with a system prepared in the ground state of
an initial Hamiltonian $H_{\mathrm{init}}$, with that ground state
being unique and having a simple form such as $|000\ldots\rangle$.
The system is then being evolved adiabatically according to $H(t) =
(1 - t/T)H_\mathrm{init} + (t/T)H_\mathrm{final}$, $t\in [0, T]$\footnote{General non-linear paths have been studied as well, see,
\emph{e.g.}, ref.~\cite{Farhi2002}.}, into a setup described by the final
Hamiltonian $H_\mathrm{final}$, \emph{whose ground state encodes the
result of the desired computation}. This has the potential advantage
of not requiring fast gate and measurement operations. Additionally,
adiabatic quantum computing is intrinsically robust against
decoherence due to environmental noise (see ref.~\cite{Kaminsky2003}
for a review). One can prove a rigorous lower bound on the value of
$T$ required to obtain a final state that is $\varepsilon$-close in
$l_2$-norm to the ground state of $H_\mathrm{final}$. $T$ depends
inversely both on $\varepsilon$ and on the minimal gap between the
ground and first excited state of $H(t)$ (called the spectral gap).
The running time $\tau$ is defined to be $T\max_t ||H(t)||$, where
the second factor makes $\tau$ invariant to rescaling of $H(t)$.
Further, $H_\mathrm{init}$ and $H_\mathrm{final}$ are restricted to
be, in the original terminology, local, meaning that they may only
allow interactions between a constant number of particles in order
to be physically realistic. This constraint also assures that the
Hamiltonians have efficient classical descriptions
\cite{Aharonov2007}.

The main difficulty to adiabatic quantum computing is the fact that
finding an $H_\mathrm{final}$ encoding the result of a computation
in its ground state is impossible, since that result is
intrinsically unknown (otherwise there would be no need for its
computation). While in the standard model an algorithm is executed
by a discrete unitary time evolution, $H_\mathrm{final}$ is
subjected to \emph{simultaneous} local constraints. This problem is
overcome in ref.~\cite{Aharonov2007} by loosening the requirement
for $H_\mathrm{final}$ to have a ground state exactly equal to the
outcome of a standard quantum computation. It is shown that it is
sufficient to obtain a ground state having nonzero overlap with the
desired state. This overlap can then be enhanced arbitrarily with
polynomial overhead.

Given a quantum circuit with $L$ gates and denoting by
$|\alpha(l)\rangle$, $l = 0, 1, \ldots, L$, the state in the circuit
after the $l$th gate $U_l$ has been applied, it is aimed at
constructing a Hamiltonian $H_\mathrm{final}$ whose ground state is
the so-called history state
\begin{equation}
|\eta\rangle = \frac{1}{\sqrt{L + 1}}\sum_{l = 0}^L
|\alpha(l)\rangle\otimes|1^l 0^{L-l}\rangle^c.
\end{equation}
The right $L$ qubits are referred to as clock qubits (superscript
`c') whose representation ``enables a local verification of correct
propagation of the computation from one step to the next, which
cannot be done without the intermediate computational steps''
\cite{Aharonov2007}. After showing that there is an
$H_\mathrm{init}$ with non-degenerate ground state $|00\ldots
0\rangle\otimes |0^L\rangle$, $H_\mathrm{final}$ is defined as
\begin{equation}
H_\mathrm{final} = \frac{1}{2}\sum_{l = 1}^L H_l + H_\mathrm{input}
+ H_\mathrm{clock},
\end{equation}
where $H_\mathrm{input}$ and $H_\mathrm{clock}$ ensure that
undesired input states and illegal clock states receive an energy
penalty. The $H_l$, $1 < l < L$ are given by
\begin{equation}
\begin{split}
H_l &= I\otimes|100\rangle\langle 100|^c_{l-1, l, l+1} -
U_l\otimes|110\rangle\langle 100|^c_{l-1, l, l+1} \\
&-U_l^\dag\otimes|100\rangle\langle 110|^c_{l-1, l, l+1} +
I\otimes|110\rangle\langle 110|^c_{l-1, l, l+1},
\end{split}
\end{equation}
and similar for $l = 1$ and $l = L$. The $H_l$ make sure that the
unitary action of each $U_l$ comes along with the correct update of
the clock register. The final Hamiltonian constructed in this way
has indeed $|\eta\rangle$ as its ground state. In
ref.~\cite{Aharonov2007} it is then further shown that the spectral
gap of $H(t)$ is lower bounded by an inverse polynomial in $L$ for
all $t$. This, together with adding identity gates (in the circuit
picture) at the end of the computation in order to arbitrarily
increase the weight of $|\alpha(L)\rangle$ in $|\eta\rangle$,
results in a running time that scales with $L^5$.

Note, however, that the Hamiltonian obtained in this way is
5-local, \emph{i.e.}, interactions between 5 arbitrarily distant particles
have to be realized. It was already shown in
ref.~\cite{Aharonov2007} that $H_\mathrm{final}$ can be made
3-local, but the running time then increases and roughly scales as
$L^{14}$. It was even demonstrated how to make an $H_\mathrm{final}$
involving only two-body nearest-neighbor interactions, although this
required the usage of particles having a six-dimensional state
space. These results have recently been extended to qubits with
2-local interactions \cite{Kempe2006, Siu2005} and to qubits on a 2D
lattice with nearest-neighbor two-body interactions
\cite{Oliveira2008}. All these constructions start from the 5-local
Hamiltonian described earlier. Very recently, a rather different
approach to adiabatic quantum computing has been taken using the
concept of `ground state quantum computation' \cite{Mizel2007}.
There, the entire temporal trajectory of an algorithm is encoded
spatially in the ground state of a suitable system. The proposal
uses qubits and requires only two-body nearest-neighbor
interactions. It has been shown that the scaling of the running time
with $N$ and $L$, where $N$ is the number of qubits, is of order
$(NL)^2$. The interested reader is referred to the original
literature \cite{Mizel2007, Mizel2002}.

\subsection{Entanglement measures}\label{sec:entanglement measures}

Quantum correlations are heavily exploited in every quantum
algorithm and form the key ingredient to the reason why quantum
computing differs from classical computation. In the circuit model
discussed in subsect.~\ref{sec:standard model of qc}, entanglement is
generated by the \textsc{cnot} gate. Typical simple examples of
entangled states are Bell states such as $|\Phi^+\rangle =
(|{\uparrow}\rangle_A|{\uparrow}\rangle_B +
|{\downarrow}\rangle_A|{\downarrow}\rangle_B)/\sqrt{2}$
\cite{Bell1964}. Imagining that two particles in such a state travel
to spatially arbitrarily separated observers $A$ and $B$, both
observers will obtain random, but perfectly correlated measurement
outcomes (assuming ideal measurement conditions). This is
inconsistent with any classical (\emph{i.e.}, local) description of the
state \cite{Bell1964, Colbeck2008}.

Entanglement hence manifests itself in the form of inter-partite
correlations in a quantum state which are not explainable by
classical means. It this context, one usually introduces the notion
of LOCC-operations \cite{Plenio2007}: If correlations observed in a
quantum state cannot be reproduced (or simulated), starting from
initially unrelated quantum subsystems, by local quantum operations
(`LO') coordinated by and influencing each other via classical
communication (`CC'), then these correlations are identified with
the presence of entanglement in that state. On the other hand, if a
state can be created by LOCC-operations alone, it is denoted
separable, \emph{i.e.}, unentangled. The fact that LOCC-operations can
neither create entanglement in a separable state, nor enhance
already present entanglement (on average\footnote{There is a
protocol known as the `Procrustean method' or `entanglement
gambling' \cite{Bennett1996a, Thapliyal2003}: One can turn any
multipartite entangled state into a Bell state shared by some pair
of parties using LOCC alone. This works however only with
probability smaller than 1. On average, LOCC cannot increase
entanglement.}), makes entanglement a resource which is sought to be
quantified.

Formally, a pure state in an $n$-partite Hilbert space $\mathcal{H}
= \bigotimes_{i = 1}^N\mathcal{H}_i$ is called entangled if it
cannot be written as a product state $|\psi\rangle = \bigotimes_{i =
1}^N|\psi_i\rangle$. For example, the state $|\phi\rangle =
(|{\uparrow\uparrow}\rangle - |{\uparrow\downarrow}\rangle +
|{\downarrow\uparrow}\rangle - |{\downarrow\downarrow}\rangle)/2$
can be written in the form $|\phi\rangle = (|{\uparrow}\rangle +
|{\downarrow}\rangle)/\sqrt{2}\otimes (|{\uparrow}\rangle -
|{\downarrow}\rangle)/\sqrt{2}$ and is thus separable, \emph{i.e.}, not
entangled, whereas such a decomposition is not possible for the Bell
state $|\Phi^+\rangle$ from above. Analogously, a mixed state $\rho$
acting on $\mathcal{H}$ is separable if it can be written in the
form $\rho = \sum_i p_i \bigotimes_{k = 1}^N \rho_k^i$, where the
$\rho_k^i$ act on $\mathcal{H}_k$ for all $k$. However, as already
indicated earlier, the story is not over after categorizing states
into a `black and white' scheme by determining whether a particular
state is separable or entangled. A simple example demonstrating that
some states can be `more entangled' than others is the fact that some
states violate Bell-type inequalities stronger than others, implying
the presence of more quantum correlations.

An entanglement measure is a function from the space of density
matrices to a closed interval in the real non-negative numbers, the
lower bound usually being $0$, and should reflect the physical
properties of entanglement. Most importantly, it should be
non-increasing under LOCC-operations on average, which is
meaningful, taking the previous discussion into account. In
particular, it should be invariant under local unitary
transformations which merely correspond to local changes of basis.
An entanglement measure satisfying the previous conditions is called
an entanglement monotone \cite{Vidal2000, Bennett1996}.
Additionally, entanglement measures are often demanded to be able to
uniquely distinguish between separable and entangled states, usually
incorporated by constructing the measures such that they are $0$ if
and only if the state examined is separable.

Entanglement in bipartite systems is the case understood by far the
most until now, in contrast to multipartite entanglement (see, \emph{e.g.},
ref.~\cite{Plenio2007}). This is also due to the existence of a
meaningful entanglement measure for bipartite states, namely, the
entanglement of formation $E_F$ \cite{Bennett1996}, defined as
\begin{equation}\label{entanglement of formation}
E_F(\rho) = \min_{\{p_i,
|\psi_i\rangle\}\in\mathfrak{D}(\rho)}\sum_i p_i E(|\psi_i\rangle),
\end{equation}
where
\begin{equation}\label{set of pure state decompositions}
\begin{split}
\mathfrak{D}(\rho) = &\Bigl\{\left\{p_i,|\psi_i\rangle\right\}_{i = 1}^K \;\big|K \geq \rank\rho,\; p_i \geq 0, \\
&\sum_{i = 1}^K p_i = 1, \;|\psi_i\rangle \in \mathcal{H},\; \rho =
\sum_{i=1}^K p_i |\psi_i\rangle\langle\psi_i| \Bigr\}
\end{split}
\end{equation}
is the set of all so-called pure-state decompositions of $\rho$, and
\begin{equation}
E(|\psi\rangle) = -\tr \left[(\tr_1{|\psi\rangle\langle\psi|})
\log_2(\tr_1{|\psi\rangle\langle\psi|)}\right]
\end{equation}
is the entropy of entanglement ($\tr_1$ denotes the partial trace
over the first subsystem). The latter is an entanglement monotone
for bipartite pure states and is closely related to the von Neumann
entropy. The numerical value of $E_F(\rho)$ is meaningful in the
following sense: It has been shown \cite{Bennett1996} that, given a
number of $N$ identical states $\rho$, one can (asymptotically)
`distill' $N\cdot E_F(\rho)$ maximally entangled Bell states (such
as $|\Phi^+\rangle$) out of them. The entanglement of formation thus
measures quantum correlations in units of the entanglement contained
in a Bell state\footnote{Note that for states in
higher-dimensional systems, the entanglement of formation can be
larger than $1$, implying that more than one Bell state is required
to create such a state.}.

There exists a vast amount of proposed entanglement measures for
multipartite pure states. The study of mixed-state entanglement is,
however, important as well, since any realistic quantum system will
eventually couple to the environment and thus decohere. The
so-called convex roof construction \cite{Uhlmann2000} (the
entanglement of formation being an early example thereof) gives a
general recipe how to extend a pure-state entanglement monotone to
mixed states: Given an arbitrary pure-state multipartite
entanglement monotone $m$, the convex roof of $m$ is given by
\begin{equation}\label{convex-roof em}
M(\rho) = \inf_{\{p_i,
|\psi_i\rangle\}\in\mathfrak{D}(\rho)}\sum_{i}p_i m(|\psi_i\rangle),
\end{equation}
where $\mathfrak{D}(\rho)$ is defined as in eq.~\eqref{set of pure
state decompositions}. $M$ has the desirable feature that it is an
entanglement monotone itself, and that it properly reduces to $m$ if
$\rho$ describes a pure state \cite{Mintert2005}. The optimization
problem coming along with eq.~\eqref{convex-roof em} is, however,
rather involved and seems impossible to be analytically solvable in
general. Remarkably, there is one major exception to this statement:
There exists a general analytical expression for the entanglement of
formation of two qubits \cite{Wootters1998}.

Nevertheless, the optimization problem in eq.~\eqref{convex-roof em}
can be tackled numerically to some extent \cite{Audenaert2001,
Rothlisberger2008, Rothlisberger2008a} by first parameterizing the
set of all pure-state decompositions $\mathfrak{D}(\rho)$. Let
$St(k, r)$ denote the set of all $k\times r$ matrices $U\in
\mathds{C}^{k\times r}$ with the property $U^\dag U = 1_{r\times
r}$. The required parametrization of $\mathfrak{D}(\rho)$ is due to
the Schr\"odinger-HJW theorem \cite{Hughston1993, Kirkpatrick2005},
stating that every decomposition of $\rho$ into $k$ states is
related to a matrix $U\in St(k, r)$, where $r = \rank\rho$,
\emph{and vice versa}. A search over $St(k, r)$ for all $k \geq r$
is thus equivalent to searching over $\mathfrak{D}(\rho)$.
Explicitly, given a matrix $U\in St(k,r)$ and the eigendecomposition
$\rho = \sum_{i = 1}^r \lambda_i |\chi_i\rangle\langle\chi_i |$ of
$\rho$, the pure-state decomposition corresponding to $U$ is given
by
\begin{equation}
p_i = \langle\tilde\psi_i|\tilde\psi_i\rangle, \qquad |\psi_i\rangle
= (1/\sqrt{p_i})|\tilde\psi_i\rangle,
\end{equation}
where
\begin{equation}
|\tilde\psi_i\rangle = \sum_{j=1}^r
U_{ij}\sqrt{\lambda_j}|\chi_j\rangle, \qquad i = 1, \ldots, k.
\end{equation}
One is hence confronted with the new optimization problem
\begin{equation}\label{general problem}
M(\rho) = \min_{k \geq r}\inf_{U\in St(k, r)} h(U, \rho),
\end{equation}
where $h(U, \rho)$ is the convex sum on the right-hand side of
eq.~\eqref{convex-roof em}. In practice, one can of course only
investigate this problem for a few values of $k$. However, numerical
studies show that a $k$ not much larger than $r$ is sufficient for
obtaining accurate results \cite{Audenaert2001, Rothlisberger2008, Rothlisberger2008a}.
General-purpose numerical algorithms tackling eq.~\eqref{general
problem} for arbitrary pure-state entanglement monotones have been
presented very recently \cite{Rothlisberger2008,
Rothlisberger2008a}, together with studies of entanglement in states
emerging from physical Hamiltonians.

\section{Spin manipulation in GaAs quantum dots}\label{sec:spin manipulation in GaAs}

As announced, we will restrict ourselves in the rest of the review
mostly to lateral GaAs quantum dots. This is motivated by the
remarkable latest achievements we have witnessed in the field. The
main features of the spin-qubit proposal of ref.~\cite{Loss1998}, in
particular the single and two-qubits gates, have by now been
realized in single and double GaAs quantum dots with various degrees
of accuracy. It is therefore meaningful to review the field in the
light of the five DiVincenzo criteria for scalable quantum computing
presented in subsect.~\ref{sec:DiVcriteria}.

\subsection{\label{sec:QbitsGaAs}Realization of well-defined spin qubits}

The qubits we are considering are obtained in a standard
two-dimensional electron gas (2DEG) formed at the interface of a
GaAs/AlGaAs heterostructure, as illustrated in fig.~\ref{fig:Qdots}.
The electron gas is then depleted by means of metallic top gates in
order to define the confinement region of the quantum dots. We refer
to refs.~\cite{Hanson2007} and \cite{Coish2007} for a detailed
discussion of the stability diagram of single and double quantum
dots. The main features thereof can be recovered from a simple
charging Hamiltonian \cite{Coish2007}
\begin{equation}\label{chargingH}
H_0=\frac{U}{2}\sum_i N_i(N_i-1)+U_{12} N_1 N_2 -e \sum_i V_i N_i +
\sum_{i,m} \epsilon_m n_{i,m},
\end{equation}
where we assumed the two dots ($i=1,2$) to be identical for
simplicity. Here, $U$ and $U_{12}$ are the on-site and
nearest-neighbor Coulomb repulsions, respectively, $V_{i}$ are the
local potentials at each dot, and the $\epsilon_m$ denote
single-particle orbital energies with occupation
$n_{i,m}=n_{i,m,\uparrow}+n_{i,m,\downarrow}$. Furthermore,
$N_i=\sum_m n_{i,m}$ is the number of electrons in dot $i$. Smaller
but still relevant corrections to $H_0$ are determined by external
fields (\emph{e.g.}, magnetic fields), spin-couplings to the environment
(\emph{e.g.}, via spin-orbit or hyperfine interaction, see
subsect.~\ref{sec:relax_short} and sect.~\ref{sec:relax}) and tunneling. The
simplest example of such an additional term is the Zeeman coupling
\begin{equation}\label{zeemanH}
H_Z=g\mu_B {\bf B}\cdot \sum_j {\bf S}_j,
\end{equation}
where ${\bf B}$ is the externally applied magnetic field and ${\bf
S}_j=\boldsymbol{\sigma}_j/2$ is the spin-$\frac12$ operator of the
$j$-th electron in the double dot, $j=1,2,\ldots, N_1+N_2$. In the
following, we generally define the quantization direction to be
along $\bf B$, with the $|{\uparrow}\rangle$ orientation having
lower Zeeman energy since $g < 0$ in GaAs.

\begin{figure}
\begin{center}
\includegraphics[width=0.8\textwidth]{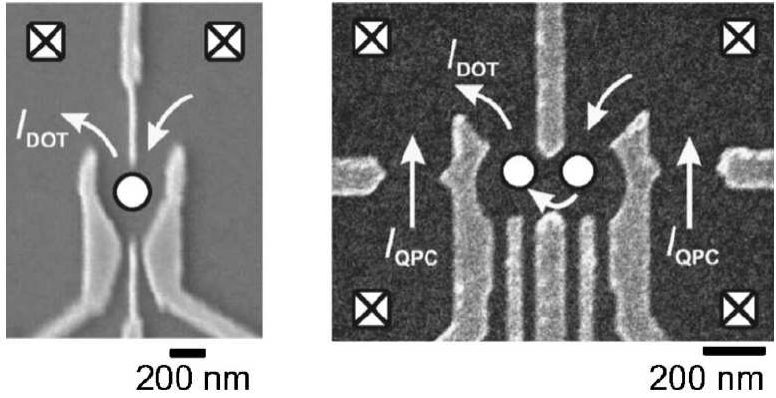}
\caption{\label{fig:devices} Scanning electron micrographs of a
single (left) and double (right) quantum dot. The electrons are
confined to the regions indicated by the white circles, surrounded
by metallic top gates (light gray areas), and the square boxes
indicate the Ohmic contacts. In the double dot setup, two quantum
point contacts used for charge sensing can be seen. (Reprinted figures with permission from ref.~\cite{Hanson2007}. \emph{Copyright (2007) by the American Physical Society.})}
\end{center}
\end{figure}

The metallic gates allow one to control the potentials $V_i$, which
determine the ground-state occupation denoted by $(N_1,N_2)$. The
spin qubits are realized by occupying each dot with exactly one
electron.
While control of the electron number down to single occupancy was
achieved early on for other types of dots (\emph{e.g.}, vertical quantum
dots \cite{Tarucha1996}), the lateral confinement tends to suppress
the tunneling rates with the reservoirs. This problem leads to
difficulties in observing the few-electron regime but can be
overcome by designing proper gating structures (figure~\ref{fig:devices}
shows two examples of actual samples). For this reason, the first
demonstrations of few-electron single \cite{Ciorga2000} and double
dots \cite{Elzerman2003, Hayashi2003, Petta2004} with lateral gating
are much more recent than for vertical dots.

\subsection{Initialization of the spin
state}\label{sec:initialization}

A straightforward procedure to initialize a systems of qubits is to
apply a sufficiently large magnetic field and wait for relaxation to
the ground state $|{\uparrow\uparrow\uparrow} \ldots \rangle$ to
occur. Experiments are usually performed in dilution refrigerators
with base temperature around 20 mK, which is smaller than typical
Zeeman splittings ($\sim 300$~mK at $B={1~\rm T}$ and using the bulk
value $g = -0.44$). The initialization time is of the order of a few
relaxation times, which in GaAs dots have been reported to be as
high as $\sim 1~{\rm s}$ (see subsubsect.~\ref{sec:time_scales} for a more
complete discussion). In the following, we discuss several other
techniques used in practice to initialize single and double GaAs
quantum dots to configurations other than
$|{\uparrow\uparrow}\rangle$. This allows, \emph{e.g.}, for more
flexibility in the subsequent manipulation of the double dot spin
state.

\subsubsection{\label{sec:singledot_states}Singlet-triplet transition in single dots}

We consider here an isolated dot (more precisely, the $i=2$ dot in
eq.~(\ref{chargingH})) and show that, if two electrons are present,
the ground state can be chosen to be either a singlet or a triplet,
depending on the value of the external magnetic field.
Initialization in the desired spin state can, in principle, be
accomplished easily by energy relaxation. We start from the lowest
energy single electron states with charge configuration (0,1).
Clearly, they only differ due to the spin and have energies
$E_\pm(0,1)=\epsilon_0-eV_2\mp \Delta E_Z/2$ (see
eqs.~\eqref{chargingH} and \eqref{zeemanH}), where $\Delta E_Z=|g
\mu_B B|$ is the Zeeman splitting. If now one more electron is
added, singlet and triplet states can be formed. The lowest lying
states are denoted by $S(0,2)$ and $T_m(0,2)$, where the subscript
$m=0, +, -$ refers to the component of the total spin parallel to
${\bf B}$. The energies are given by $E_{S}(0,2)=2(\epsilon_0- e
V_2)+U$ and $E_{T_m}(0,2)=E_{S}(0,2)+\Delta E_{ST}- m \Delta E_Z$,
where the single-triplet splitting at zero magnetic field is given
by $\Delta E_{ST}=\Delta \epsilon_{orb}-\Delta U$. Here, $\Delta
\epsilon_{orb}=\epsilon_1-\epsilon_0$ is the difference in orbital
energies, which would be the only contribution to $\Delta E_{ST}$
according to the simple charging Hamiltonian eq.~(\ref{chargingH}).
However, the splitting is experimentally found to be smaller than
$\Delta \epsilon_{orb}$ due to a change $\Delta U$ in the charging
energy \cite{Hanson2007}. The splitting $\Delta E_{ST}$ is generally
still positive, resulting in a singlet ground state at zero magnetic
field.

Interestingly, the Zeeman energy is smaller than $\Delta E_{ST}$ for
typical values of the magnetic field. Therefore, an in-plane
magnetic field cannot induce a singlet-triplet transition since, as
a good approximation, it does not affect the orbital states.
Instead, a significant decrease of $\Delta \epsilon_{orb}$ and
increase of $\Delta U$ is produced by a magnetic field $B_\perp$
perpendicular to the 2DEG due to orbital effects. The energy
crossing of singlet and triplet is typically realized around
$B_\perp^0\simeq 1$ T \cite{Kyriakidis2002,Zumbuhl2004}. This
condition is required for the realization of the parity gate
discussed in subsubsect.~\ref{sec:measurement based qc}. Furthermore,
$B_\perp^0$ can be tuned via electric gates \cite{Zumbuhl2004}. In
the following, however, we will usually neglect orbital effects of
the magnetic field, supposing $\bf B$ to be either in-plane or
sufficiently small.

\subsubsection{\label{sec:pauli_blockade}Pauli spin blockade in double dots}

\begin{figure}
\begin{center}
\includegraphics[width=0.25\textwidth]{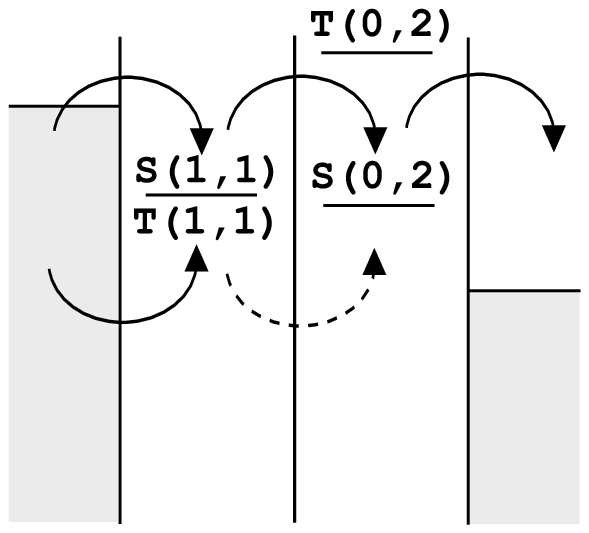}
\hspace{2cm}
\includegraphics[width=0.25\textwidth]{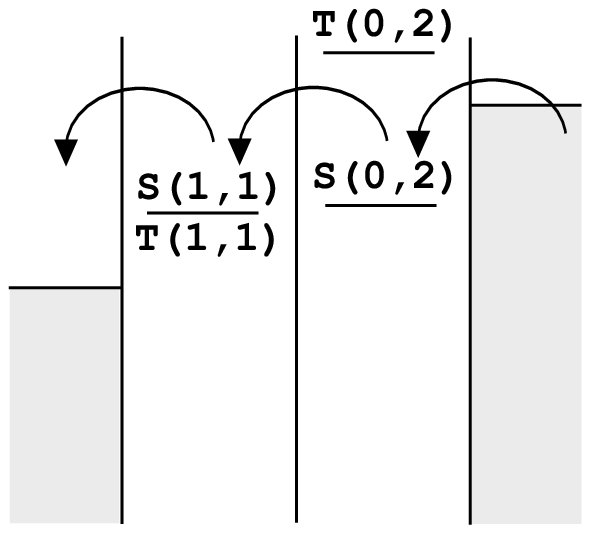}
\vspace{0.5cm} \caption{\label{fig:blockade} In the left panel, the
Pauli spin blockade is illustrated. At positive bias, an electron
from the left reservoir can tunnel to the right contact only if it
enters the (0,1) double dot in a singlet configuration $S(1,1)$,
while the triplet $T(1,1)$ is blocked due to spin conservation
(dashed transition). At negative bias the electron can only tunnel
in the $S(0,2)$ singlet and transport is always possible. The position of 
the chemical potentials $\mu_\alpha(1,1)$ ($\alpha=S, T_m$)
is indicated in the first dot, while $\mu_S(0,2)$ and $\mu_{T_m}(0,2)$ 
are shown in the second one. The detailed structure of
the (1,1) and $T(0,2)$ levels is not specified, since it is of an energy scale
much smaller than the $S(0,2)$-$T(0,2)$ singlet-triplet splitting
(see subsubsect.~\ref{detuning} for a more detailed discussion).}
\end{center}
\end{figure}

The lowest-lying (1,1) spin states are the singlet $S(1,1)$ and the
triplets $T_m(1,1)$. The energies $E_{\alpha}(1,1)$ (where
$\alpha=S$, $T_m$) are degenerate in first approximation and
eq.~(\ref{chargingH}) gives $E_{\alpha}(1,1)\simeq
E(1,1)=2\epsilon_0-e(V_1+V_2)+U_{12}$. It is however possible to
selectively prepare the system in a triplet state via Pauli spin
blockade \cite{Ono2002, Johnson2005}. This is realized at
\emph{positive} bias if the chemical potentials $\mu_{\alpha}(1,1)$
and $\mu_{\alpha}(0,2)$ of the double dot are adjusted as shown in
the left panel of fig.~\ref{fig:blockade}. The chemical potentials
with respect to the $(0,1)$ occupation are defined as
$\mu_{\alpha}(0,2)=E_{\alpha}(0,2)-E(0,1)$ and
$\mu_{\alpha}(1,1)=E_{\alpha}(1,1)-E(0,1)$, and are shown in
fig.~\ref{fig:blockade} (for simplicity, we neglect the presence of
small Zeeman splittings and assume $E_+(0,1)\simeq E_-(0,1) \equiv E(0,1)$). Tunneling occurs from the left reservoir
which is connected to the first dot, to the right reservoir
connected to the second dot. The sequence $(0,1) \rightarrow (1,1)
\rightarrow S(0,2) \rightarrow (0,1)$ would be energetically allowed
both through $S(1,1)$ and $T_m(1,1)$, but the transition
$T_m(1,1)\rightarrow S(0,2)$ is forbidden due to spin conservation.
Therefore, as soon as an electron tunnels from the left reservoir to
$T_m(1,1)$, the double dot is blocked in the triplet state.
Transport is only possible after relaxation into $S(1,1)$, which can
occur on a millisecond time scale.

Note that at \emph{negative} bias (cf. the right panel of
fig.~\ref{fig:blockade}) a finite current can flow through the dots
following the sequence $(0,1) \rightarrow S(0,2) \rightarrow S(1,1)
\rightarrow (0,1)$. In this case electrons can only tunnel from the
right reservoir to a singlet state, since the triplet is too high in
energy, and the $T_m(1,1)$ states are never involved. The Pauli spin
blockade effect thus leads to current rectification.

\subsubsection{\label{detuning} Singlet-triplet and charge states mixing in double dots}

In the presence of an external magnetic field, the triplet energies
are $E_{T_m}=E(1,1)-m \Delta E_Z$. Therefore, the states $T_0(1,1)$
and $S(1,1)$ remain degenerate. However, the magnetic fields on the
two dots generally differ slightly due to different nuclear
configurations (see subsubsect.~\ref{sec:hyperfine_intro} on the hyperfine
interaction). We assume that the value of the magnetic field is
$B\pm\Delta B_N$ at the left/right dot, where $\Delta B_N \sim
2~{\rm mT}$. This causes the relevant eigenstates to have spin
configuration $|{\uparrow\downarrow}\rangle$ or
$|{\downarrow\uparrow}\rangle$ (singlet-triplet mixing), with
energies $E_{\uparrow\downarrow(\downarrow\uparrow)}=E(1,1)\mp
g\mu_B \Delta B_N$. A first consequence of this fact is that the
Pauli spin blockade discussed above occurs only for a mixture of
$T_\pm(1,1)$, but not for $T_0(1,1)$. The reason is that $T_0(1,1)$
rotates to $S(1,1)$ due to the field inhomogeneity $\Delta B_N$,
which removes the spin blockade since $S(1,1)$ can tunnel to
$S(0,2)$. A second consequence is the possibility to initialize the
system into the spin configurations $|{\uparrow\downarrow}\rangle$
or $|{\downarrow\uparrow}\rangle$. This requires a more
sophisticated procedure relying on mixing of charge states, as
described in the following.

\begin{figure}
\begin{center}
\includegraphics[width=0.45\textwidth]{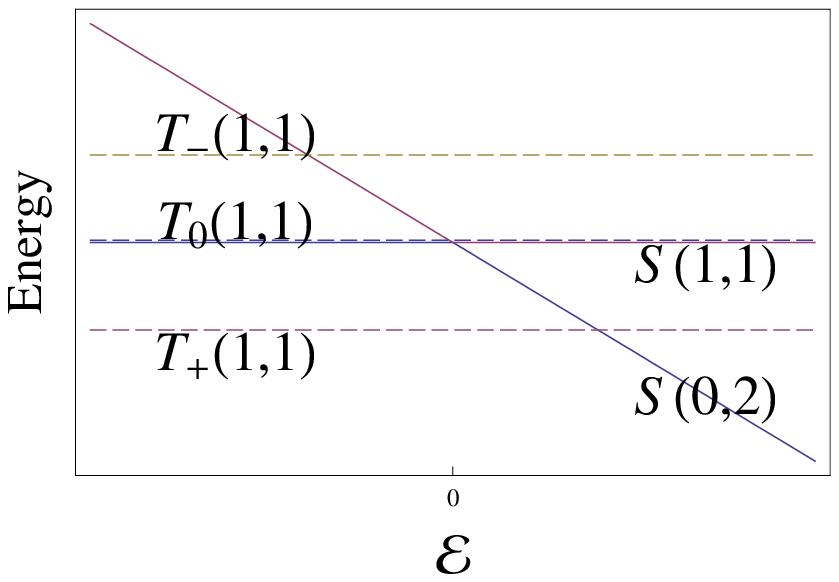}
\includegraphics[width=0.45\textwidth]{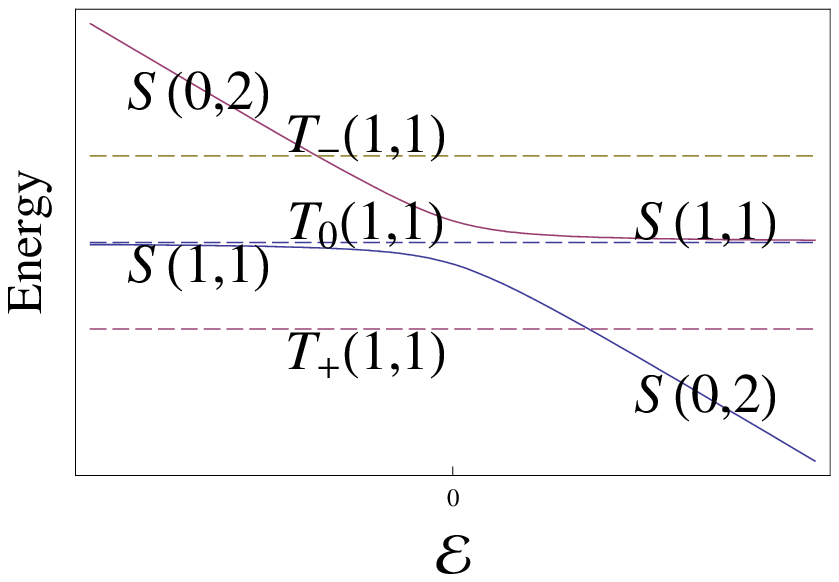}
\caption{\label{fig:detuning} Double-dot energy diagram as a
function of the detuning $\varepsilon$, close to $\varepsilon =0$.
Without tunneling (left panel) $S(1,1)$ and $T_0(1,1)$ are always
degenerate, while the $T_\pm(1,1)$ are split by a finite magnetic
field. At $\varepsilon=0$, the two singlet states $S(1,1)$ and
$S(0,2)$ are also degenerate. Inter-dot tunneling (right panel)
causes mixing of the $S(1,1)$ and $S(0,2)$ states. The effect of a
small difference $\Delta B_N$ in the nuclear fields of the two dots
is not shown here, and is only relevant when the splitting between
$S(1,1)$ and $T_0(1,1)$ is small (\emph{i.e.}, in the left panel
(independent of $\varepsilon$) and at large detuning
($|\varepsilon|\gg 0$) in the second one). Then, the appropriate
eigenstates have spin $|{\uparrow\downarrow}\rangle$ and
$|{\downarrow\uparrow}\rangle$, as discussed in the text.}
\end{center}
\end{figure}

For simplicity, we neglect for the moment the small field
inhomogeneity due to $\Delta B_N$. We also define the detuning
$\varepsilon$ of the local potentials at the two dots as follows:
$eV_1=e \tilde V_1-\varepsilon/2$ and $eV_2=e\tilde
V_{2}+\varepsilon/2$, where the $\tilde V_i$ are constant potentials
such that $S(1,1)$ and $S(0,2)$ are degenerate at $\varepsilon=0$.
The detuning $\varepsilon$ changes the energy of the (0,2) singlet,
$E_S(0,2)=E_S(1,1)-\varepsilon$, while  $E_S(1,1)$ and other (1,1)
states are not affected. These double dot levels are shown in the
left panel of fig.~\ref{fig:detuning}, while the triplet states
$T_m(0,2)$ have much higher energy and are thus not depicted.
Consider next the effect of tunneling in the vicinity of
$\varepsilon=0$. The tunneling Hamiltonian $H_T$ has a matrix
element $\langle S(1,1)|H_T | S(0,2)\rangle = \sqrt{2} t_c$ and
causes mixing of the singlets with different charging
configurations. At $\varepsilon=0$ perfect mixing is realized, with
energy splitting $2\sqrt{2}t_c$, while at large $\varepsilon$ the
unperturbed eigenstates $S(0,2)$ and $S(1,1)$ are recovered. At large positive detuning,
$S(0,2)$ is lower in energy than the $T_+(1,1)$ state, and
initialization can hence be performed via energy relaxation. If one
then slowly changes $\varepsilon$ toward negative values, the
system evolves adiabatically along the lower singlet branch into the
$S(1,1)$ state (cf. the right panel of fig.~\ref{fig:detuning}). The
leakage to the $S(0,2)$ state is estimated in ref.~\cite{Coish2007}.
Note also that $T_+(1,1)$ is not mixed with the singlet in this
simple model\footnote{In reality, small spin perturbations cause
anticrossing of the singlet branch with $T_+(1,1)$. In experiment,
$\varepsilon$ is swept faster around the $S-T_+$ degeneracy in order
to avoid the $T_+(1,1)$ state \cite{Petta2005}.}.

We now consider the effect of $\Delta B_N$, which is important
whenever the $S-T_0$ splitting is small. Since it usually holds that $t_c
\gg |g\mu_B \Delta B_N|$, this is only the case if the detuning
becomes large in magnitude. In this limit the splitting goes to zero
(see fig.~\ref{fig:detuning}) and the inhomogeneity becomes the
dominant effect. If the detuning $\varepsilon$ is decreased from
large positive to large negative values faster than the time scale
determined by $\Delta B_N$, the system will be initialized to
$S(1,1)$ and will then begin to oscillate between $S(1,1)$ and
$T_0(1,1)$ with frequency $2g \mu_B \Delta B_N/h$. Instead, by
adiabatically reducing the value of $\varepsilon$, the system can be
initialized to the spin configuration with lower energy,
$|{\uparrow\downarrow}\rangle$ or $|{\downarrow\uparrow}\rangle$,
depending on the sign of $\Delta B_N$.

\subsection{\label{sec:relax_short}Relaxation and decoherence in GaAs dots}

The requirement of sufficiently long coherence times is perhaps the
most challenging aspect for quantum computing architectures in the
solid state. It requires a detailed understanding of the different
mechanisms that couple the electron's spin to its environment. We
introduce here the main concepts relevant for GaAs dots, while a
detailed discussion is postponed to sect.~\ref{sec:relax}.

\subsubsection{\label{sec:SOC}Spin-orbit coupling}

While fluctuations in the electrical environment do not directly
couple to the electron spin, they become relevant for spin
decoherence in the presence of spin-orbit interaction. In GaAs 2DEGs
two types of spin-orbit coupling (Dresselhaus and Rashba) are
present. The Dresselhaus spin-orbit coupling originates from the
bulk properties of GaAs \cite{Dresselhaus1955}. The zinc-blend
crystal structure has no center of inversion symmetry and a term of
the type $H^{3D}_D\propto
p_x(p_y^2-p_z^2)\sigma_x+p_y(p_z^2-p_x^2)\sigma_y+p_z(p_x^2-p_y^2)\sigma_z$
is allowed in three dimensions, where ${\bf p}$ is the momentum
operator and $\boldsymbol{\sigma}$ are the Pauli matrices. Due to
the confining potential along the $z$-direction, we can substitute
the $p_z$ operators with their expectation values. Using $\langle
p_z^2 \rangle\neq 0$ and $\langle p_z\rangle=0$, one obtains
\begin{equation}
H_D=\beta (p_y \sigma_y - p_x \sigma_x).
\end{equation}
Smaller terms cubic in ${\bf p}$ have been neglected, which is
justified by the presence of strong confinement.

The Rashba spin-orbit coupling is due to the asymmetry of the
confining potential \cite{Bychkov1984} and can be written in the
suggestive form $H_R\propto (\boldsymbol{\mathcal{E}} \times {\bf
p})\cdot \boldsymbol{\sigma}$, where $
{\boldsymbol{\mathcal{E}}}=\mathcal{E} \hat z$ is an effective
electric field along the confining direction:
\begin{equation}
H_R=\alpha (p_x \sigma_y - p_y \sigma_x).
\end{equation}

The Rashba and Dresselhaus terms produce an internal magnetic field
linear in the electron momentum defined by ${\bf B}_{SO} = -2[(\beta
p_x+\alpha p_y){\bf e}_x -(\beta p_y+\alpha p_x){\bf e}_y]/g\mu_B$.
If $\beta=0$, the magnitude of ${\bf B}_{SO}$ is isotropic in ${\bf
p}$ and the direction is always perpendicular to the velocity. While
moving with momentum ${\bf p}$, the spin precesses around ${\bf
B}_{SO}$ and a full rotation is completed over a distance of order
$\lambda_{SO}=|\hbar/(\alpha m^*)|=1-10~\mu{\rm m}$, where $m^*$ is
the effective mass. Generally, Rashba and Dresselhaus spin-orbit
coupling coexist, their relative strength being determined by the
confining potential. This results in the anisotropy of the
spin-orbit coupling in the 2DEG plane (\emph{e.g.}, of the spin splitting
as function of ${\bf p}$). In this case, two distinct spin-orbit
lengths can be introduced
\begin{equation}\label{lambdaSO}
\lambda_\pm=\frac{\hbar}{m^* (\beta\pm \alpha)}.
\end{equation}

For GaAs quantum dots, the spin-orbit interaction is usually a small
correction that can be treated perturbatively since the size of the
dot (typically $\sim 100$ nm) is much smaller than the spin-orbit
coupling lengths $\lambda_\pm$. The qualitative effect introduced by
the spin-orbit coupling is a small mixing of the spin eigenstates.
As a consequence, the perturbed spin eigenstates can be coupled by
purely orbital perturbation even if the unperturbed states have
orthogonal spin components. Relevant charge fluctuations are
produced by lattice phonons, surrounding gates, electron-hole pair
excitations, etc. with the phonon bath playing a particularly
important role (see subsect.~\ref{sec:SOC_complete}).

\subsubsection{Hyperfine interaction \label{sec:hyperfine_intro}}

The other mechanism for spin relaxation and decoherence that has
proved to be effective in GaAs dots, and ultimately constitutes the
most serious limitation of such systems, is due to the nuclear spins
bath. All three nuclear species ${}^{69}$Ga, ${}^{71}$Ga, and
${}^{75}$As of the host material have spin 3/2 and interact with the
electron spin via the Fermi contact hyperfine interaction
\begin{equation}\label{hyperfine}
H_{HF}={\bf S}\cdot \sum_i A_i {\bf I}_i ,
\end{equation}
where $A_i$ and ${\bf I}_i$ are the coupling strengths and the
nuclear spin operator at site $i$, respectively. The density of
nuclei is $n_0=45.6~{\rm nm}^{-3}$ and there are typically $N\sim
10^6$ nuclei in a dot. The strength of the coupling is proportional
to the electron density at site $i$, and one has $A_i=A |\psi({\bf
r}_i)|^2/n_0$, where $\psi({\bf r})$ is the orbital envelope
wave function of the electron and $A \approx 90~\mu$eV\footnote{This
value is a weighted average of the three nuclear species
${}^{69}$Ga, ${}^{71}$Ga, and ${}^{75}$As, which have abundance 0.3,
0.2, and 0.5, respectively. For the three isotopes we have
$A=\frac{8\mu_0}{9} \mu_B \mu_I \eta n_0$, where
$\mu_I=(2.12,~2.56,~1.44)\times\mu_N$, while $\eta_{\rm
Ga}=2.7~10^3$ and $\eta_{\rm As}=4.5~10^3$ \cite{Paget1977}.}.

The study of the hyperfine interaction (\ref{hyperfine}) represents
an intricate problem involving subtle quantum many-body correlations
in the nuclear bath and entangled dynamical evolution of the
electron's spin and nuclear degrees of freedom. While these topics
will be discussed much more deeply in subsect.~\ref{sec:HF_complete}, it
is nevertheless useful to present here a qualitative picture based
on the expectation value of the Overhauser field ${\bf B}_N = \sum_i
A_i {\bf I}_i/g \mu_B$. This field represents a source of
uncertainty for the electron dynamics, since the precise value of
${\bf B}_N$ is not known. Due to the fact that the nuclear spin bath
is in general a complicated mixture of different nuclear states (see
subsubsect.~\ref{sec:InitialState} for a more detailed discussion of the
nuclear density matrix), the operator ${\bf B}_N$ in the direction
of the external field ${\bf B}$ does not correspond to a
well-defined eigenstate, but results in a statistical ensemble of
values.
These fluctuations have an amplitude of order
$B_{N,max}/\sqrt{N}\sim 5$~mT since the maximum value of $B_N$ (with
fully polarized nuclear bath) is about 5~T.

Finally, even if it were possible to prepare the nuclei in a
specific configuration (\emph{e.g.},
$|{\uparrow\uparrow\downarrow\uparrow}\ldots\rangle$), the nuclear
state would still evolve in time to a statistical ensemble on a time
scale $t_{nuc}$. Although direct internuclear interactions are
present (\emph{e.g.}, magnetic dipole-dipole interactions between nuclei),
the most important contribution to the bath's time evolution is in
fact due to the hyperfine coupling itself, causing the back action
of the electron spin on the nuclear bath. Estimates of the nuclear
bath timescale lead to $t_{nuc}=10-100~\mu$s or longer at higher
values of the external magnetic field ${\bf B}$ \cite{Hanson2007}.

\subsubsection{\label{sec:time_scales}Relevant time scales}

We provide here a summary of the relevant time scales for spin
decoherence in GaAs dots. In the Bloch phenomenological description
of the time evolution, the spin density matrix $ \rho= (1+{\bf
P}\cdot \boldsymbol{\sigma})/2$ (where ${\bf P}$ is the spin
polarization) satisfies
\begin{equation}\label{eq:bloch}
\dot{\bf P}=\frac{g\mu_B}{\hbar} {\bf B}\times {\bf
P}-\boldsymbol{\Gamma}({\bf P}- {\bf P}_0),
\end{equation}
where the tensor $\Gamma_{ij}$ is diagonal in a reference frame with
the $z$-axis along $\bf B$. With this choice, the equilibrium
polarization is ${\bf P}_0=P_0 {\bf e}_z$. The time
$T_1=\Gamma_{zz}^{-1}$ is the longitudinal spin decay time, or
spin-flip time, and describes the energy relaxation to the ground
state. In GaAs quantum dots $T_1$ has a strong magnetic field
dependence and can be very long, ranging from $1$ ms around $5~{\rm
T}$ to more than 1 s at $1~{\rm T}$ \cite{Amasha2008}. This
dependence originates entirely from the spin-orbit interaction
since, at such high values of the magnetic field, the hyperfine
coupling plays no role for energy relaxation (due to the large
mismatch between the nuclear and electron Zeeman energies). At small
magnetic fields the spin-orbit coupling becomes ineffective and, in
fact, does not cause any relaxation at $B=0$ \cite{Golovach2004}.
Nevertheless, the hyperfine interaction contributes to the reduction
of $T_1$ to much smaller values (down to $10 - 100$ ns, due to
electron-nuclear flip-flops \cite{Hanson2007}).

The transverse spin decay time
$T_2=\Gamma_{xx}^{-1}=\Gamma_{yy}^{-1}$  describes the decay of the
transverse polarization components $P_x$ and $P_y$. The $T_2$ time
cannot be larger than $2T_1$. This maximal value is obtained if only
the spin-orbit coupling were present \cite{Golovach2004}. However,
$T_2$ is dominated by the hyperfine interaction and is much shorter
than $T_1$. Due to the fluctuations of the Overhauser field in the
nuclear bath's initial state, a transverse decay time of order $10$
ns is obtained (see subsubsect.~\ref{sec:InitialState}). In this case, it
is clear that the much longer timescale $t_{nuc}=10-100~\mu$s does
not play a role for the transverse electron spin evolution. This
decay time is usually denoted as $T_2^{*}$ and referred to as
'ensemble-averaged' transverse spin decay time. We note that the
decoherence process is generally non-exponential (see
subsubsect.~\ref{sec:InitialState}).

If the initial nuclear state is prepared in an eigenstate of the
Overhauser field in the ${\bf B}$ direction, an 'intrinsic' decay
time $T_2$ is obtained. A technique for narrowing the initial
nuclear state was proposed in ref.~\cite{Klauser2006} and is
discussed in subsubsect.~\ref{sec:narrow}. The decay time $T_2$ is
determined in this case by the coupled dynamics of the electron spin
and the nuclear bath. It is comparable to the $t_{nuc}$ time scale
(estimates give $T_2\sim 1-100~\mu$s) and therefore much longer than
$T_2^*$. However, $T_2$ is clearly very difficult to access
experimentally. A quantity more easily measured is the spin echo
decay time $T_{echo}$. We refer to ref.~\cite{Abragam1961} for a
description of the spin echo technique, and to ref.~\cite{Petta2005}
for its application to GaAs double dots. This method can be used to
perfectly refocus an ensemble of spins in the idealized case where
decoherence is only due to static fluctuations of the environment.
However, in reality the initial polarization cannot be completely
recovered due to the time evolution of the nuclear bath. A decay
time $T_{echo}>1~\mu$s is reported in ref.~\cite{Petta2005} at
100~mT.

\subsection{\label{sec:gates}Universal quantum gates}

Both single- and two-qubit gates have been demonstrated in GaAs
quantum dots. The single gate was realized in
ref.~\cite{Koppens2006} by means of the well-known electron spin
resonance (ESR), which we briefly describe here (for a more extended
discussion see, \emph{e.g.}, ref.~\cite{Abragam1961}). An oscillating
magnetic field is applied in the transverse direction (perpendicular
to $\bf B$) at the resonant frequency $\omega=\Delta E_Z/\hbar$.
This ESR field can be seen as a sum of two contributions, rotating
clockwise and counterclockwise around $\bf B$ at the same frequency
$\omega$. However, only the contribution precessing in resonance
with the electron spin is of relevance. We denote this component by
${\bf B}_1$, while the counter-propagating field is neglected in the
following.

Consider now the effect on the electron spin. Without the ESR
signal, the spin simply precesses around ${\bf B}$ with angular
frequency $\omega$. It is useful to introduce a reference frame
rotating around ${\bf B}$ in which the precessing spin appears
static. We now apply ${\bf B}_1$, which also appears static in the
rotating frame. The effect is to induce a precession of the spin
around ${\bf B}_1$ \emph{in the rotating frame}.
In particular, if the spin is initialized along $\bf B$, a complete
spin-flip is realized after a time $\pi \hbar/g\mu_B B_1$. Typical
fields in the experimental setup \cite{Koppens2006} are up to $\sim
100$ mT, which gives a $\sim 15$ ns switching time.

On the other hand, the two-qubit \textsc{swap} operation was
implemented with a much faster gate duration $\sim 0.5$~ns
\cite{Petta2005}. The gate is realized in a similar way to the
original proposal of ref.~\cite{Loss1998} (see subsect.~\ref{sec:Loss DiVi
Proposal}), based on the control of the exchange coupling. In
practice, the detuning $\varepsilon$ of the double dot is changed,
since this modifies the splitting between the lower singlet branch
and the triplet $T_0$, as described in subsubsect.~\ref{detuning} and
illustrated in fig.~\ref{fig:detuning}. If a pulse from a large
negative value $\varepsilon_0$ to some value $\varepsilon_c$ around
zero and back to $\varepsilon_0$ is applied, a finite energy
splitting $J(\varepsilon_c)$ between triplet $T_0$ and singlet $S$
exists for the duration $\tau$ of the pulse. This causes the spin
state $|{\uparrow\downarrow}\rangle$ to rotate to
$|{\downarrow\uparrow}\rangle$ if the pulse has length
$\tau=\pi\hbar/J(\varepsilon_c)$, which realizes the $\textsc{swap}$
operation. The $\sqrt{\textsc{swap}}$ operation is obtained if the
time $\tau$ is half of that required by the $\textsc{swap}$ gate.

\subsubsection{\label{sec:EDSR}Electrical manipulation of individual spins}

While standard ESR is useful for single spin manipulation
\cite{Koppens2006} and can in principle be applied to the individual
dots of a large array (see subsect.~\ref{sec:Loss DiVi Proposal}), it is
much more convenient to perform coherent spin rotations through the
electric gates at the individual dots. An example of such a
technique is the electric-dipole-induced spin resonance (EDSR),
which is well known in two dimensions \cite{Rashba2003,Duckheim2006,
Kato2004} and was also studied in lower dimensional-systems
\cite{Levitov2003,Golovach2006}. EDSR in quantum dots was
investigated theoretically in ref.~\cite{Golovach2006}, which
discusses in detail the effect on the electron spin of an external
ac \emph{electric} field mediated by the spin-orbit interaction. In
the following, we review the main results of this analysis.

The single dot is described by the two-dimensional Hamiltonian
\begin{equation}\label{EDSRhmilt}
H=\frac{{\bf p}^2}{2m^*}+U({\bf r})  + H_{SO}+H_Z- e {\bf E}_0\cdot
{\bf r}\sin(\omega t),
\end{equation}
where ${\bf r}=(x,y)$ is the electron's coordinate. The second term
is the lateral confining potential of the dot and the third term is
the spin-orbit coupling (Rashba and Dresselhaus) discussed in
subsubsect.~\ref{sec:SOC}. For the present section, it is convenient to
define new axes ${\bf e}_x=({\bf a}_x+{\bf a}_y)/\sqrt{2}$, ${\bf
e}_y=-({\bf a}_x-{\bf a}_y)/\sqrt{2}$, and ${\bf e}_z={\bf a}_z$
(instead of ${\bf e}_i={\bf a}_i$, as in subsubsect.~\ref{sec:SOC}), where
${\bf a}_{i}$ are unit vectors along the cubic axes of the crystal.
With this choice, $H_{SO}$ takes the particularly simple form
\begin{equation}\label{HSOrotated}
H_{SO}=\frac{\hbar}{m^*}\left(\frac{p_y\sigma_x}{\lambda_-}+\frac{p_x\sigma_y}{\lambda_+}\right),
\end{equation}
where $\lambda_\pm$ are the spin-orbit lengths defined in
eq.~(\ref{lambdaSO}). The third term in eq.~(\ref{EDSRhmilt}) is the
usual Zeeman coupling $H_Z=g \mu_B{\bf B}\cdot {\bf S}$ and the last
term is the external electric perturbation. The electric field is
assumed to be spatially uniform on the small region of the dot.

The unperturbed states are the eigenstates $\psi_m({\bf
r})|\pm\rangle$ of the dot Hamiltonian $H_d={\bf p}^2/2m^*+U({\bf
r})$. To calculate the effect of the oscillating electric field, one
has to resort to third order perturbation theory since the final
result has to be proportional to the spin-orbit coupling as well as
to the electric field and to the Zeeman splitting $\Delta E_Z$. No
spin-electric coupling can be obtained at $B=0$, a property related
to the invariance of $H_{SO}$ upon time-reversal. It is convenient
to approach the problem by making use of the unitary
Schrieffer-Wolff transformation $e^S H e^{-S}$ described in the
Appendix. If the confining potential is harmonic, \emph{i.e.}, $U({\bf
r})=\frac12 m^* \omega_0^2 r^2$, the final result is obtained
explicitly as $S=i \boldsymbol{\xi}\cdot
\boldsymbol{\sigma}-i\frac{g\mu_B}{m^*\hbar \omega_0^2}({\bf
B}\times \boldsymbol{\zeta})\cdot \boldsymbol{\sigma}$, where
$\boldsymbol{\xi}=(y/\lambda_-,x/\lambda_+,0)$,
$\boldsymbol{\zeta}=(p_y/\lambda_-,p_x/\lambda_+,0)$, and the spin
Hamiltonian for the ground state reads
\begin{equation}\label{EDSRresult}
H_{\rm eff}=g\mu_B{\bf B}\cdot {\bf S}+g\mu_B{\bf h}_0\cdot {\bf
S}\sin(\omega t),
\end{equation}
where
\begin{equation}\label{EDSRfield}
{\bf h}_0=2{\bf B}\times{\bf \Omega}=\frac{2e}{m^*\omega_0^2} {\bf
B}\times\left(\frac{E_{0,y}}{\lambda_-},\frac{E_{0,x}}{\lambda_+},0\right).
\end{equation}
Equation~(\ref{EDSRresult}) clearly reveals the possibility to perform
ESR-type spin manipulation, since the electric field induces an
effective oscillating magnetic field ${\bf h}_0 \sin(\omega t)$.
From the above expression, we estimate ${\bf h}_0\simeq 2~{\rm mT}$,
using $B=1~{\rm T}$, $\lambda_\pm= 10~\mu$m, and $E_0=100~{\rm
V/cm}$. This value is in agreement with a recent experiment
\cite{Nowack2007}, in which a spin-flip time around $100$ ns has
been found. Corrections to the linear spin-orbit coupling and to the
harmonic approximation of the confining potential are also
considered in ref.~\cite{Golovach2006} and are responsible for an
additional contribution to the EDSR signal which is not discussed
here. This additional term only exists in combination with orbital
effects of the magnetic field and is absent for an in-plane field
${\bf B}$ (as realized in ref.~\cite{Nowack2007}).

Finally, we note that EDSR is not the only method for spin
manipulation via oscillating electric fields. Spin-electric coupling
can be also realized by oscillating the position of the dot in the
presence of a static but inhomogeneous magnetic field, \emph{e.g.},
provided by the stray field of a nearby micromagnet
\cite{Tokura2006}. This proposal  was recently realized in
ref.~\cite{Pioro-Ladri`ere2008}. Gate-induced coherent single
spin-rotations were also reported in \cite{Laird2007} with a  setup
very similar to that of the EDSR experiment of
ref.~\cite{Nowack2007}. In that case however, the magnetic field is
applied in the $[1\bar10]$ direction, perpendicular to the electric
field modulation, and eq.~(\ref{EDSRfield}) gives ${\bf h}_0=0$.
Therefore, the spin-electric coupling in ref.~\cite{Laird2007} was
attributed to the inhomogeneous Overhauser field produced by the
hyperfine interaction.

\subsection{Readout of electron spin states}\label{sec:readout}

\begin{figure}
\begin{center}
\includegraphics[width=0.65\textwidth]{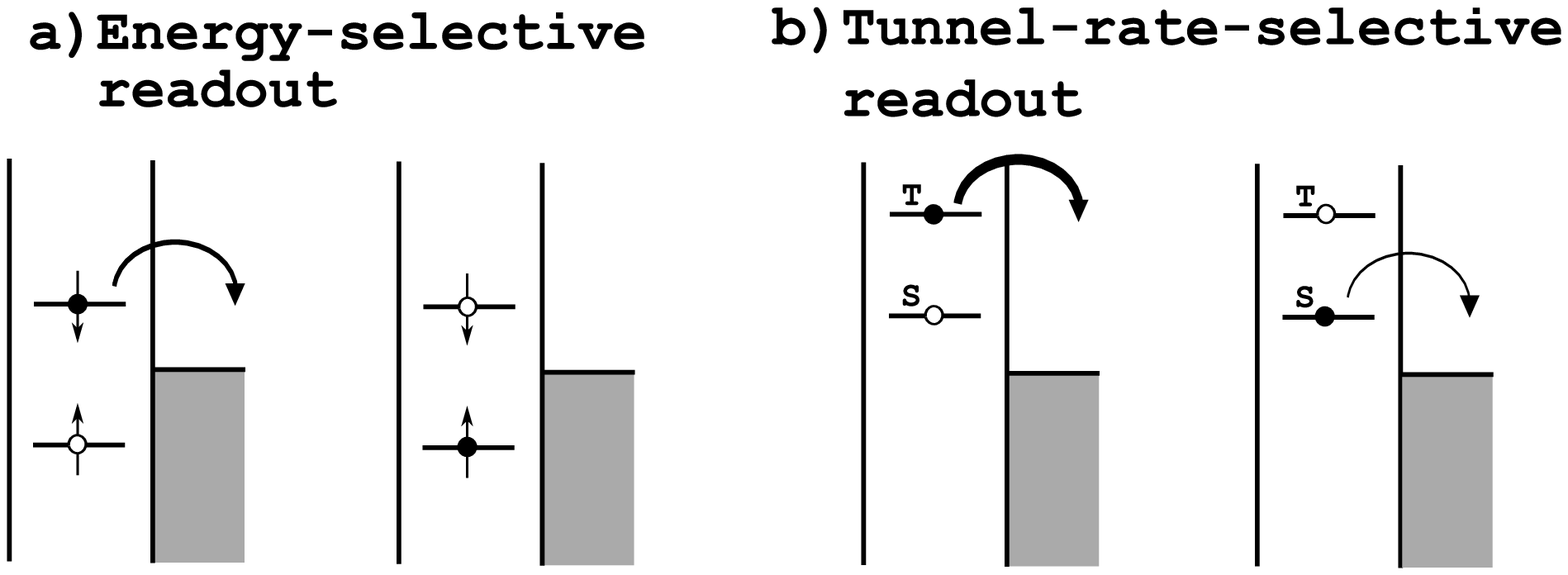}\hspace{0.5cm}
\raisebox{-0.5cm}{\makebox[0.3\textwidth]{\includegraphics[width=0.3\textwidth]{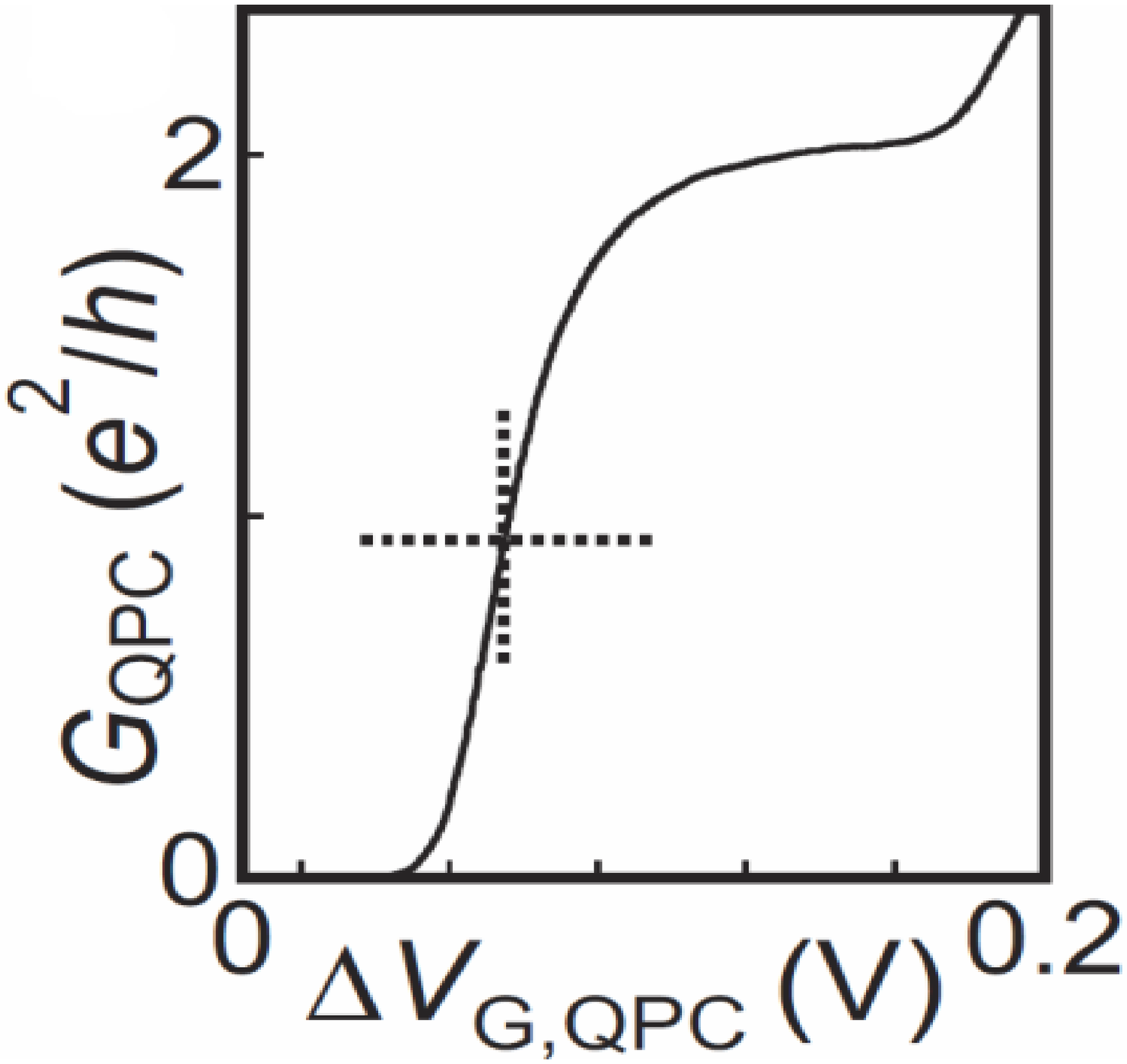}}}
\raisebox{0cm}{ \makebox[0.6\textwidth]{
\includegraphics[width=0.45\textwidth]{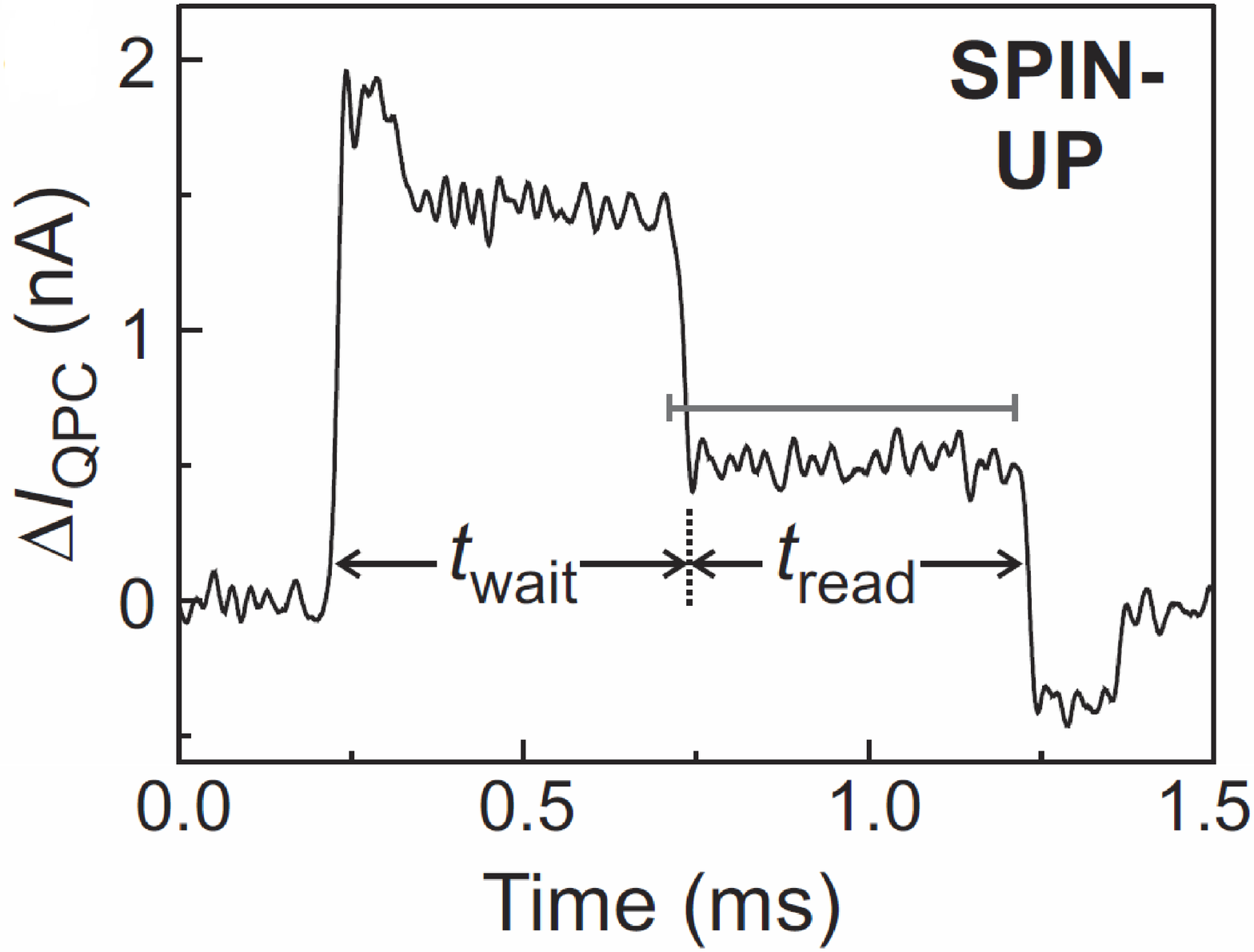}
\includegraphics[width=0.45\textwidth]{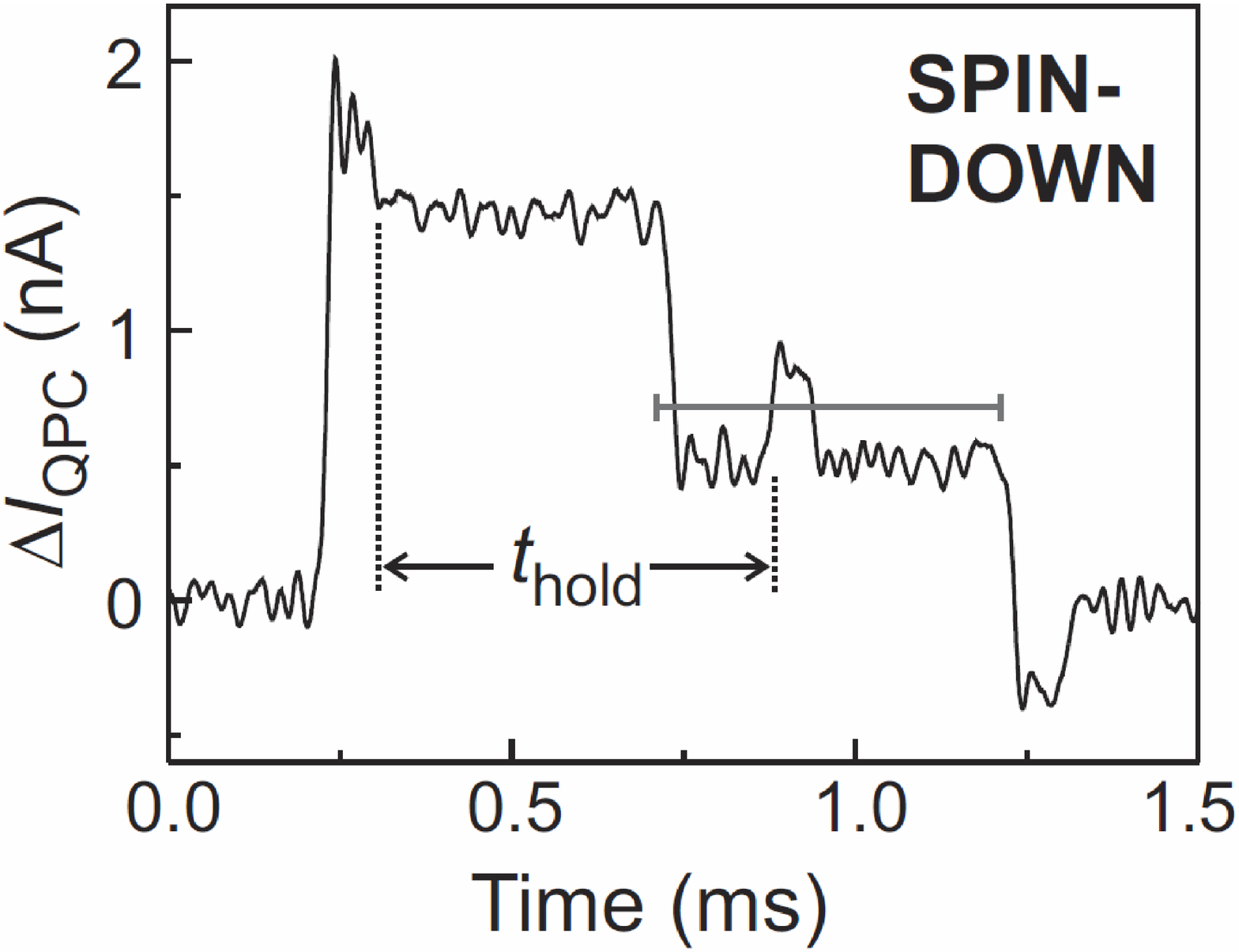}}}
\caption{\label{fig:readout} Spin readout in a single dot. In the
upper left panels, two possible configurations of the dot are shown,
which allow to distinguish between the spin states of the ground and
excited levels. In (a) only an electron in the excited state can
tunnel out, which is useful for the readout of Zeeman-split states
in a single electron dot. In (b) one of the two states (here, the
excited state) happens to have a much larger tunneling rate. This
configuration is useful for discriminating singlet and triplet
states in two-electron dots. In the upper right panel, the
conductance of a point contact is shown as a function of the voltage
of an adjacent gate. The cross marks the most sensitive position for
charge detection. In the bottom panels, an example of the point
contact current signal for energy selective readout is shown: The
actual measurement is performed in the $t_{\rm read}$ interval
(marked in both panels by the horizontal threshold bar), during
which the dot is as in (a). In the bottom right panel, a signal
above threshold is detected as long as the dot is empty. This
corresponds to the spin-$|{\downarrow}\rangle$ electron tunneling
out of the dot, followed by tunneling in of an
$|{\uparrow}\rangle$ electron. In the left panel, conversely, the
dot is always occupied by an electron in the $|{\uparrow}\rangle$
state. The point contact signal before $t_{\rm read}$ refers to the
initialization of the dot. See, \emph{e.g.}, ref. \cite{Hanson2007} for
greater detail. (The top right and the two bottom figures are reprinted with permission from ref.~\cite{Hanson2007}. \emph{Copyright (2007) by the American Physical Society.})}
\end{center}
\end{figure}

Several methods are available for reading out the spin state of
single and double quantum dots and all of them rely on the mechanism
of spin-to-charge conversion. While the electron's magnetic moment
is too small to be directly detected, the charge configuration of
the single or double dot system can be measured accurately. This is
usually accomplished by means of one or more quantum point contacts
adjacent to the dots (these are narrow constrictions of the 2DEG
through which current can flow), as shown in the right sample in
fig.~\ref{fig:devices}. The conductance of a point contact is
quantized and is, at the transition between two plateaus, highly
sensitive to the electrostatic environment, in particular to the
charge distribution in the quantum dots (see the upper right panel
in fig.~\ref{fig:readout}). For example, a general strategy for spin
readout consists in tuning the system to a configuration in which
tunneling between different charge states is allowed or suppressed,
depending on the particular spin state. The occurrence of the
tunneling process is monitored by the point contact signal, and the
corresponding spin state is inferred. We discuss below some more
specific examples.

\subsubsection{Single dot readout}

The first method for single-shot spin detection we will address here
is the so called energy-selective readout \cite{Elzerman2004}. The
chemical potentials of the single dot are aligned with the 2DEG
reservoir such that the $|{\uparrow}\rangle$ ground state is blocked
while a $|{\downarrow}\rangle$ electron can still tunnel out (see
fig.~\ref{fig:readout}). Whether the electron leaves the dot or not
is ascertained through charge sensing. It was proposed in
\cite{Engel2001,Engel2002} to demonstrate the ESR rotation of a
single spin by making use of this method. Note, however,  that the
Zeeman splitting must be much larger than the thermal broadening,
which implies large ESR excitation frequencies in the microwave
range. This has caused additional problems, \emph{e.g.}, photon-assisted
tunneling out of the dot, that are difficult to overcome
experimentally and prevented this type of experiment from
succeeding. Instead, the ESR experiment was performed in double dots
using another type of spin readout, which we discuss in the next
section. We would also like to mention that another detection method
exists for single dots in the case where there is a large difference
between the tunneling rates of two states \cite{Hanson2005}. This
method is useful to discriminate singlet and triplet states in a
two-electron dot (\emph{i.e.}, the $S(0,2)$ and $T_m(0,2)$ discussed in
subsubsect.~\ref{sec:singledot_states}), since the excited triplet has a
more extended orbital wave function and better contact with the 2DEG
reservoir.

\subsubsection{Spin readout in double dots}\label{sec:spin readout
dd}

The mechanisms used for readout of double-dot states are identical
to those described for initialization in
subsubsects.~\ref{sec:pauli_blockade} and \ref{detuning}. For example, if a
double dot is in the charge configuration (1,1), tunneling to
$S(0,2)$ is only possible if the spin state is
$|{\uparrow\downarrow}\rangle$ or $|{\downarrow\uparrow}\rangle$.
Again, the (1,1) and (0,2) states are easily distinguished by charge
sensing. This method of spin readout is used in the ESR experiment
of ref.~\cite{Koppens2006}: First, the double dot is initialized to
a mixture of $T_+(1,1)$ and $T_-(1,1)$ via Pauli spin blockade (see
left panel of fig.~\ref{fig:blockade}). The system is then brought
to the Coulomb blockade regime by decreasing the detuning
$\varepsilon$. As a consequence, the chemical potential $\mu_S(0,2)$
becomes higher than $\mu_\alpha(1,1)$ and tunneling is not allowed.
The ESR signal can now be applied to one of the two electrons\footnote{Because of the hyperfine shift $\Delta B_N$ of the
magnetic field in the two dots, only one of the two electrons is
usually in resonance with the ESR signal.}, thereby rotating the
initially parallel spin configuration ($|{\uparrow\uparrow}\rangle$
or $|{\downarrow\downarrow}\rangle$) to the antiparallel states
$|{\uparrow\downarrow}\rangle$ or $|{\downarrow\uparrow}\rangle$.
The double dot is finally brought again to the Pauli spin blockade
regime and tunneling can now occur. Repeating the procedure many
times, the probability that the single spin was rotated during the
application of the ESR signal is determined.

Another detection method can be used in the (1,1) charge
configuration to distinguish $|{\uparrow\downarrow}\rangle$ from
$|{\downarrow\uparrow}\rangle$: Consider fig.~\ref{fig:detuning},
and suppose that at large negative detuning the system is in one of
the two states $|{\uparrow\downarrow}\rangle$ or
$|{\downarrow\uparrow}\rangle$. By adiabatically changing the
detuning to large positive values, $|{\uparrow\downarrow}\rangle$
evolves to $S(0,2)$ and $|{\downarrow\uparrow}\rangle$ to $T_0(1,1)$
(for definiteness, we assume here and in the following $\Delta
B_N>0$) and the two different charge configurations can be
distinguished. We can now describe the experiment realizing the
$\sqrt{\textsc{swap}}$ operation \cite{Petta2005} in greater detail:
The system is first initialized in $|{\uparrow\downarrow}\rangle$ as
described in subsubsect.~\ref{detuning}. A pulse in $\varepsilon$ is then
applied, which introduces a large singlet-triplet splitting, as
discussed in subsect.~\ref{sec:gates}. The $\textsc{swap}$ or
$\sqrt{\textsc{swap}}$ operations are realized for appropriate pulse
lengths, but for an arbitrary pulse the double dot is brought, at
large negative detuning, into a superposition of
$|{\uparrow\downarrow}\rangle$ and $|{\downarrow\uparrow}\rangle$.
Finally, the detection method described above is applied and,
repeating this scheme many times, the probabilities of the two spin
states are measured. Singlet-triplet spin echo experiments can be
also performed with a similar procedure \cite{Petta2005}.

\section{Relaxation and spin decoherence in GaAs quantum dots \label{sec:relax}}

Electron spins in GaAs quantum dots are inevitably coupled to the
surrounding environment. This coupling results in decoherence, which
is the process leading to the loss of information stored in a qubit.
While an introduction to the mechanisms behind decoherence of
electron spins in GaAs quantum dots has already been presented in
subsect.~\ref{sec:relax_short}, we will review this topic here in much
greater detail. For the reader's convenience, we will initially
repeat some of the basic concepts with additional remarks required
for the subsequent treatment.

As already mentioned in subsubsect.~\ref{sec:time_scales}, two time scales
describing the decoherence process of a single spin can be
distinguished. The spin-flip time (also called longitudinal spin
decay time) $T_{1}$ describes the time scale for random spin-flips
$\left|\uparrow\right>\leftrightarrow\left|\downarrow\right>$,
whereas the transverse spin decay time $T_{2}$ describes the decay
of superpositions of spin-up and spin-down states
$\alpha\left|\uparrow\right>+\beta\left|\downarrow\right>$. The
first time scale $T_{1}$ is important if the qubit is operated as a
classical bit. For quantum computing, also the spin decoherence time
$T_{2}$ plays a major role and must thus be sufficiently long.

The relaxation time $T_{1}$ and the decoherence time $T_{2}$ are not
unrelated. Naively, one might expect $T_{2}\ll T_{1}$, but as shown
in \cite{Golovach2004} and as we will see below, this is not
necessarily the case for arbitrarily large Zeeman splittings. In
general, the electron spin $\bm{S}$ couples both to the external
magnetic field $\bm{B}$, and to the fluctuating internal field
$\bm{h}(t)$ with the time-averaged value $\left<\bm{h}(t)\right>=0$
(actual sources of this volatile internal field will be discussed
later on). We also assume $\left<h_{i}(t)h_{j}(t')\right>\propto
\delta_{ij}$, where $h_{i}(t)$ with $i=x,y,z$ are the components of
the vector $\bm{h}(t)$. The Hamiltonian for the single-electron spin
reads
\begin{equation}\label{eq:H_magnetic_fluctuations}
    H = g\mu_{B}\bm{S}\cdot\bm{B}+ \bm{S}\cdot\bm{h}(t),
\end{equation}
where $g$ is the $g$-factor ($g=-0.44$ in bulk GaAs) and
$\mu_{B}=9.27\times10^{-24}$~J/T is the magnetic moment of the
single electron spin. The relaxation time can be expressed in the
weak coupling limit as \cite{Slichter1980}
\begin{equation}\label{eq:T1_general}
    \frac{1}{T_{1}}=\frac{1}{2 \hbar^2} \int_{-\infty}^{\infty}dt~ \tm{Re}\left[
    \left<h_{x}(0)h_{x}(t)\right>+\left<h_{y}(0)h_{y}(t)\right>\right]e^{-i\omega_Z t},
\end{equation}
where $\omega_Z=g \mu_B B/\hbar$ is the Zeeman frequency. On the
other hand, the expression for the $T_2$ time is
\begin{equation}
    \frac{1}{T_{2}}=\frac{1}{2T_{1}}+\frac{1}{2 \hbar^2} \int_{-\infty}^{\infty}
    dt~\tm{Re}\left<h_{z}(0)h_{z}(t)\right>.
\end{equation}
Notably, the relaxation contribution $(2T_{1})^{-1}$ has been
separated from the dephasing part incorporated in the integral. It
was proven that for the spin-orbit interaction (discussed in the
next section), up to linear order in momentum,
$\bm{h}(t)\cdot\bm{B}=0$ and the effective magnetic field can only
have fluctuations transverse to the applied $\bm{B}$-field. As a
result, the integral yields zero and the upper bound on $T_{2}$ is
realized, \emph{i.e.}, $T_{2}=2T_{1}$, independent of the origin of the
fluctuations \cite{Golovach2004}.

In contrast to the single-electron case, experiments performed on an
ensemble of systems with different environments are subject to
additional decoherence. It is therefore required to introduce an
ensemble-averaged transverse spin decay time $T_{2}^{*}$, which is
typically much shorter than the transverse single-spin decay time
$T_{2}$. One finds various other symbols for $T_{2}^{*}$ in the
literature, such as $\tau_{c}$ (the correlation time) and $T_{M}$
(the magnetization envelope decay time), used to emphasize the
non-exponential character of the decay. See
subsubsect.~\ref{sec:time_scales} for a summary of theoretical estimates
and experimental results for these various decoherence time scales.


For quantum computers to work on a large scale, it is crucial to
understand the microscopic mechanisms underlying dissipation and
decoherence, and to devise effective methods to reduce their impact
on the electron's spin dynamics. This would allow to achieve longer
coherence times and, in turn, reduce qubit errors. Two main sources
of decoherence in GaAs are to be identified, namely the spin-orbit
and the hyperfine interaction. (i) The spin-orbit interaction
couples the electron's spin to its orbital degrees of freedom. The
orbital motion is influenced by lattice phonons, which provide a
large dissipative bosonic reservoir. In this way, an effective
coupling between the electron's spin and the phonon bath is
established leading to energy dissipation and decoherence. (ii) The
Fermi contact hyperfine interaction couples the electron's spin
directly to the surrounding bath of fluctuating nuclear spins. In a
typical GaAs quantum dot the electron wave function overlaps with
wave functions of approximately $10^{5}$ nuclei. The electron spin
dynamics is thus strongly affected by the nuclear spin bath. These
two decoherence mechanisms will be discussed extensively in the
remaining part of this section.

\subsection{\label{sec:SOC_complete}Spin-orbit interaction}

The main features of the spin-orbit interaction have already been
discussed in subsubsect.~\ref{sec:SOC}. Here, we would like to emphasize
that this interaction originates, in fact, from the relativistic
Dirac equation and provides a direct coupling between the spin
$\bm{S}$ and the momentum $\bm{p}$. In vacuum, the spin-orbit term
derived from the Dirac equation turns out to be
\begin{equation}
    H_{SO} = -\frac{\hbar e}{2m^{2}c^{2}}\bm{S}\cdot\left(\bm{p}\times{\boldsymbol \nabla} V\right),
\end{equation}
where $c$ is the speed of light, $m$ is the free electron mass, $e$ the electron charge,
$\bm{S}=\bmg{\sigma}/2$ with $\bmg{\sigma}$  being the vector of
Pauli matrices, $\bm{p}$ is the canonical momentum, and $V$ is an
electric potential. Accounting for the lack of spatial inversion
symmetry in the bulk GaAs crystal and assuming the presence of an
asymmetric confining potential originating in the GaAs/AlGaAs
heterostructure, the two-dimensional spin-orbit Hamiltonian reduces
to the sum of the Rashba \cite{Rashba1960} and Dresselhaus
\cite{Dresselhaus1955} contributions
\begin{equation}
    H_{SO} = H_{R} + H_{D} = \alpha\left(p_{x}\sigma_{y}-p_{y}\sigma_{x}\right)
                   + \beta\left(p_{x}\sigma_{x}-p_{y}\sigma_{y}\right),
\end{equation}
where $\alpha$ is tunable by external gates while $\beta$ is a
material constant stemming from the bulk inversion asymmetry. The
tunability of $\alpha$ allows one, in principle, to achieve
$\alpha=\pm\beta$. In this particular case,
$H_{SO}=\alpha\left(p_{x}\mp
p_{y}\right)\left(\sigma_{x}\pm\sigma_{y}\right)$ and
$\left(\sigma_{x}\pm\sigma_{y}\right)$ is conserved. As a result,
the electron's spin decouples from its momentum degrees of freedom.

\subsubsection{Relaxation through phonons}

\begin{figure}
\begin{center}
    \includegraphics[width=.6\textwidth]{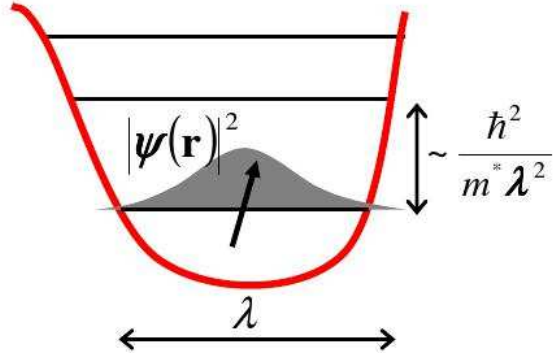}
    \caption{\label{fig:spin}
An electron in its orbital ground state, as determined by the
two-dimensional confining potential (thick red curve). The
probability to find the electron becomes negligible at distances
greater than dot's diameter $\lambda\sim 100~{\rm nm}$. The ground
state is separated from the first excited state by an energy gap
$\sim \hbar^{2}/m^{*}\lambda^{2}$. }
\end{center}
\end{figure}

We consider in this section the following two-dimensional model
Hamiltonian for an electron in the conduction band
\begin{equation}\label{eq:phonon_relax_hamiltonian}
    H = H_{d} + H_{Z} + H_{SO} + H_{el-ph}(t).
\end{equation}
Equation~(\ref{eq:phonon_relax_hamiltonian}) above only differs from
eq.~(\ref{EDSRhmilt}) by the last term. $H_{d} = {\bf p}^{2}/2m^{*}
+ U(\bm{r})$ describes the electron in the presence of a lateral
confining potential, which is assumed parabolic, \emph{i.e.},
$U(\bm{r})=m^*\omega_0^2 r^2/2$, and determines the ground state
orbital wave function $\psi_0({\bf
r})=\exp(-r^2/2\lambda^2)/\lambda\sqrt{\pi}$, where
$\lambda^{-2}=\hbar^{-1}\sqrt{(m^*\omega_0)^2+(eB_z/2c)^2}$. The
probability density $|\psi_0(\bm{r})|^{2}$ to find the electron
outside a circle of radius $\lambda/2$ is negligible, where a
typical dot diameter is $\lambda\sim100$~nm (see
Figure~\ref{fig:spin}). The second term of
eq.~(\ref{eq:phonon_relax_hamiltonian}) is the Zeeman Hamiltonian
$H_{Z}=g\mu_{B}\bm{B}\cdot{\bf S}$ and the third one is the
spin-orbit interaction $H_{SO}$, discussed briefly in the previous
section. As discussed for eq.~(\ref{EDSRhmilt}), it is convenient to
express $H$ in a rotated coordinate system, such that $H_{SO}$ takes
the simpler form eq.~(\ref{HSOrotated}). Note that, for a typical
GaAs quantum dot, the spin orbit lengths
$\lambda_{\pm}\sim1-10~\mu$m are much larger than the diameter of
the dot and the orbital level spacing
$\hbar^{2}/m^{*}\lambda^{2}\sim1$~meV~$\approx10$~K far exceeds
typical experimental temperatures $k_{B}T$ and Zeeman energies
$g\mu_{B}B$. The last contribution in
eq.~(\ref{eq:phonon_relax_hamiltonian}) takes into account two
different types of electron-phonon interactions
\cite{Gantmakher1987}
\begin{equation}\label{eq:H_el_ph}
    H_{el-ph}(t) = \sum_{\bm{q},j}\frac{F(q_{z})e^{i\bm{q}_{\parallel}\cdot\bm{r}}}
    {\sqrt{2\rho_{c}\omega_{\bm{q},j}/\hbar}}\left(e\beta_{\bm{q},j}-iq\Xi_{\bm{q},j}\right)
    \left(b_{-\bm{q},j}^{\dagger}+b_{\bm{q},j}\right),
\end{equation}
where the time-dependence is due to the phonon operators. In the
above equation, $b_{\bm{q},j}^{\dagger}$ creates an acoustic phonon
with wave vector $\bm{q}=\left(\bm{q}_{\parallel},q_{z}\right)$,
branch index $j$, and dispersion $\omega_{\bm{q},j}$.  Furthermore,
$\rho_{c}$ is the sample density and
$F(q_{z})=\int_{-\infty}^{+\infty}|\varphi(z)|^2e^{iq_z z} dz$,
where $\varphi(z)$ is the electron wave function in the direction $z$
of the confinement (the full wave function is $\varphi(z)\psi({\bf
r})$, where ${\bf r}=(x,y)$). Note that $F(q_{z})$ equals unity
for $|q_{z}|\ll d^{-1}$ and vanishes for $|q_{z}|\gg d^{-1}$, where
$d$ is the size of the quantum well along the $z$ axis. The
couplings $\beta_{\bm{q},j}$ are determined by the piezoelectric
electron-phonon interaction as follows
\begin{equation}
    \beta_{\bm{q},j} = \frac{2\pi}{q^{2}\kappa}\beta^{\mu\nu\eta}q_{\mu}
    \left(q_{\nu}e_{\eta,j}(\bm{q})+q_{\eta}e_{\nu,j}(\bm{q})\right),
\end{equation}
where $\beta^{\mu\nu\eta}$ is the electro-mechanical tensor,
$\kappa$ is the dielectric constant, and $e_{\mu,j}(\bm{q})$ is the
phonon polarization unit vector for branch $j$. For GaAs,
$\beta^{\mu\nu\eta}=h_{14}$ if the indices $\mu\nu\eta$ are a cyclic
permutation of $xyz$, where $h_{14}\approx0.16$~C/m$^{2}$.
Otherwise, $\beta^{\mu\nu\eta}=0$ \cite{Yu2001}. Finally, the deformation potential
electron-phonon interaction gives
\begin{equation}
    \Xi_{\bm{q},j} = \frac{1}{2q}\Xi^{\mu\nu}\left(q_{\mu}e_{\nu,j}(\bm{q})+q_{\nu}e_{\mu,j}(\bm{q})\right),
\end{equation}
where $\Xi^{\mu\nu}$ is the deformation tensor. For GaAs one has
$\Xi^{\mu\nu}=\Xi_{0}\delta_{\mu\nu}$, where $\Xi_{0}\approx7$~eV
\cite{Yu2001}. Therefore, the above expression simply becomes
$\Xi_{\bm{q},j} = \Xi_{0}\delta_{j,1}$, since only the longitudinal
branch $j=1$ gives a non-vanishing contribution.

As in subsubsect.~\ref{sec:EDSR}, it is convenient to approach the
perturbative treatment of the Hamiltonian
(\ref{eq:phonon_relax_hamiltonian}) by making use of the unitary
Schrieffer-Wolff transformation described in the Appendix.
The final result of this procedure is an effective spin Hamiltonian
of the form
\begin{equation}
    H_{\tm{eff}} = \left<\psi_0({\bf r})\right|e^S H e^{-S}\left|\psi_0({\bf r})\right>
    = g\mu_{B} \bm{B}\cdot {\bf S}+g\mu_B {\bf S}\cdot\delta\bm{B}(t) +
    \ldots,
\end{equation}
where spin-independent terms are omitted. In the above equation,
$\delta\bm{B}(t) = 2\bm{B}\times\bm{\Omega}(t)$ where
eq.~(\ref{eq:Omega_appendix}) immediately gives
$\bm{\Omega}(t)=\left<\psi_0({\bf r})\right|[((1-\hat
P)\hat{L}_{d}^{-1}\bmg{\xi}),H_{el-ph}(t)]\left|\psi_0({\bf
r})\right>$. The vector $\bmg{\xi}$ and the superoperators $\hat P$
and $\hat L_d^{-1}$ are defined in the Appendix.
Note that the form of $H_{\rm eff}$ above is the same as
eq.~(\ref{eq:H_magnetic_fluctuations}). However, as pointed out in
the beginning of sect.~\ref{sec:relax}, there can be only transverse
fluctuations of the effective magnetic field, \emph{i.e.},
$\delta\bm{B}(t)\cdot\bm{B}=0$, to first order in the spin-orbit
interaction. This property holds not only for phonons, but is valid
regardless of the nature of the charge fluctuations, as is seen from
the general form of eq.~(\ref{eq:Heff_appendix}).

\subsubsection{Energy relaxation}

If the scattering events are not correlated, \emph{i.e.}, if the phonons
emitted and absorbed by the electron leave the dot in a time
$\tau_{c}$ which satisfies $d/s\lesssim\tau_{c}\lesssim\lambda/s$
(with $s$ being the sound velocity), then the expectation value
$\left<\bm{S}\right>$ obeys the Bloch equation
\begin{equation}
    \dot{\left<\bm{S}\right>} = g\mu_{B}\bm{B}\times\left<\bm{S}\right>
    - \bm{\Gamma}\left<\bm{S}\right> + \bm{\Upsilon}.
\end{equation}
In this formula, the decay tensor $\bm{\Gamma}$ and the
inhomogeneous term $\bm{\Upsilon}$ can be derived in the Born-Markov
approximation for a generic $\delta\bm{B}(t)$ which fulfills
$\left<\delta\bm{B}(t)\right>=0$, and are expressed in terms of the
spectral function
\begin{equation}
   J_{ij}(\omega) = \frac{g^{2}\mu_{B}^{2}}{2\hbar^{2}}\int_{0}^{\infty}
    \left<\delta B_{i}(0)\delta B_{j}(t)\right>e^{-i\omega t}dt.
\end{equation}
In general, besides a term proportional to the spectral function
$\bm{J}$, the tensor $\bm{\Gamma}$ receives an additional
contribution from elastic scattering of the electron spin. However,
because of the transverse nature of the magnetic field fluctuations,
this contribution vanishes identically. The final result for $T_1$
and $T_2$ reads
\begin{equation}
\frac{1}{T_1}=\frac{2}{T_2}={\rm
Re}[J_{xx}(\omega_Z)+J_{xx}(-\omega_Z)+J_{yy}(\omega_Z)+J_{yy}(-\omega_Z)],
\end{equation}
where $\omega_Z=g \mu_B B/\hbar$ is the Zeeman frequency, as in
eq.~(\ref{eq:T1_general}). The explicit expression for
$\tm{Re}J_{xx}(\omega)$ reads \cite{Golovach2004}
\begin{eqnarray}\label{eq:relaxation_phonons}
    \tm{Re}J_{xx}(\omega) &=&
    \frac{\omega_Z^{2}\omega^{3}(2N_{\omega}+1)}{(2\Lambda_{+}m^{*}\omega_{0}^{2})}
    \sum_{j=1}^3\frac{\hbar}{\pi\rho_{c}s_{j}^{5}}\int_{0}^{\pi/2}d\theta\sin^{3}\theta\nonumber \\
    &&\times e^{-(\omega\lambda\sin\theta)^{2}/2s_{j}^{2}}\left|F\left(\frac{|\omega|}{s_{j}}\cos\theta\right)\right|^{2}
    \left(e^{2}\overline{\beta}_{j,\theta}^{2}+\frac{\omega^{2}}{s_{j}^{2}}\overline{\Xi}_{j}^{2}\right)
\end{eqnarray}
where $N_{\omega}=(e^{\hbar \omega/T}-1)^{-1}$ is the Bose
distribution and $s_j$ is
the velocity of sound for the branch $j$. In GaAs, the $s_j$ have
values $s_{1}\approx4.7\times10^{3}$~m/s and
$s_{2}=s_{3}\approx3.37\times10^{3}$~m/s. Furthermore,
$\overline{\Xi}_{j}=\Xi_{0}\delta_{j,1}$ with $\Xi_{0}\approx7$~eV,
$\overline{\beta}_{1,\theta}=3\sqrt{2}\pi
h_{14}\kappa^{-1}\sin^{2}\theta\cos\theta$,
$\overline{\beta}_{2,\theta}=\sqrt{2}\pi
h_{14}\kappa^{-1}\sin2\theta$,
$\overline{\beta}_{3,\theta}=3\sqrt{2}\pi
h_{14}\kappa^{-1}(3\cos^{2}\theta-1)\sin\theta$ with
$h_{14}\approx0.16$~C/m$^{2}$ and $\kappa\approx13$. The result for
$J_{yy}(\omega)$ is obtained by substituting $\Lambda_{+}\to
\Lambda_{-}$ in the above expression. Here $\Lambda_{\pm}$ are
effective spin-orbit lengths given in \cite{Golovach2004}. For a
magnetic field $\bf B$ in the $z$ direction the simple result
$\Lambda_{\pm}=\lambda_{\pm}$ is obtained, where $\lambda_{\pm}$ are
defined in eq.~(\ref{lambdaSO}). Note that both $J_{xx}$ and
$J_{yy}$ are multiplied by $\omega^{3}$, thereby exhibiting a
super-Ohmic behavior.

By explicitly evaluating this rather cumbersome expression, a
relaxation time $T_{1}\approx(825\pm275)~\mu$s is found at $B=8$~T.
This result is in very good agreement with the experimental value
$T_{1}^{\tm{exp}}=800~\mu$s at the same strength of the magnetic field
\cite{Elzerman2004}. The rather large uncertainty of the theoretical
prediction is due to the measured value of the $g$-factor.

\subsubsection{Magnetic field dependence of the relaxation rate}

\begin{figure}
    \includegraphics[width=.5\textwidth]{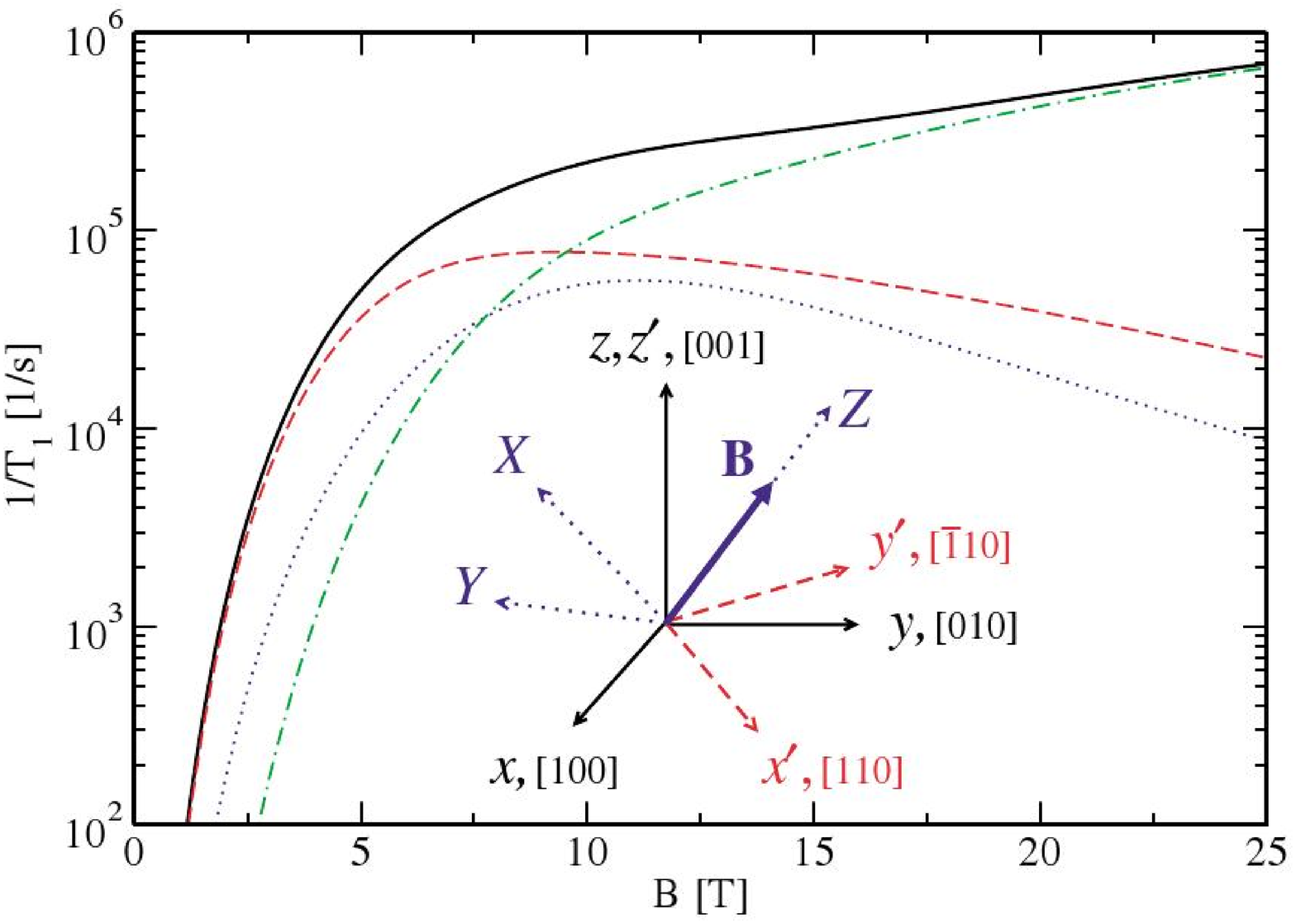}
    \includegraphics[width=.5\textwidth]{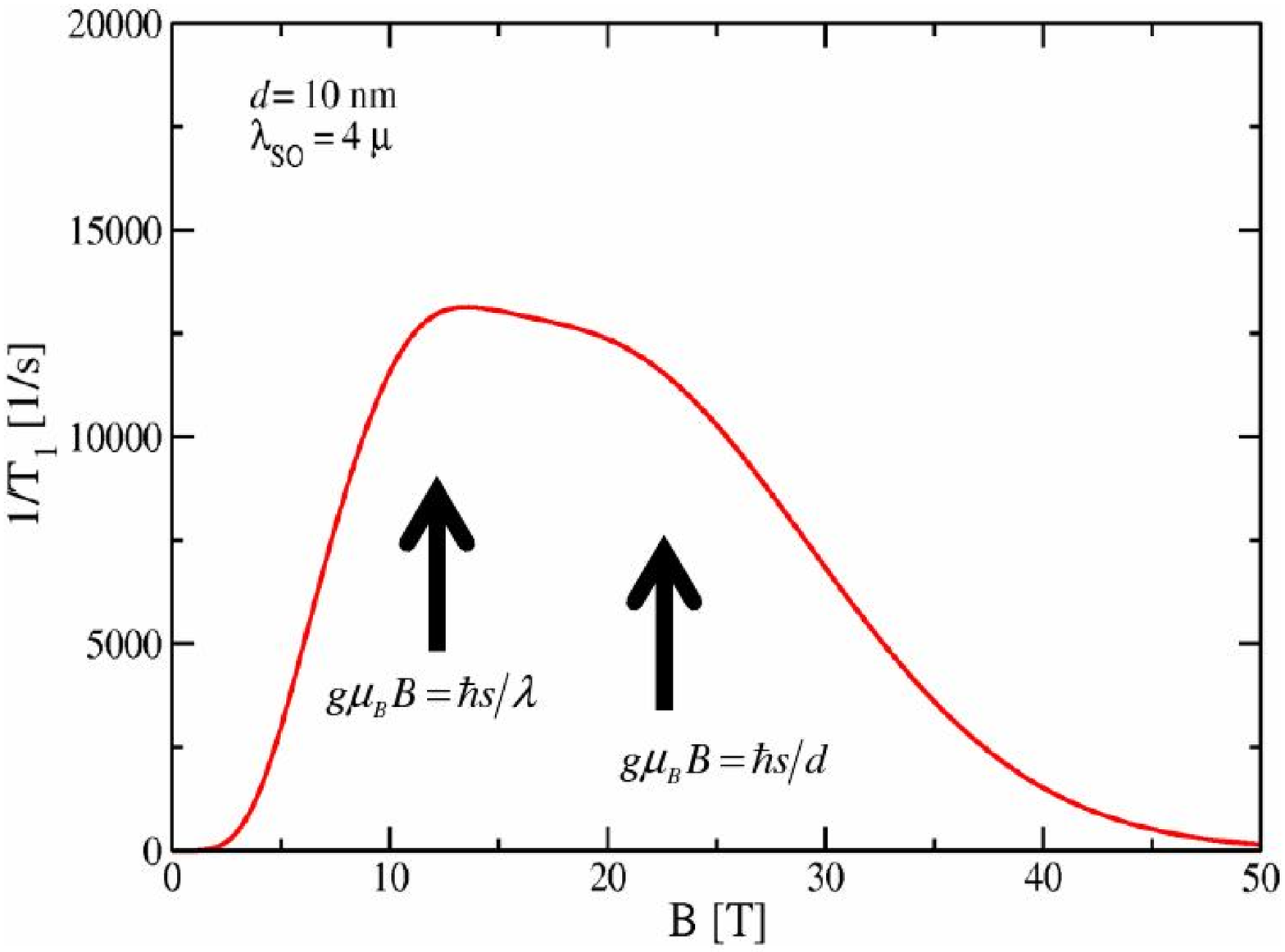}
    \caption{
Left panel: The relaxation rate $1/T_{1}$ given by
eq.~\ref{eq:relaxation_phonons} is plotted as a function of the
magnetic field $B$. Contributions due to the piezoelectric effect
with transverse and longitudinal phonons are plotted as dashed and
dotted curves, respectively. The contribution due to the deformation
potential mechanisms is shown by a dot-dashed curve. Right panel:
The relaxation rate $1/T_{1}$ is plotted for large magnetic fields,
where the $1/T_{1}\propto B^{5}$ behavior is
suppressed.\label{fig:golovach}}
\end{figure}

Another important result is obtained in the range of magnetic field below $\sim 3$~T, when the
relaxation rate $1/T_1$ increases with the magnetic field as $B^{5}$. This can be inferred from
the presence in eq.~(\ref{eq:relaxation_phonons}) of the prefactor $\omega_Z^2\omega^3$ (note that,
in the final expression for $1/T_1$, one has to set $\omega=\omega_Z$).
There are three contributions to this power-law behavior:
(i) The fluctuating magnetic field is proportional to the external
magnetic field, \emph{i.e.}, $\delta B^{2}\propto B^{2}$. (ii) The phonon
velocity is proportional to the phonon dispersion, \emph{i.e.},
$v_{ph}(\omega)\propto\omega^{2}$ with $\omega=g\mu_{B}B$. (iii) The
spin-orbit Hamiltonian couples to the magnetic field through its
momentum, \emph{i.e.}, $H_{SO}\propto p_{\alpha}\propto B$. The final
result $\propto B^5$ was confirmed experimentally in
ref.~\cite{Amasha2008}. In the opposite limit, \emph{i.e.} for magnetic
fields larger than $\sim 12$~T, the power law is suppressed, since
phonons are averaged to zero over the dot size
$\lambda_{ph}^{B}=s/h\mu_{B}B\ll\lambda$. These two regimes have been summarized in
fig.~\ref{fig:golovach}.

\subsubsection{Magnetic field angular dependence}

For an arbitrary direction of the magnetic field
$\bm{B}=B(\sin\theta\cos\varphi,$
$\sin\theta\sin\varphi,\cos\theta)$ it was found \cite{Golovach2004}
that
\begin{equation}
    \frac{1}{T_{1}} = \frac{\left(\alpha^{2}+\beta^{2}\right)\left(1+\cos^{2}\theta\right)
+2\alpha\beta\sin^{2}\theta\sin2\varphi}{\beta^{2}
T_{1}(\theta=\pi/2,\alpha=0)}.
\end{equation}
Noteworthy, there is an interference between the Rashba and
Dresselhaus terms which leads to a diverging relaxation time
$T_{1}\rightarrow\infty$ when $\alpha=\beta$, $\theta=\pi/2$, and
$\varphi=3\pi/4$. This result is valid to all orders in the
spin-orbit interaction and is an effect of spin conservation, which
occurs for the special condition  $\alpha=\beta$ of the spin-orbit
couplings (see also the discussion of $H_{SO}$, earlier in this
section).

\subsection{\label{sec:HF_complete}Hyperfine interaction}

As discussed in the previous section, the relaxation process of the
spin polarization in GaAs quantum dots is dominated by the
spin-orbit interaction, which couples the electron spin to the
phonon bath. If no other effect were present, the upper bound for
the decoherence time $T_{2}=2T_{1}$ would be satisfied
\cite{Golovach2004}. Measurements of $T_1$ reveal that ultra-long
relaxation times $\sim1$~s are achievable for magnetic fields
$B\sim1$~T \cite{Amasha2008}. Hence, long decoherence times might be
expected as well. Unfortunately, measured spin decoherence times are
considerably shorter and range from $1~\mu$s \cite{Petta2005} to
roughly $10~\mu$s \cite{Koppens2006,Koppens2007}. Thus, the spin
decoherence in GaAs must be dominated by other effects.

The major source of decoherence in GaAs was investigated
theoretically in \cite{Burkard1999} and was attributed to the
hyperfine interaction with the nuclear spins. In fact, this is the
cause of decoherence in numerous candidate systems for quantum
information processing applications such as quantum dots
\cite{Ono2004,Petta2005,Koppens2006}, Si:P donors \cite{Abe2004}, NV
centers in diamond \cite{Childress2006,Hanson2006}, and molecular
magnets \cite{Ardavan2007,Bertaina2008}.

A possible way to limit this decoherence problem in GaAs might be to
use holes instead of electrons \cite{Bulaev2005,
Bulaev2007,Fischer2008}, since the Fermi contact hyperfine
interaction vanishes in this case. The detailed form of the
hyperfine interaction for holes was recently studied in
ref.~\cite{Fischer2008} and receives contributions from the
dipole-dipole interaction and the coupling of the electron orbital
angular momentum to the nuclear spins. It is indeed found to be
smaller than for electrons (but still sizable) and of Ising type,
differently from the electron's isotropic Heisenberg interaction.
Another strategy against decoherence would be to employ materials such as C, Si, Ge and
others, which do not host any nuclear magnetic moment. Finally, one
can insist on GaAs, which is still the most common material for
quantum information processing applications, and deal with the
nuclear spins. In this section we follow this last approach and
focus on the decoherence process caused by the hyperfine
interaction. Based on its detailed understanding, we suggest
possible schemes for its reduction or elimination.

The following Hamiltonian describes the hyperfine interaction in a
single quantum dot in the presence of a magnetic field applied in
$z$ direction $\bm{B}=B\bm{e}_{z}$ and dipole-dipole interaction
$H_{dd}$ between nuclear spins
\begin{equation}\label{eq:HamHF}
    H = \bm{h}\cdot\bm{S} + bS_{z} + \epsilon \sum_i I^{z}_i + H_{dd},
\end{equation}
where $\bm{h}=\sum_{i}A_{i}\bm{I}_{i}$ is the nuclear magnetic field
also known as the Overhauser field, $A_i$ is the hyperfine coupling
strength at site $i$, $b=g\mu_{B}B$ is the electron Zeeman splitting
and $\epsilon=g_{I}\mu_{N}B$ is the nuclear Zeeman splitting. The
nuclear (Bohr) magneton is denoted by $\mu_{I}$ ($\mu_{B}$) and
nuclear (electron) $g$-factor by $g_{I}$ ($g$). Typically, the
electron orbital energy level spacing for lateral quantum dots
containing one electron is much larger than the energy scale of the
hyperfine interaction. Therefore, the electron remains in its
orbital ground state and no orbital excitations due to the
interactions with nuclei are possible.

The hyperfine coupling is non-uniform since we have $A_{i}\propto
|\psi(\bm{r}_{i})|^{2}$. For GaAs, the average coupling strength
weighted by the natural abundance of each isotope (cf.
subsubsect.~\ref{sec:hyperfine_intro}) is $A\approx90~\mu$eV. The typical
energy associated with the hyperfine interaction is then
$A/N\approx10^{6}$~s$^{-1}$, where $N\approx10^{5}$ is the typical
number of nuclei in contact with the electron spin. This scale has
to be compared with the dipole-dipole interaction energy
$\left<(\delta H_{dd})^{2}\right>^{1/2}\approx10^{4}$~s$^{-1}$,
which is much smaller and will be neglected in the following.

The nuclear Zeeman term can be formally eliminated from the
Hamiltonian by transforming to a rotating reference frame
\cite{Coish2004}. One can also separate the longitudinal part
$H_{0}$ from the transverse flip-flop part $V$ of the hyperfine
Hamiltonian, which leads to the following expression
\begin{equation}\label{eq:HFHamiltonianCoish}
    H' = H_{0} + V = (b'+h_{z})S_{z} + \frac{1}{2}(h_{+}S_{-}+h_{-}S_{+}),
\end{equation}
where $h_{\pm}=h_{x}\pm i h_{y}$ and $S_{\pm}=S_{x}\pm i S_{y}$ are
ladder operators for the nuclear field and the electron spin,
respectively, and $b'=b-\epsilon$. From this representation, it
becomes clear that the longitudinal component $V$ describes
flip-flop processes: If the electron spin and the spin of a
neighboring nucleus are opposite, their direction can be
simultaneously swapped. This flip-flop mechanism allows for electron
mediated diffusion of the nuclear spins and determines fluctuations
of the nuclear spins polarization on a timescale of order
$100~\mu$s.

\subsubsection{Ensemble averaged decoherence time}\label{sec:InitialState}

A central question to address is how the initial state of the nuclei
affects the evolution of the electron spin. Assume that for times
$t<0$ the electron spin and the nuclear bath are decoupled and
described by the density operators $\rho_{S}(0)$ and $\rho_{I}(0)$
respectively. At $t=0$, they are brought into contact over a
switching time scale that is sufficiently short, \emph{e.g.}, the time taken
to inject an electron into a quantum dot. The initial state of the
entire system $\rho(t=0)$ is continuous at $t=0$ and, hence,
$\rho(0^{-})=\rho(0^{+})=\rho_{S}(0)\otimes\rho_{I}(0)$. Following
ref.~\cite{Coish2004}, we assume for simplicity nuclei with
spin-$\frac12$ and consider three types of initial spin
configurations, namely
\begin{eqnarray}
    &&\rho_{I}^{(1)}(0) = \left|\psi_{I}\right>\left<\psi_{I}\right| \hs{10pt}\tm{with}\hs{10pt}
    \left|\psi_{I}\right>=\bigotimes_{k=0}^{N}\left(\sqrt{f_{\uparrow}}\left|\uparrow_{k}\right>
    +e^{i\phi_{k}}\sqrt{1-f_{\uparrow}}\left|\downarrow_{k}\right>\right),\\
    &&\rho_{I}^{(2)}(0) = \sum_{N_{\uparrow}}\binom{N}{N_{\uparrow}}
    f_{\uparrow}^{N_{\uparrow}}\left(1-f_{\uparrow}\right)^{N-N_{\uparrow}}
    \left|N_{\uparrow}\right>\left<N_{\uparrow}\right|,\\
    &&\rho_{I}^{(3)}(0) = \left|n\right>\left<n\right| \hs{20pt}\tm{with}\hs{10pt}
    h_{z}\left|n\right> = \sum_{k}A_{k}I_{k}^{z}\left|n\right> =
    \frac{pA}{2}\left|n\right>,
\end{eqnarray}
where $\left|\uparrow_{k}\right>$ and $\left|\downarrow_{k}\right>$
are the spin-up and spin-down eigenstates of the $k$-th nucleus,
$f_{\uparrow}$ determines the nuclear polarization
$p=2f_{\uparrow}-1$, $\phi_{k}$ is an arbitrary site-dependent
phase, $N$ is the total number of nuclei, and $N_{\uparrow}$ is the
number of the nuclei in the spin-up state.
$\left|N_{\uparrow}\right>$ thus denotes any product state of the
form $\left|\uparrow\downarrow\uparrow\uparrow\cdots\right>$ with
$N_{\uparrow}$ nuclear spins up and $N-N_{\uparrow}$ spins down, and
$A=\sum_k A_k$.

We note that $\rho_{I}^{(1)}(0)$ and $\rho_{I}^{(3)}(0)$ are both
pure states but in the first one $\left|\psi_{I}\right>$ is chosen
to render the $z$ component of nuclear spin translationally
invariant:
$\left<\psi_{I}\right|I_{k}^{z}\left|\psi_{I}\right>=\left(2f_{\uparrow}-1\right)/2=p/2$,
while in the second one $\left|n\right>$ is chosen to be an
eigenstate of $h_{z}$ with eigenvalue $pA/2$. On the other hand,
$\rho_{I}^{(2)}(0)$ is a~mixed state, which corresponds to
an~ensemble of product states where the $N$ spins in each product
state are selected from a bath of polarization $p$.

Now we evaluate the nuclear spin dynamics under $H_{0}$, \emph{i.e.}, when
the flip-flop term $V$ can be neglected. This is justified, for
instance, at large magnetic fields, because the energy gap between
spin-up and spin-down states of the electron is much larger than the
nuclear splitting between spin-up and spin-down states, thus making
the flip-flop transition energetically forbidden. We also assume the
simplified case of uniform coupling constants $A_{k}=A/N$. Since
$\left[H_{0},S_{z}\right]=0$, $\left<S_{z}\right>_{t}$ is constant.
Instead, $[H_{0},S_{\pm}]\neq0$ and the transverse components,
$\left<S_{\pm}\right>_t=\left<S_{x}\right>_{t}\pm
i\left<S_{y}\right>_{t}$, have a nontrivial time dependence, which
can be evaluated by tracing out the electron and nuclear degrees of
freedom
$\left<S_{\pm}\right>_{t}=\tm{Tr}\left[e^{iH_{0}t}S_{\pm}e^{-iH_{0}t}\rho(0)\right]$.
The final results read
\begin{eqnarray}
\label{eq:Spm_rho12}
    &&\left<S_{\pm}\right>_{t}^{(1,2)} = \left<S_{\pm}\right>_{0}\sum_{N_{\uparrow}}\binom{N}{N_{\uparrow}}
    f_{\uparrow}^{N_{\uparrow}}\left(1-f_{\uparrow}\right)^{N-N_{\uparrow}}e^{\pm i\left(b'+A M(N_\uparrow)/2N\right)t},\\
    \label{eq:Spm_rho3}
    &&\left<S_{\pm}\right>_{t}^{(3)} = \left<S_{\pm}\right>_{0}e^{\pm
    i\left(b'+pA/2\right)t},
\end{eqnarray}
where $M(N_\uparrow)=2N_{\uparrow}-N$ is the nuclear magnetization on
a dot with $N_{\uparrow}$ nuclear spins up and we generally set $\hbar=1$ from now on. Note the similarity
between randomly correlated pure states and mixed states, which yield the same final result eq.~(\ref{eq:Spm_rho12}). It is also seen that no decay in eq.~\eqref{eq:Spm_rho3} is obtained for the eigenstate $|n\rangle$, but a finite transverse relaxation time follows from eq.~\eqref{eq:Spm_rho12}, due to the average over the binomial distribution. One can further evaluate the expression in eq.~\eqref{eq:Spm_rho12}, by direct application of the central limit theorem, and obtain
\begin{equation}
\label{eq:gaussian_decay}
    \left<S_{\pm}\right>_{t}^{(1,2)} = \left<S_{\pm}\right>_{0}
    e^{-t^{2}/2t_{c}^{2}\pm i\left(b'+pA/2\right)t} \hs{10pt}\tm{with}\hs{10pt}
    t_{c} = \frac{2}{A}\sqrt{\frac{N}{1-p^{2}}}.
\end{equation}
This Gaussian decay occurs on a timescale
$t_{c}\approx5$~ns, for a GaAs quantum dot with
$p^{2}\ll1$ and $N=10^{5}$.

\subsubsection{Electron Spin Resonance}\label{sec:EPR}

To study the electron spin dynamics, the electron spin resonance
(ESR) technique is very fruitful. In order to achieve the resonance
condition, an alternating magnetic field
$\bm{B}_{\tm{ac}}=B_{\tm{ac}}\bm{e}_{x}\cos\omega t$ is applied in the
transverse direction, in addition to the static out-of-plane
magnetic field $\bm{B}=B\bm{e}_{z}$. The Hamiltonian of this
system is
\begin{equation}\label{eq:HamiltonianESR}
    H_{ESR} = (h_{z}+b)S_{z} + b_{1}\cos(\omega t)S_{x},
\end{equation}
where $b_{1}=g\mu_{B}B_{\tm{ac}}$. The first term is $H_{0}$ of eq.~(\ref{eq:HFHamiltonianCoish}),
but we neglected the nuclear Zeeman splitting $\epsilon$. We have again assumed
that spin-flip processes are not important since the static magnetic field is
large. As discussed in the previous section, the nuclear bath is in a
superposition of eigenstates $|n\rangle$ of the $h_z$ operator, with corresponding
eigenvalues $h_z^n$. The distribution of the $h_z^n$ eigenvalues is Gaussian,
with mean $h_0={\rm Tr}\{\rho_I h_z \}$ and variance $\sigma \sim A/\sqrt{N}$, defined by
$\sigma^2={\rm Tr}\{\rho_I (h_z-h_0)^2 \}$. As in the previous section,
$\rho_I$ is the density matrix of the nuclear system.

The decay of the driven Rabi oscillations is found in a
rotating-wave approximation (valid for $(b_{1}/b)^{2}\ll1$)
and is given by \cite{Klauser2006}
\begin{equation}
    P_{\uparrow}(t) \sim \frac{1}{2} + C + \sqrt{\frac{b_{1}}{8\sigma^{2}t}}
    \cos\left(\frac{b_{1}}{2}t+\frac{\pi}{4}\right) + o\left(\frac{1}{t^{3/2}}\right)
\end{equation}
where $C=1/2-\left(\sqrt{2\pi
b_{1}^{2}}/8\sigma\right)\exp\left(b_{1}^{2}/8\sigma^{2}\right)
\tm{erfc}\left(b_{1}/\sqrt{8\sigma^{2}}\right)$ is a
time-independent constant. The above formula holds for
$t\gg\tm{max}\left(1/\sigma,1/b_{1},b_{1}/2\sigma^{2}\right)$ and
$h_{0}+b=\omega$. Interestingly, the decay is a slow power
law $\propto 1/\sqrt{t}$ and exhibits a universal phase shift of
$\pi/4$. Even if the Rabi period exceeds the timescale
$\tau\sim15$~ns for the transverse spin decay, this result implies
that Rabi oscillations are visible, due to the fact that the power
law becomes valid after a short time $\tau$ for $b_{1}\approx\sigma$.
The universal phase shift originates from the off-resonant
contributions, which possess a higher Rabi period and shift the
average oscillation in phase.
\begin{figure}
\begin{center}
    \includegraphics[width=.7\textwidth]{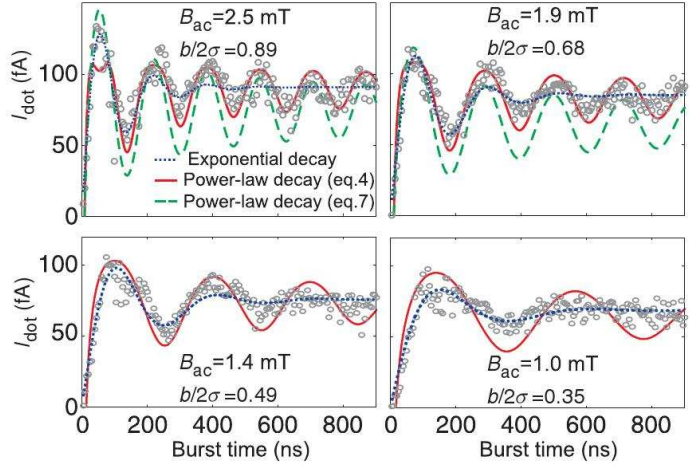}
    \caption{Rabi oscillations for four different driving fields $B_{\tm{ac}}$.
The gray circles represent the experimental data obtained in the
transport measurement as described in the main text. The best fit of
eqs.~(\ref{eq:Podd2}) and (\ref{eq:Podd1}) is represented by solid and
dashed lines, respectively. Two fitting parameters were used: the
saturation value and the phase shift. The latter turns out to be
close to the theoretical value $\pi/4$. See ref.~\cite{Koppens2007} for more information.\label{fig:klauser}}
\end{center}
\end{figure}
These predictions, \emph{i.e.}, both the non-exponential decay and the
universal phase shift, have been recently confirmed experimentally
\cite{Koppens2007}. The electron spin state was detected by a
transport measurement in a double dot configuration. Electrons are
transported through the double quantum dot via transitions from the
state with one electron in each dot to the singlet state with two
electrons in the right dot, as extensively described in
subsubsects.~\ref{sec:pauli_blockade} and~\ref{sec:spin readout dd}. The
states with even spin parity (parallel spins)
block transport, whereas spins with odd spin parity (antiparallel
spins) allow for transport. Given that the system is initialized in
an even spin-parity state, the resonant oscillating transverse
magnetic field rotates at least one of two electron spins and lifts
the blockade. The time evolution of the probability to measure an
odd parity state after a time $t$ has been calculated for the
following two cases. When both electrons are in resonance with the
alternating magnetic field, one finds
\begin{equation}
\begin{split}\label{eq:Podd2}
    P_{\tm{odd}}^{(2)}(t) = &\frac{1}{2} - 2C^{2} - C\sqrt{\frac{2b_{1}}{\sigma^{2}t}}
    \cos\left(\frac{b_{1}t}{2}+\frac{\pi}{4}\right)\\
    &-\frac{b_{1}}{8\sigma^{2}t}\left[1+\cos\left(b_{1}t+\frac{\pi}{4}\right)\right]
    +o\left(\frac{1}{t^{3/2}}\right).
\end{split}
\end{equation}
Secondly, when only one electron is on resonance, one obtains
\begin{equation}
    P_{\tm{odd}}^{(1)}(t) = \frac{1}{2} - C - \sqrt{\frac{b_{1}}{8\sigma^{2}t}}
    \cos\left(\frac{b_{1}t}{2}+\frac{\pi}{4}\right)+o\left(\frac{1}{t^{3/2}}\right)\label{eq:Podd1}.
\end{equation}
The first result eq.~\eqref{eq:Podd2} is valid for times
$t\gtrsim\tm{max}\left(1/\sigma,1/b_{1},b_{1}/2\sigma^{2}\right)\sim20$~ns
for a $1.4$~mT nuclear field and $b_{1}\leq2\sigma$. Note that the
$1/t$ term, which becomes important for $b_{1}>\sigma$, oscillates
with twice the Rabi frequency. This is the result of the
simultaneous rotation of both spins. The term $t^{-1/2}$, which is
dominant for $b_{1}<\sigma$, oscillates with the Rabi frequency and
stems from the rotation of one spin only. As might be expected from
these considerations, $P_{\tm{odd}}^{(1)}$ contains exclusively the
latter term. The comparison of these theoretical results with the
experimental data is presented in fig.~\ref{fig:klauser}.

It should be emphasized that the power-law decay and the universal,
\emph{i.e.}, independent of all parameters, phase shift $\pi/4$ are
obtained with the nuclear field $h_{z}$ being static during a time
much longer than the Rabi period. Therefore, the good agreement
between the experiment \cite{Koppens2007} and theory confirms the
assumption of a static nuclear bath.

\subsubsection{Narrowing of the nuclear state}\label{sec:narrow}

The evolution of the electron spin governed by
eq.~(\ref{eq:HamiltonianESR}) depends on the value of the nuclear
field since the effective Zeeman splitting is given by
$h_{z}^{n}+b$, where
$h_{z}\left|n\right>=h_{z}^{n}\left|n\right>$. This means that the
resonance condition $b+h_{z}^{n}-\omega=0$ for ESR also
depends on the nuclear field. Therefore, a measurement of the
electron spin state determines $h_{z}^{n}$ and the related state of
the nuclei.

The eigenvalues of the nuclear field in equilibrium obey a Gaussian
distribution, as discussed in the previous section. That is,
the diagonal elements of the nuclear spin density matrix
$\rho_{I}\left(h_{z}^{n},0\right)=\left<n\right|\rho_{I}\left|n\right>=\left(\sqrt{2\pi}\sigma\right)^{-1}\exp\left[-\left(h_{z}^{n}-h_{0}\right)^{2}/2\sigma^2\right]$
are Gaussian with mean $h_{0}$ and variance
$\sigma$. After the electron spin is initialized in a state
$\left|\uparrow\right>$ at time $t=0$, the system evolves under the
Hamiltonian $H_{ESR}$ until a measurement of the spin is performed
at $t=t_{m}$. Of special interest is the probability to find the
electron spin in the orthogonal spin state $\left|\downarrow\right>$
and a given nuclear eigenstate $\left|n\right>$
\begin{equation}
    P_{\downarrow}^{n}(t)=\frac{1}{2}\frac{b_{1}^{2}}{b_{1}^{2}+4\delta_{n}^{2}}
    \left[1-\cos\left(\frac{t\sqrt{b_{1}^{2}+4\delta_{n}^{2}}}{2}\right)\right],
\end{equation}
where $b_{1}=g\mu_{B}B_{1}$ as defined before, and
$\delta_{n}=b+h_{z}^{n}-\omega$ is the measure of deviation
from the resonance condition. On the other hand, the probability to
find the electron spin in the state $\left|\downarrow\right>$ with
arbitrary configuration of the nuclei is easily obtained by
integrating out the nuclear field $P_{\downarrow}(t)=\int
dh_{z}^{n}\rho_{I}\left(h_{z}^{n},0\right)P_{\downarrow}^{n}(t)$.
The measurement with outcome $\left|\downarrow\right>$ performed on
the system results in the collapse of the diagonal part of the
nuclear spin density matrix into
\begin{equation}
    \rho_{I}\left(h_{z}^{n},0\right)\xrightarrow{\left|\downarrow\right>}
    \rho_{I}^{(1,\downarrow)}\left(h_{z}^{n},t_{m}\right) = \rho_{I}\left(h_{z}^{n},0\right)
    \frac{P_{\downarrow}^{n}(t_{m})}{P_{\downarrow}(t_{m})},
\end{equation}
according to the basic rules of quantum mechanics. If the
measurement has a time resolution smaller than $1/b_{1}$ (\emph{i.e.},
giving time-averaged values), the probability of outcome
$\left|\downarrow\right>$ as a function of the nuclear field
eigenvalue $h^z_n$ reads
$P_{\downarrow}^{n}=b_{1}^{2}/2(b_{1}^{2}+4\delta_{n}^{2})$. In
turn, the nuclear spin density matrix is multiplied by a Lorentzian
with width $b_{1}$ and mean $h_{z}^{n} = \omega-b$. As a
result, for $b_{1}<\sigma$, the nuclear spin distribution becomes
narrowed and prolongation of the electron spin coherence is
achieved. Analogously, if the measured outcome is
$\left|\uparrow\right>$, the Gaussian nuclear spin distribution is
modified as follows:
\begin{equation}
    \rho_{I}\left(h_{z}^{n},0\right)\xrightarrow{\left|\uparrow\right>}
    \rho_{I}^{(1,\uparrow)}\left(h_{z}^{n},t_{m}\right) = \rho_{I}\left(h_{z}^{n},0\right)
    \frac{1-P_{\downarrow}^{n}(t_{m})}{1-P_{\downarrow}(t_{m})}.
\end{equation}
Hence, the probability to match the resonance condition
$b+h_{z}^{n}-\omega=0$ is considerably reduced. It has already been
proven (cf. subsubsect.~\ref{sec:EPR}) that the nuclear spin evolution is
slow enough to allow for multiple measurements of the electron spin
(each of them performed after re-initializion to
$\left|\uparrow\right>$) over a timescale on which the nuclear spins
may be considered static. Repeating the initialization and
measurement scheme $M$ times under the assumption of the static
nuclear field, we arrive at
\begin{equation}
    \rho_{I}\left(h_{z}^{n},0\right)\longrightarrow
    \rho^{(M,\alpha_{\downarrow})}\left(h_{z}^{n}\right) =
    \frac{1}{N}\rho_{I}\left(h_{z}^{n},0\right)
    \left(P_{\downarrow}^{n}\right)^{\alpha_{\downarrow}}
    \left(1-P_{\downarrow}^{n}\right)^{M-\alpha_{\downarrow}},
\end{equation}
where $\alpha_{\downarrow}$ denotes the number of times the state
$\left|\downarrow\right>$ was obtained. From the experimental point
of view it should be easiest to narrow the nuclear field
distribution by performing measurements with $b_{1}\ll\sigma$.
Provided that the electron spin was projected to
$\left|\downarrow\right>$, the narrowing has been achieved.
Otherwise, the additional initialization-measurement cycles should
be repeated until the narrowing is observed. The driving frequency
should be adjusted after each initialization-measurement cycle to
match the resonance condition in order to systematically move
towards the narrowed state. Such an adaptive scheme is described in
refs. \cite{Stepanenko2006} and \cite{Klauser2006}.

\subsubsection{Exact solution with a fully polarized nuclear bath}

\begin{figure}
\begin{center}
    \includegraphics[width=.5\textwidth]{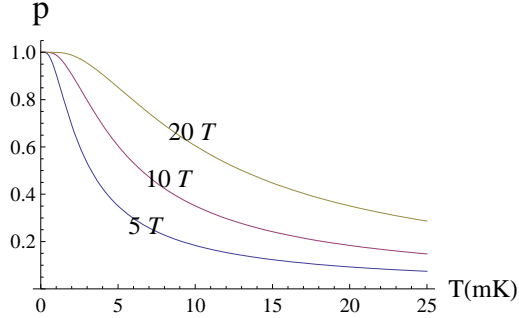}
    \caption{\label{direct_pol}  Average nuclear polarization in GaAs ($p=\frac{2}{3N}\sum_{i}\langle I_i \rangle$, where $i$ runs over the $N$ nuclei) as a function of temperature at different values of the magnetic field.}
\end{center}
\end{figure}

The idea of fully polarizing the nuclear system has attracted considerable attention, 
since in this special configuration the hyperfine-induced decoherence problem can be overcome. 
Furthermore, the system of an electron spin confined to a quantum dot with a 
completely polarized nuclear bath is solvable exactly \cite{Khaetskii2002,Khaetskii2003}. 
Despite the fact that the complete polarization of the nuclear spins is not accessible with current 
experimental methods (to reduce decoherence by an order of magnitude, polarization 
of above $99\%$ is required \cite{Coish2004}, whereas the current world record is about $60\%$
\cite{Bracker2005}), some schemes to achieve full polarization of the nuclei
have been recently proposed. Figure~\ref{direct_pol} illustrates the temperature range 
that has to be accessed to directly polarize the nuclei in the presence of the external
magnetic field \cite{Chesi2008a}. Cooling to sub-mK temperatures is an experimental challenge which is 
currently actively pursued. At such low temperatures refs.~\cite{Simon2007} and \cite{Simon2008}
predict a phase transition to an ordered nuclear state, even in the absence of the external field. 
Of crucial importance is in this case the long range nature of the electron-mediated RKKY 
interaction, which is determined by nonanalytic corrections in the momentum dependence of the electronic spin-susceptibility \cite{Simon2008,Chesi2008b}.

Consider a single electron confined to a quantum dot in its ground
state. As discussed earlier, coupling of an external field $\bm B$
to the nuclear spins and direct dipole-dipole interactions among
them are negligibly small. Hence, the Hamiltonian can be written as
in eq.~(\ref{eq:HamHF}) with $\epsilon=0$ and $H_{dd}=0$.
Additionally, we assume that the system is initially in the state
$\left|\Psi_{0}\right>=\left|\Downarrow\uparrow\uparrow\cdots\right>
=\left|\Downarrow\right>\otimes\left|\uparrow\uparrow\uparrow\cdots\right>$,
\emph{i.e.}, the electron spin $|{\Downarrow}\rangle$ is opposite to the
fully polarized nuclear bath state
$\left|\uparrow\uparrow\uparrow\cdots\right>$. Due to conservation
of total angular momentum, the time evolution of the wave function
is fully described by
$\left|\Psi(t)\right>=\alpha(t)\left|\Psi_{0}\right>+\sum_{k}\beta_{k}(t)\left|\Uparrow\uparrow\uparrow\cdots\downarrow_{k}\cdots\right>$
with the time-dependent coefficients $\alpha(t)$ and $\beta_k(t)$
obeying the normalization condition
$\left|\alpha(t)\right|^{2}+\sum_{k}\left|\beta_{k}(t)\right|^{2}=1$.
The second term in $\left|\Psi(t)\right>$ is a coherent
superposition of states with exactly one nuclear spin flipped and
thus resembles a magnon excitation.

We will study the correlator $C_{0}(t)=-\left<\Psi_{0}\right|\delta
S_z(t) {S}_{z}\left|\Psi_{0}\right>
=\left(1-|\alpha(t)|^{2}\right)/2$, which describes the decay of the
electron spin from its initial state $\left|\Downarrow\right>$.
Here, $\delta S_z(t) = S_z(t) - S_z$, where $S_z(t)$ is the
Heisenberg representation of $S_z$. Inserting $\left|\Psi(t)\right>$
into the Schr\"{o}dinger equation, we obtain a set of coupled
differential equations of the form
\begin{eqnarray}
    2i\frac{d\alpha(t)}{dt}=-A\alpha(t)/2+\sum_{k}A_{k}\beta_{k}(t)-b\alpha(t),\\
    2i\frac{d\beta_{l}(t)}{dt}=\left(A/2-A_{l}\right)\beta_{l}(t)+A_{l}\alpha(t)+b\beta_{l}(t),
\end{eqnarray}
with $A=\sum_{k}A_{k}$ and $b=g\mu_{B}B$. These equations can be
solved in the usual way by performing a Laplace transform
$\alpha(u)=\int_{0}^{\infty}dt\alpha(t)e^{-ut}$. One obtains
\begin{equation}
    \alpha(u)=i\frac{\alpha(0)}{D(u)}+\frac{i}{2D(u)}\sum_{k}\frac{A_{k}\beta_{k}(0)}{iu-(A+2b)/4+A_{k}/2},
\end{equation}
where
\begin{equation}
    D(u)=iu+\frac{A+2b}{4}-\frac{1}{4}\sum_{k}\frac{A_{k}^{2}}{iu-(A+2b)/4+A_{k}/2}.
\end{equation}
Defining $iu=i\omega+(A+2b)/4$, using the initial conditions
$\alpha(0)=1$ and $\beta_{k}(0)=0$, and replacing the sum over $k$
by an integral, \emph{i.e.},
\begin{equation}
    \sum_{k}\frac{A_{k}^{2}}{i\omega+A_{k}/2} = 2\left[A-2\pi iN\omega
    \int dz\ln \left(1-\frac{iA\chi_{0}^{2}(z)}{2\pi N\omega}\right)\right],
\end{equation}
where $\chi_{0}(z)$ is the normalized transverse wave function
defined through the envelope wave function
$|\Psi(r,z)|^{2}=\exp(-r^{2}/a^{2})\chi_{0}^{2}(z)/(\pi
a^{2}a_{z})$, where $a$ and $a_{z}$ are dot's dimensions in the
lateral and transverse direction, respectively, one arrives at the
following expression for $\alpha(t)$ (required to calculate the
correlator $C_{0}(t)$)
\begin{equation}
    \alpha(t) = \frac{e^{-iA't/4}}{2\pi}\int_{\gamma-i\infty}^{\gamma+i\infty}d\omega
    \frac{e^{i\omega t}}{i\omega+b+\pi N i\omega
    \int dz\ln \left(1-\frac{iA\chi_{0}^{2}(z)}{2\pi
    N\omega}\right)}.
\end{equation}
Here, $A'=A+2b$, and $z$ is dimensionless in units of $a_z$. The
integration contour is a vertical line in the complex $\omega$ plane
such that all singularities lie to its left. There are two types of
singularities: two branch points $\omega=0$ and
$\omega_{0}=iA\chi_{0}^{2}(0)/2\pi N$, and first order poles which
lie on the imaginary axis ($\omega=i\upsilon$). For $b>0$ there is
one pole, while for $b<0$ there are two poles; for $b=0$ there is
one pole at $\omega_{1}\approx iA/2+iA\int dz\chi_{0}^{4}(z)/4\pi
N$. The contribution from the branch cut between $\omega=0$ and
$\omega=\omega_{0}$ is
\begin{equation}\label{eq:BranchCut}
    \tilde{\alpha}(t) = \frac{e^{-iA't/4}}{\pi N}\int_{0}^{1}d\kappa
    \frac{2z_{0}\kappa e^{i\tau'\kappa}}{\left|\kappa\int dz
    \left|\chi_{0}^{2}(z)/\chi_{0}^2(0)\kappa-1\right| + \kappa/\pi N
    -2b/A\chi_{0}^{2}(0)\right|^{2}+\left(2\pi
    z_{0}\right)^{2}\kappa^{2}},
\end{equation}
where $\tau'=\tau\chi_{0}^{2}(0)$ with $\tau=At/2\pi N$, and
$z_{0}=z_{0}(\kappa)$ is defined through
$\chi_{0}^{2}(z_{0})=\chi_{0}^{2}(0)\kappa$. $N=n_0 a_z a^2$ is a
number of nuclear spins of density $n_{0}$ and we have introduced
the dimensionless variable $\kappa=\omega/\omega_{0}\leq1$.

The physical picture is the following. At time $t=0$ the system has
some energy corresponding to the pole. It then starts oscillating
back and forth, while each time visiting different frequencies
within the branch cut. This corresponds to the flip-flop processes
with the nuclei located at different sites. The contribution from
the branch cut therefore describes the electron spin decoherence. At
time $\tau\sim1$ (where the decay mainly takes place) the
decoherence is due to the interaction with the nuclei located at
distances of order of the dot radius where the derivative of the
coupling constant is maximal. For longer times, $\tau\gg1$, the
asymptotic behavior is determined either by the interaction with the
nuclei located far from the dot or near the dot center depending on
the Zeeman field value.

First, consider the limit of large Zeeman field ($b\gg A$) and large
times ($\tau\gg1$). The main contribution to the integral comes from
$\kappa\rightarrow1$, \emph{i.e.}, by the interaction with the nuclei
located near the dot center. The asymptotic behavior of
$\tilde\alpha(t)$ is
\begin{equation}
    \tilde{\alpha}(\tau\gg1) = \frac{e^{-iA't/4}e^{i\tau'}}{\pi N}
    \frac{\chi_{0}^{2}(0)}{\sqrt{\left(\chi_{0}^{2}\right)''}}\frac{A^{2}}{b^{2}}
    \frac{(1-i)\sqrt{\pi}}{4i\tau^{3/2}}
    \propto \frac{1}{N}\left(\frac{A}{b}\right)^{2}.
\end{equation}
Here, $(\chi_0^2)''$ denotes the second derivative of $\chi_0^2$
evaluated at $z = 0$. Remember that the correlation function is
given by $C_{0}(t)=\left(1-|\alpha(t)|^{2}\right)/2$ and that
$\alpha(t)$ contains, besides the branch cut contribution
$\tilde\alpha(t)$ (which is a decaying function of time), an
oscillating term due to the pole contribution. Therefore, the
leading term in the full correlator for $\tau\gg 1$ is a constant,
given by the square of the modulus of the pole contribution. The
next higher correction is the product of the pole contribution and
$\tilde\alpha(t)$ as stated above.

Secondly, for the magnetic field turned off ($b=0$) the asymptotic
behavior of the integrand in eq.~(\ref{eq:BranchCut}) for $\tau\gg1$
is determined by $\kappa\ll1$. Taking for instance
$\chi_{0}^{2}(z)/\chi_{0}^{2}(0)=e^{-z^{2}}$, we find
\begin{equation}
    \tilde{\alpha}(\tau\gg1)\propto1/\ln^\frac{3}{2}\tau.
\end{equation}
This result is non-universal. It depends on the form of the electron
wave function at distances larger than the dot size, since the
decoherence is due to the interaction with the nuclei located far
from the dot. The disturbance of the nuclear spins propagates from
the center of the dot outwards.

To summarize this part, we have shown that the decaying part of the
correlator $C_0(t)$ is strongly affected by the magnetic field
strength. However, the characteristic time scale for the onset of
the non-exponential decay is the same for all cases and is given by
$(A/N)^{-1}$. For GaAs quantum dots, this is in the range of
microseconds.

\subsubsection{\label{sec:non_markov} Time evolution with arbitrary nuclear polarization}

In the previous section we discussed the exactly solvable case of an
electron spin coupled to a fully polarized nuclear bath. However,
for an initial nuclear spin configuration which is not completely
polarized, no exact solution exists and standard time-dependent
perturbation theory fails \cite{Khaetskii2002,Khaetskii2003}. In this section,
we review a systematic approach to the electron spin
dynamics in the presence of the Fermi contact hyperfine interaction
eq.~(\ref{eq:HFHamiltonianCoish}). This theory is valid in the limit
of high magnetic fields and for arbitrary spin polarization and
nuclear spin $I$ ($I = 3/2$ in GaAs). Furthermore, the initial state
of the system is $\rho(0)=\rho_{S}(0)\otimes\rho_{I}(0)$ and it is assumed that
$\rho_{I}(0)=|n \rangle\langle n|$, where $|n\rangle$ is an eigenstate of $h_z$.
In practice, this is not usually the case and an appropriate `narrowing' procedure
as the one described in subsubsect.~\ref{sec:narrow} has to be applied.
Throughout this section and in the next one we also assume isotropic
hyperfine couplings (in $d$ dimensions)
\begin{equation}\label{eq:psi_m}
A_k=A_0\exp\left[-\left(\frac{r_k}{l_{0}}\right)^{q}\right],
\end{equation}
where $r_k$ is the radial coordinate of the $k$-th nuclear site.
Note that $\sum_k A_k=A$, and therefore
$A_0\sim A/N$, where  $N$ is defined as the number of nuclei within the radius $l_{0}$ and $A\sim 90~\mu{\rm eV}$ in GaAs.
In the following, we choose the unit of energy such that $A_{0}=2$.

We follow ref.~\cite{Coish2004} and start from the
generalized master equations describing the exact dynamics of the
reduced electron spin polarization
\begin{eqnarray}
&\langle \dot S_z \rangle_t& = N_z(t)-i \int_0^t dt' \Sigma_{zz}(t-t')\langle S_z \rangle_{t'}, \label{eq:Sz_evolution}\\
&\langle \dot S_+ \rangle_t &= i\omega_n \langle S_+ \rangle_t -i \int_0^t dt' \Sigma_{++}(t-t')\langle S_{+} \rangle_{t'}, \label{eq:Sp_evolution}
\end{eqnarray}
where $\omega_n=b'+\langle n|h_z |n\rangle$ and $S_\pm=S_x\pm i S_y$.
It is also useful to introduce the reduced self-energy superoperator, which acts on a generic
operator $O$ as follows \cite{Coish2004}
\begin{equation}\label{eq:Sigma_S_def}
   \hat \Sigma_{S}(t) O = -i\tm{Tr}_{I}\left\{ \hat L e^{-i\hat Q\hat Lt}\hat
   L_{V}[\rho_{I}(0)O]\right\},
\end{equation}
where $\hat L=\hat L_{0}+\hat L_{V}$ is the full Liouvillian superoperator, defined by $\hat
L_{0}O=[H_{0},O]$ and $\hat L_{V}O=[V,O]$, the superoperator $\hat Q$ is defined by $\hat Q
O=O-\rho_{I}(0)\tm{Tr}_{I}O $, and $\tm{Tr}_{I}$ is the partial trace over the nuclear degrees of freedom.
By using eq.~\eqref{eq:Sigma_S_def} above, we can write the kernel of eq.~\eqref{eq:Sp_evolution} as $\Sigma_{++}(t) = \tm{Tr}_S [ S_- \hat \Sigma_{S}(t) S_+]$, where $\tm{Tr}_{S}$ is the partial trace over the electron spin. We also define $\Sigma_{\mu\nu}(t) = \tm{Tr}_S [ \rho_\mu \hat \Sigma_{S}(t) \rho_\nu]$ where $\mu,\nu=\uparrow,\downarrow$ and $\rho_{\uparrow(\downarrow)}=(1\pm \sigma_z)/2$. In terms of these quantities, the kernel of eq.~\eqref{eq:Sz_evolution} is given by $\Sigma_{zz}(t)=\Sigma_{\uparrow\uparrow}(t)-\Sigma_{\uparrow\downarrow}(t)$. Finally, the inhomogeneous term of eq.~\eqref{eq:Sz_evolution} is most simply defined in terms of its Laplace transform $f(s)=\int_{0}^{\infty}dt f(t)e^{-st}$, which gives
$N_{z}(s)=(2is)^{-1}\left[\Sigma_{\uparrow\uparrow}(s)+\Sigma_{\uparrow\downarrow}(s)\right]$.

\begin{figure}
\begin{center}
    \includegraphics[width=.5\textwidth]{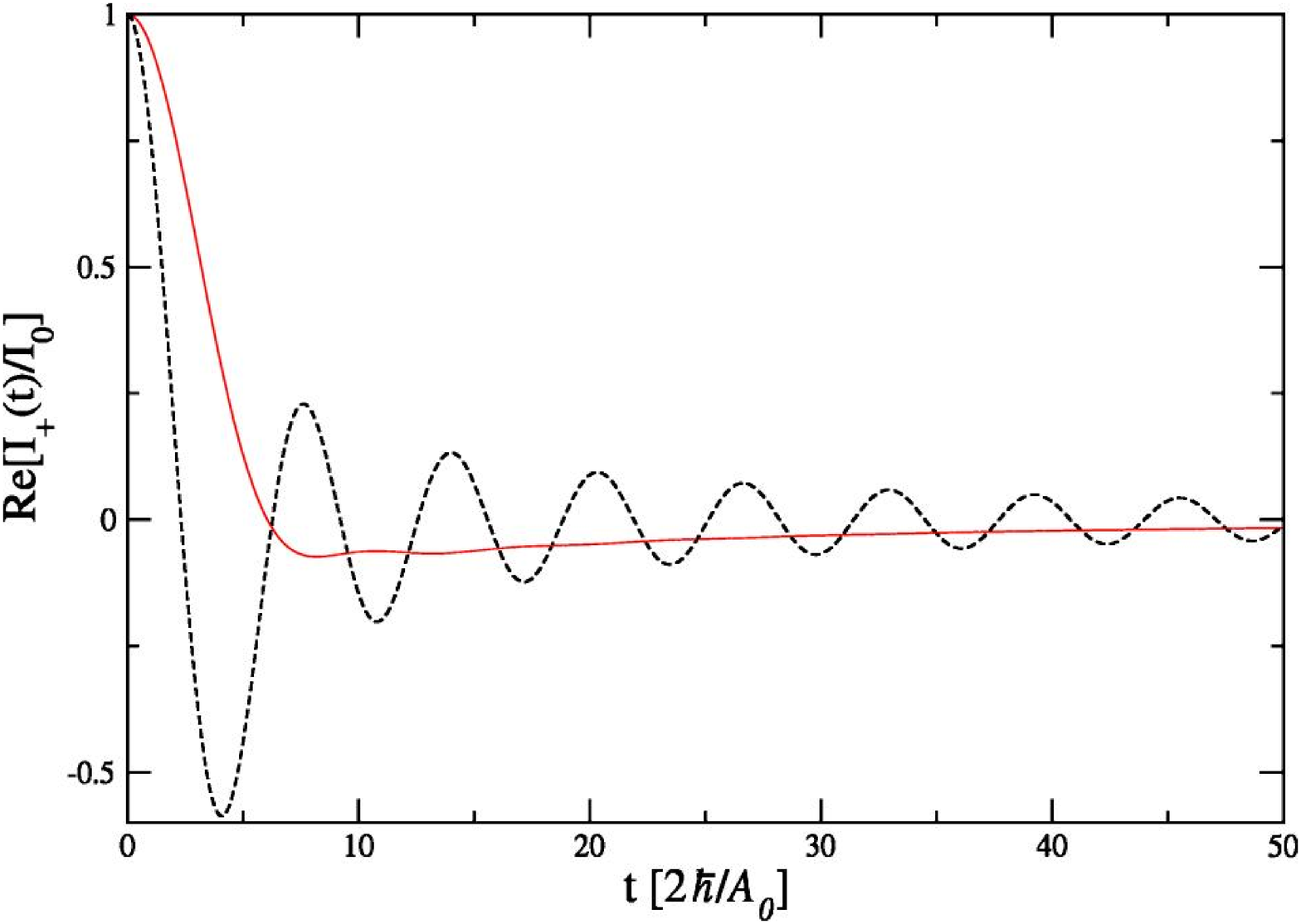}
    \raisebox{0.2cm}{\includegraphics[width=.47\textwidth]{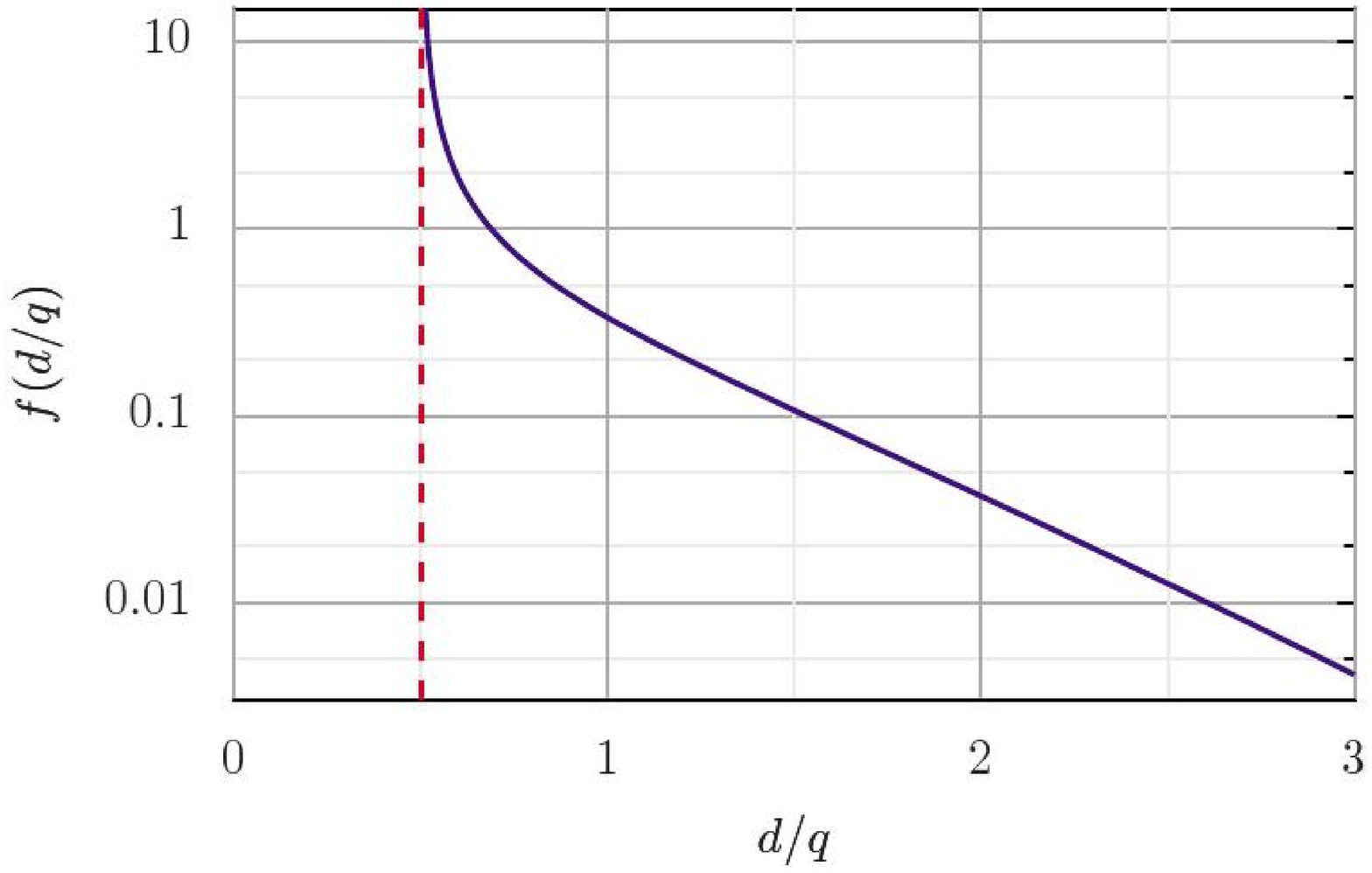}}
    \caption{\label{fig:ReI_plots} Left panel: Plot of ${\rm Re}[I_{+}(t)/I_{+}(0)]$, as obtained from eq.~\eqref{eq:integralI}.
    The solid line was calculated with $d=3$ and $m=1$, which is appropriate for a hydrogen $s$-type envelope wave function. The dashed curve
    refers to a two-dimensional Gaussian envelope wave function ($d=m=2$). The asymptotic behaviors in the two cases are given in
    eqs.~(\ref{eq:I_pm_31}) and (\ref{eq:I_pm_22}) respectively. Right panel: Geometrical factor $f(d/q)$ defined by eq.~(\ref{eq:geom}), with $d$
    being the dimensionality and $q$ is related to the shape of the ground-state wave function (see eq.~\eqref{eq:psi_m}). }
\end{center}
\end{figure}

Solutions to eqs.~\eqref{eq:Sz_evolution}
and~\eqref{eq:Sp_evolution} cannot be obtained exactly and, to
proceed further, an expansion in powers of the flip-flop term $V$
(see eq.~\eqref{eq:HFHamiltonianCoish}) is performed. The
self-energy superoperator \eqref{eq:Sigma_S_def} can be expressed as
a series in terms of $\hat L_V$, where only even powers appear,
\emph{i.e.}, $\Sigma_{S}(t)=\Sigma^{(2)}_{S}(t)+\Sigma^{(4)}_{S}(t)+\ldots\,\,$. At
order $2(k+1)$, the self-energy contribution is suppressed at least
by a factor $\Delta^k$, where $\Delta=N/\omega_n$ becomes small at
large values of the external magnetic field. Therefore, the Born
approximation $\Sigma_{S}(t)=\Sigma^{(2)}_{S}(t)$ is justified in
this limit. The time evolution can be determined from the Laplace transforms of
eqs.~\eqref{eq:Sz_evolution} and~\eqref{eq:Sp_evolution}, which involve the following 
lowest-order self-energies
\begin{eqnarray}
   & \Sigma_{\uparrow\uparrow}^{(2)}(s)
   & =-iNc_{+}\left[I_{+}(s-i\omega_{n})+I_{-}(s+i\omega_{n})\right],\\
   & \Sigma_{\uparrow\downarrow}^{(2)}(s)
   & =iNc_{-}\left[I_{-}(s-i\omega_{n})+I_{+}(s+i\omega_{n})\right],\\
   & \Sigma_{++}^{(2)}(s)
   & =-iN\left[c_{-}I_{+}(s)+c_{+}I_{-}(s)\right],
\end{eqnarray}
where $c_{\pm}=I(I+1)-\left<\right<m(m\pm1)\left>\right>$ and
$\left<\right<F(m)\left>\right>=\sum_{m=-I}^{I}P(m)F(m)$. Here,
$P(m)$ is the probability to find a nuclear spin with
$z$-projection $m$ and $F(m)$ is an arbitrary function (for example
$\left<\right<m\left>\right>=pI$ gives the polarization $p$ of the initial state).
The explicit expression of $I_{\pm}(s)$ reads
\begin{equation}\label{eq:integralI}
    I_{\pm}(s) = \frac{1}{4N}\sum_{k}\frac{A_{k}^{2}}{s\mp iA_{k}/2}
   = \frac{d}{m}\int_{0}^{1}dx\frac{x|\ln x|^\nu}{s\mp ix},
\end{equation}
where $\nu=d/q-1$ and we performed the continuum limit using eq.~\eqref{eq:psi_m}.
Two examples obtained inverting eq.~\eqref{eq:integralI} are plotted in fig.~\ref{fig:ReI_plots} as functions of
time. Note that $t$ is in units of $2/A_0$, which corresponds to a timescale of order $N/A \sim 1~\mu{s}$.

We describe next the case $\Delta \ll 1$, in which the solution has a simple form in terms of the small
parameter $\delta=N/\omega_n^2$. The longitudinal component is given by $\langle S_z\rangle_t=\langle S_z\rangle_\infty +\sigma_z^{\rm dec}(t)$, where
\begin{eqnarray}
&&\langle S_z\rangle_\infty =[1-2 \, \delta \, I_0 (c_++c_-)]\langle S_z\rangle_0+2\, pI \,\delta I_0,\\
&&\sigma_z^{\rm dec}(t)=2\, \delta \, {\rm Re}\left[ \left(C_+ I_- (t)+
 C_- I_+(t)\right) e^{-i\omega_nt}\right].
\end{eqnarray}
Here, $I_0=I_\pm(t=0)$, $c_\pm$ were defined before, and $C_\pm=c_\pm(\langle S_z\rangle_0\pm \frac12)$. The transverse solution is
$\langle S_+\rangle_t=\sigma_+^{\rm osc}(t)+\sigma_+^{\rm dec}(t)$, where
\begin{eqnarray}
&&\sigma_+^{\rm osc}(t) =[1-\delta \, I_0 (c_++c_-)]\langle S_+\rangle_0e^{i\omega_n t},\label{eq:sigma_osc}\\
&&\sigma_+^{\rm dec}(t)=\delta \left[c_+ I_- (t)+
 c_- I_+(t)\right]\langle S_+\rangle_0.\label{eq:sigma_dec}
\end{eqnarray}
In both cases, the solution is expressed as a sum of a term which is constant in absolute value ($\langle S_z\rangle_\infty$ and $\sigma_+^{\rm osc}(t)$
respectively), and a decaying part ($\sigma_z^{\rm dec}(t)$ and $\sigma_+^{\rm dec}(t)$ respectively). The former is almost equal to the initial polarization, $\langle S_z\rangle_0$ or $\langle S_+\rangle_0$, except a correction of order $\delta$. Reintroducing dimensional units (\emph{i.e.}, multiplying by $A_0^2/4$), one obtains that the fraction of polarization which decays is $\delta\sim A^2/\omega_n^2 N$. The time dependence of this decaying contribution is determined by the $I_\pm(t)$ functions. As illustrated in fig.~\ref{fig:ReI_plots}, the $I_\pm(t)$ can have a very different form, depending on the ratio $d/q$. In particular
\begin{eqnarray}
&&I_\pm (t\gg 1)\propto \frac{e^{\pm i t}}{t^{d/q}} \qquad\qquad {\rm for}~\frac{d}{q}<2, \label{eq:I_pm_22} \\
&&I_\pm (t\gg 1)\propto \frac{(\ln t)^{\frac{d}{q}-1}}{t^2}\qquad {\rm for}~\frac{d}{q}\geq 2, \label{eq:I_pm_31}
\end{eqnarray}
where the first expression is determined by the nuclei close to the
origin, and the second one from distant nuclei with a small coupling
$A_k$. These two asymptotic forms are relevant in different physical
situations, \emph{e.g.}, the first one for $s$-type hydrogenic functions of
Si impurities ($d=3$ and $q=1$) and the second for a parabolic
quantum dot in two dimensions ($q=d=2$). Note that in
eq.~\eqref{eq:I_pm_22} the decay has an oscillatory character,
differently from eq.~\eqref{eq:I_pm_31}.

Within the second-order Born approximation scheme discussed here, the time
evolution at generic values of the expansion parameter $\Delta<1$
offers a rich scenario, including terms with exponential decay,
non-exponential decay, and undamped oscillations. 
The same method also reproduces the exact solution
of the previous section, at $p=1$, and can be extended to higher
order in $\hat L_V$. We refer to ref.~\cite{Coish2004} for the
detailed treatment of these topics. As a final comment we note
that the electron spin dynamics, as described at this level of
approximation, is non-Markovian. For large times the electron spin
polarization does not display an exponential decay but is constant
in magnitude. The interesting part of the time dependence is given
by the small decaying functions $\sigma_z^{\rm dec}(t)$ and
$\sigma_+^{\rm dec}(t)$, which also have a non-exponential charater,
see eqs.~\eqref{eq:I_pm_22} and~\eqref{eq:I_pm_31}. This picture is
reexamined in the next section, where a different result (\emph{i.e.},
Markovian decay) is found at very large time scales ($T_2 \sim
\omega_n^2 N /A^3$, see also fig.~\ref{fig:bill_final}).

\subsubsection{\label{sec:markov}Markovian dynamics}

At large magnetic fields, it is in principle possible to carry on
the perturbative treatment of the previous section in terms of $V$ 
to an arbitrary accuracy. Higher order corrections to the self-energy
are multiplied by a small factor $\Delta$ and are expected to be small.
However, it will become clear in the following that these initially small 
perturbative corrections can grow with time and become significant. Therefore, we 
reexamine the problem of the previous section (\emph{i.e.}, the time evolution 
of the spin polarization if the initial state $|n\rangle$ of the 
nuclear bath is an eigenstate of $h_z$) with special attention to the time evolution
at large time scales. We focus on the transverse 
component, generally described by eq.~\eqref{eq:Sp_evolution}. Full 
decay of this polarization component at large times, of order $\omega_n N/A^2\gg N/A$, is found
in ref.~\cite{Liu2005} and an asymptotic dependence $\propto 1/t^2$
is derived in Ref~\cite{Deng2006}. It is therefore necessary to
improve the treatment yielding eqs.~\eqref{eq:sigma_osc}
and~\eqref{eq:sigma_dec} to describe the full decay of the
transverse polarization. We outline here a different method of
solution, which in fact predicts a Markovian spin decay at large
times \cite{Coish2008}.

As a first step, consider the following second order effective Hamiltonian derived from a
Schrieffer-Wolff transformation of eq.~\eqref{eq:HamHF}
\begin{equation}\label{eq:Heff_HF}
    H_{\tm{eff}}=\left(b+h_{z}+\frac{1}{2}\sum_{k\neq l}\frac{A_{k}A_{l}}{\omega}
    I_{k}^{-}I_{l}^{+}\right)S_{z}+b\sum_{k}\gamma_{k}I_{k}^{z},
\end{equation}
where $\omega=b+h_{z}$ and $I_{k}^{+}$ ($I_{k}^{-}$) is the raising (lowering) ladder
operator for the nuclear spins at site $k$. Differently from eq.~\eqref{eq:HamHF}, 
a site-dependent nuclear $g$-factor $g_{I_{k}}$ is taken into account, such that $\gamma_{k}=g_{I_{k}}\mu_{N}/g\mu_{B}$, and the term $H_{dd}$ is omitted.
The Schrieffer-Wolff transformation was performed assuming that corrections to the diagonal part of
$H_{\tm{eff}}$ of order of $\sim A^{2}/Nb$ can be neglected, but
corrections of the same order to the non-diagonal part should be
retained. This is justified by the fact that the bath correlation time
$\tau_{c}\sim N/A\ll Nb/A^{2}$ is small compared to the time scale
on which the diagonal corrections become relevant for $b\gg A$.

In this effective Hamiltonian, the perturbing term is $V_{\rm
eff}=XS_z$, where $X$ is given by $X=1/2\sum_{k\neq
l}A_{k}A_{l}I_{k}^{-}I_{l}^{+}/\omega$. We can now proceed as in the
previous section, and approximate the memory kernel $\Sigma_{++}(t)$
of eq.~\eqref{eq:Sp_evolution} to lowest order in $V_{\rm eff}$.
However, the result obtained in this case is more accurate: the
second order Born approximation in $V_{\rm eff}$, applied to
\eqref{eq:Heff_HF}, retains contributions up to fourth order in the
hyperfine couplings $A_k$. Differently from the previous section,
the Markov approximation can be applied in this case, and leads to a
non-vanishing decay. The decoherence rate
\begin{equation}
    \frac{1}{T_{2}}=\tm{Re}\int_{0}^{\infty}dte^{-i\Delta\omega t}\left<X(t)X(0)\right>
\end{equation}
is given in terms of the non-diagonal part dynamics
$X(t)=e^{-i\omega t}Xe^{i\omega t}$. The
expectation value is taken with respect to the initial `narrowed'
nuclear spin state and $\Delta\omega$ is a shift of the precession frequency 
(\emph{i.e.}, $\langle S_+ \rangle_t =\frac{x_t}{2} e^{i(\omega_n+\Delta\omega)t}$, where $x_t$ is a 
slowly varying envelope) which has to 
be determined self-consistently, see ref.~\cite{Coish2008}. Provided that the initial nuclear spin
polarization is smooth on the scale of the electron wave function,
the matrix elements of operators such as $I_{k}^{\pm}I_{k}^{\mp}$
can be replaced by their average values and the correlator
$\left<X(t)X(0)\right>$ takes the form
\begin{equation}
    \left<X(t)X(0)\right> = \frac{c_+c_-}{4\omega_n^{2}}\sum_{k\neq l}A_{k}^{2}A_{l}^{2}
    e^{-i(A_{k}-A_{l})t},
\end{equation}
where $c_\pm$ are as in the previous section and $\omega_n=\langle n|\omega |n\rangle$ .
We consider now an initially uniform unpolarized state with equal
populations of all nuclear Zeeman levels. This means that
$\left<\right<m\left>\right>=0$ and
$\left<\right<m^{2}\left>\right>=I(I+1)/3$, while $\omega_n=b$.
The final result takes the particularly simple form
\begin{equation}
    \frac{1}{T_{2}} =\pi\left(\frac{I (I+1)A}{3b} \right)^2 f\left(\frac{d}{q}\right)
    \frac{A}{N},
\end{equation}
where
\begin{equation}\label{eq:geom}
    f\left(\frac{d}{q}\right)=\frac{q}{d}\left(\frac{1}{3}\right)^{2\frac{d}{q}-1}
    \frac{\Gamma(2d/q-1)}{(\Gamma(d/q))^{3}},
\end{equation}
which is valid for any dimensionality provided that $d/q>1/2$. The
geometrical factor dependence is shown in the right plot of
fig.~\ref{fig:ReI_plots}. In 3D it may describe a donor impurity, in
2D a lateral gated quantum dot, while in 1D a nanotube or a
nanowire. Note that for $q/d=2$ the decoherence rate diverges so
that the decoherence time tends to zero, which signals the failure of
the Makov approximation. Furthermore, $1/T_{2}\propto I^{4}$ depends
strongly on the nuclear spin so that systems with large
magnetic moment like In (with $I=9/2$) exhibit faster decay than,
for instance, GaAs (with $I=3/2$). Finally, the validity condition
for the Markov approximation, $T_{2}>\tau_{c}\sim N/A$ is satisfied
if $A/b<1$, and coincides with the requirement for the Born
approximation.

In heteronuclear systems the decoherence time is given by the sum of
decoherence rates weighted by the natural abundance of each isotope
$\nu_{i}$ squared, \emph{i.e.}, $1/T_{2}=\sum_{i}\nu_{i}^{2}\Gamma_{i}$, if
interspecies flip-flops are neglected. The quadratic dependence on
isotopic concentration is particularly striking. In spite of the
fact that all isotopes in GaAs have the same nuclear spin and
nominally the same hyperfine coupling constant, the decay is
dominated by intraspecies flip-flops between As spins. This effect
may both explain why only Ga spins have been seen to contribute to
coherent effects in transport experiments through (In/Ga)As quantum
dots \cite{Ono2004} and why polarization seems to be transferred
more efficiently from electron to As -- rather than Ga -- in GaAs
quantum dots \cite{Foletti2008}.

\begin{figure}
\begin{center}
    \includegraphics[width=.6\textwidth]{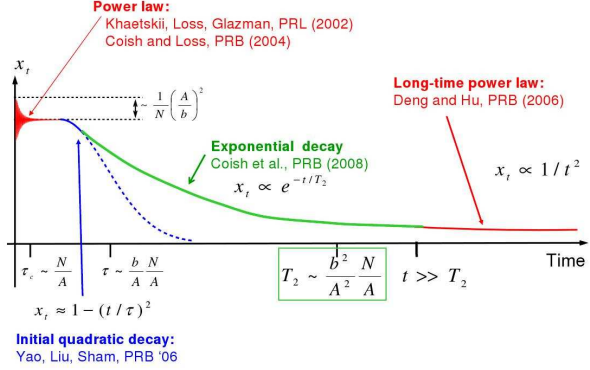}
    \caption{\label{fig:bill_final} Schematic representation of the decay of the transverse polarization.
The  different regimes are discussed in subsubsects.~\ref{sec:non_markov} and~\ref{sec:markov}.\label{fig:geom_hf}}
\end{center}
\end{figure}

Finally, taking into account the discussion of the present section,
we summarize in fig.~\ref{fig:geom_hf} the functional dependence of
the transverse spin polarization. There, four regimes are
schematically depicted : (i) If $t<\tau_{c}\sim N/A$, the power law
decay discussed in the previous section applies
\cite{Khaetskii2002,Khaetskii2003,Coish2004}. (ii) If $\tau_{c}\ll
t\ll \tau\sim bN/A^{2}$, a quadratic correction $\propto
-(t/\tau)^{2}$ is obtained in ref.~\cite{Liu2005}. (iii) If
$\tau<t\ll T_{2}\sim (b/A)^{2}(N/A)$, the exponential decay $\propto
e^{-t/T_{2}}$ discussed in this section is valid \cite{Coish2008}.
(iv) If $t\gg T_{2}$, the long-time power law decay is $\propto
1/t^{2}$ \cite{Deng2006}.

\appendix

\section*{The Schrieffer-Wolff transformation}

The Schrieffer-Wolff transformation is a very useful method to treat
the coupling of external orbital perturbations to the spin states in
the presence of spin-orbit interaction. It is applied to the
theoretical treatment of EDSR in quantum dots (see
Section~\ref{sec:EDSR}) and spin relaxation due to phonons and other
charge fluctuations (see Section~\ref{sec:relax}). We give here some
additional details on the subject (see also ref.~\cite{Golovach2006}).

The Hamiltonian treated here is similar to eq. (\ref{EDSRhmilt}) and
(\ref{eq:phonon_relax_hamiltonian}). More specifically, we consider
$H=H_d+H_Z+H_{SO}+V(t)$, where the unperturbed dot Hamiltonian is
$H_d={\bf p}^2/2m^*+U({\bf r})$ as before (the eigenstates are
$|n,\pm\rangle=\psi_n({\bf r})|\pm \rangle$, with energy
$\epsilon_n$). The only difference is that $V(t)$ refers here to a
general orbital perturbation, whereas in  eq.~(\ref{EDSRhmilt}) and
eq.~(\ref{eq:phonon_relax_hamiltonian}) $V(t)$ specifically is the
external ac electric field and the electron-phonon coupling,
respectively.

Consider now a unitary transformation of $H$ of the form
\begin{equation}
\tilde H= e^S H e^{-S}\simeq H_d+H_Z+V(t)+[S,V(t)]+\ldots
\end{equation}
where $S$ satisfies $[H_d+H_Z,S]=H_{SO}$, such that the spin-orbit
perturbation $H_{SO}$ does not appear in $\tilde H$. The transformed
Hamiltonian $\tilde H$ is accurate to first order in the spin-orbit
coupling, which determines  the last term $[S,V(t)]$. The solution
for $S$ can be formally written in terms of the Liouvillian
superoperators $\hat L_{d(Z)}$  defined by $\hat L_{d(Z)} A
=[H_{d(Z)},A]$, where $A$ is a generic electron operator. Note that
the spin-orbit coupling in the form of eq.~(\ref{HSOrotated}) can be
written as
$H_{SO}=i[H_d,\boldsymbol{\sigma}\cdot\boldsymbol{\xi}]=i\hat L_d
\boldsymbol{\sigma}\cdot\boldsymbol{\xi}$, where
$\boldsymbol{\xi}=(y/\lambda_-,x/\lambda_+,0)$ and $S$ satisfies
\begin{equation}\label{Seq}
(\hat L_d+\hat L_Z)S=i\hat L_d
\boldsymbol{\sigma}\cdot\boldsymbol{\xi}.
\end{equation}
To every order in $\hat L_Z$, eq.~(\ref{Seq}) is solved by
\begin{equation}\label{eq:LZexpansion}
S= (1-\hat P)\sum_{n=0}^\infty  (- \hat L_d^{-1}\hat L_Z)^n ~
i\boldsymbol{\sigma}\cdot\boldsymbol{\xi},
\end{equation}
where $\hat L_d^{-1}=-i\int_0^\infty e^{i(\hat L_d+i\eta)t}{\rm d}t
$, with $\eta=0^+$, and $\hat P$ is such that 
$\langle n \sigma |\hat P A|n' \sigma' \rangle=\langle n \sigma |
A|n' \sigma' \rangle$ if $\epsilon_n=\epsilon_{n'}$ and $\langle n
\sigma |\hat P A|n' \sigma' \rangle=0$ otherwise. Therefore, $\hat P
A$ commutes with $H_d$ and $\hat L_d \hat P=0$. Furthermore, $\hat
L_d$ and $\hat L_Z$ commute (since $H_d$ and $H_Z$ do) and $\hat
L_d^{-1}$ and $\hat P$ commute with them as well, following directly
from their definitions. By making use of these properties, one can
verify that (\ref{Seq}) is satisfied by (\ref{eq:LZexpansion}).

We can now calculate the effective potential $[S,V(t)]$. The first
non-vanishing contribution to the commutator is obtained to first
order in $\hat L_Z$ and can be written in the following form
\begin{equation}
[S,V(t)]=[g\mu_B \boldsymbol{\sigma}\cdot ({\bf B}\times (1-\hat
P)\hat L_d^{-1} \boldsymbol{\xi}),V(t)]+\ldots.
\end{equation}
Finally, we derive an effective spin Hamiltonian by calculating the
expectation value $\langle \psi_0({\bf r})|\tilde H | \psi_0({\bf
r})\rangle$ with respect to the ground state orbital wave function of
$H_d$. Neglecting spin-independent terms we obtain
\begin{equation}\label{eq:Heff_appendix}
H_{\rm eff}=\frac{g\mu_B}{2}{\bf B}\cdot \boldsymbol{\sigma}+g\mu_B
({\bf B}\times \boldsymbol{\Omega}(t))\cdot \boldsymbol{\sigma},
\end{equation}
where the first term is the usual Zeeman coupling and an additional
effective magnetic field is produced by $V(t)$ through
$\boldsymbol{\Omega}(t)$, which is defined by
\begin{equation}\label{eq:Omega_appendix}
\boldsymbol{\Omega}(t)= \langle \psi_0({\bf r})|[(1-\hat P)\hat
L_d^{-1}\boldsymbol{\xi},V(t)]| \psi_0({\bf r})\rangle.
\end{equation}


\end{document}